\documentclass[preprint,usenatbib]{aastex61}
\usepackage[utf8]{inputenc}
\usepackage{amsmath}
\usepackage{mathtools}
\usepackage{amsfonts}
\usepackage{amssymb}
\usepackage{makeidx}
\usepackage{indentfirst}
\usepackage{graphicx}
\usepackage{color}
\usepackage{lmodern}
\usepackage{url}
\usepackage{mathrsfs}
\usepackage{longtable}
\usepackage{multirow}
\usepackage [english]{babel}
\usepackage [autostyle, english = american]{csquotes}
\MakeOuterQuote{"}
\usepackage{chemformula}
\usepackage{textcomp}
\usepackage{aas_macros}
\usepackage{natbib}
\usepackage{tabularx}
\usepackage{comment}
\usepackage{xcolor}
\definecolor{voilet}{RGB}{127,0,255}
\definecolor{vikas1}{RGB}{255,198,0}
\definecolor{vikas2}{RGB}{0, 255, 0}
\definecolor{vikas3}{RGB}{0, 0, 255}
\definecolor{vikas}{RGB}{255,0,127}
\definecolor{KA}{rgb}{0.76, 0.13, 0.28}

\shorttitle{Effect of Metallicity on Hydrogen Dominated Exoplanet Atmosphere}
\shortauthors{Soni \& Acharyya}

\begin{document}
\title{The Effect of Metallicity on the Non-Equilibrium Abundance of Hydrogen Dominated Exoplanet Atmosphere}
\correspondingauthor{Kinsuk Acharyya}
\email{acharyya@prl.res.in}

\author[0000-0001-9273-9694]{Vikas Soni}
\affil{Planetary Sciences Division, Physical Research Laboratory, Ahmedabad, 380009, India}
\affil{Indian Institute of Technology, Gandhinagar, 382355, India}

\author[0000-0002-0603-8777]{Kinsuk Acharyya}
\affil{Planetary Sciences Division, Physical Research Laboratory, Ahmedabad, 380009, India}

\nocollaboration

\newpage
\begin{abstract}
The atmospheric metallicity greatly influences the composition of exoplanet atmospheres. The effect of metallicity 
on the thermochemical equilibrium is well studied, though its effect on the disequilibrium abundance is loosely 
constrained. In this study, we have used the quenching approximation to study the effect of metallicity on the 
quenched abundance for a range of parameters (temperature: 500-2500 K, pressure: 10$^{-4}$-10$^3$ bar, 
metallicity: 0.1-1000 $\times$ solar metallicity). We determine the chemical timescale by finding rate 
limiting steps in a reduced chemical network with a network analysis tool and the thermochemical equilibrium 
abundance. The equilibrium abundance results are similar to the literature. The \ch{CO}, \ch{H2O}, and \ch{CO2} 
abundances increase with metallicity in the parameter range considered. The \ch{CH4} abundance increases with metallicity 
for \ch{CO/CH4} $<$ 1 and is unaffected for \ch{CO/CH4} $>$ 1.  The chemical timescale of CO shows minimal change 
with the metallicity, while the \ch{CH4} chemical timescale is inversely proportional to atmospheric metallicity. 
The quench level of \ch{CO} shifts into the high-pressure region, and the quench level of \ch{CH4} shows complex 
behavior with metallicity. We benchmarked the quenching approximation with the 1D photochemistry-transport model 
for two test exoplanets (GJ 1214 b and HD 189733 b) and found it to be in good agreement. We also found that the 
quenching approximation is 
a powerful tool to constrain atmospheric parameters. We demonstrated this by constraining the metallicity and 
transport strength for the test exoplanets HR 8799 b, HD 189733 b, GJ 436 b, and WASP-39 b.
\end{abstract}

\section{Introduction}
Since the first discovery of exoplanets in 1992 \citep{Wolszczan1992}, the database for exoplanets has been 
expanding rapidly. Characterizing the atmospheres of exoplanets ranging from Earth-like to gas-giants like 
Jupiter and finding atmospheric compositions and potential bio-signatures have become the fundamental facets 
of modern-day planetary sciences \citep{Madhusudhan2016, Sing2018, Fortney2018, Helling2019, Madhusudhan2019}. 
Molecules such as \ch{CO}, \ch{CO2}, \ch{CH4}, \ch{H2O}, \ch{NH3}, \ch{HCN}, which are the building blocks for 
more complex organic molecules and major reservoirs of elemental N, C, and O have already been detected in 
exoplanet atmospheres \citep{Madhusudhan2019}. Observational spectra have been used in the retrieval 
models to constrain atmospheric abundances of these atmospheres \citep{Madhusudhan2009, Madhusudhan2011, 
Line2012, Kreidberg2014, Ranjan2014, Haynes2015, Benneke2015, Stevenson2017, Rajpurohit2020}. However, the 
abundances and detailed understanding of the formation of these molecules are most often model-dependent. 
The parameter space for reproducing the atmospheric composition of an exoplanet is degenerate due to 
uncertainty in factors such as the effects of interior processes, the T-P profile of the atmosphere, the 
efficiency of atmospheric dynamics, the presence of clouds and hazes, bulk metallicity, elemental ratios, 
among many others \citep{Seager2010, Madhusudhan2016}.

The atmospheric elemental abundances, i.e., the presence of elements such as H, O, C, N, S, etc., vary 
significantly from one planet to another. If we look at the gas giants of the solar system, a large variation 
can be seen, and the general trend is that the atmospheric metallicity increases with decreasing mass. 
Metallicities for Jupiter, Saturn, Neptune, and Uranus are 3.3 - 5.5, 9.5 - 10.3, 71 - 100, and 67 - 111 $\times$ 
solar metallicity, respectively, although large uncertainties exist in the abundances of individual elements 
\citep{Atreya2018}. For exoplanets, at the present sensitivity level, there is a large uncertainty, although 
several studies were made to discern metallicities from high-precision spectral analysis. Exoplanet metallicities 
vary from sub-solar (e.g., HAT-P-7 b, \citealp{Mansfield2018}), to comparable to solar (e.g, WASP-43 b, 
\citealp{Stevenson2017}), to moderately enriched (e.g., WASP-103 b, \citealp{Kreidberg2018}; WASP-127 b, 
\citealp{spake2021}; WASP-121 b, \citealp{Mikal-Evans2019}; WASP-39 b, \citealp{Ahrer2022}) to significantly 
enriched (e.g., GJ 436 b, \citealp{Knutson2014}). Thus, even though only a small number of exoplanets are studied, 
the metallicity space appears to be diverse and can range between 0.1 to more than 1000 times solar metallicity 
\citep{Wakeford2020}. 

The wide variation in atmospheric elemental abundances for exoplanets significantly affect the 
atmospheric compositions. However, due to the considerable variation in the various physical quantities 
such as pressure, temperature, transport strength, irradiation, and surface gravity, the signature of 
enriched/decreased metallicity vary significantly from one planet to the other. For example, the T-P 
profile of the planet decides the efficiency of a chemical reaction and the formation of chemical 
species. Besides various disequilibrium processes such as transport and photochemistry greatly alter 
the equilibrium compositions \citep{Seager2010, Madhusudhan2016}. Studies have been undertaken to understand how 
varying elemental abundances affect the atmospheric properties and the equilibrium composition \citep{Line2011, 
Madhusudhan2011, Moses2011, Madhusudhan2012, Moses2013, Moses2013a, Venot2014, Zahnle2014, Charnay2015, Heng2016, 
Drummond2018}. These studies are primarily divided into two classes;  in one, the carbon to oxygen ratio (C/O) is 
varying \citep{Moses2013, Moses2013a}, and in the other, the elemental abundances of O, C, N, etc., are increased/decreased 
compared to the solar values \citep{Venot2014, Drummond2018, Charnay2015}. \cite{Moses2013} studied the effect of 
metallicity from 0.1 to $10^4 \times$ solar metallicity for solar C/O ratio on the equilibrium abundance. The focus 
of their study is at a pressure of 100 mbar, since the infrared photospheres are present at this pressure. The study 
by \cite{Moses2013} shows that the abundances of \ch{H2O}, CO, and \ch{CO2} 
increase with metallicity, although the extent of this increase is different. In the CO dominant region, \ch{CO2} increases 
as a square of the metallicity and becomes the dominant species at very high metallicity. In contrast, \ch{H2O} 
and CO increase linearly with metallicity. At very high metallicity, the \ch{H2O} abundance decreases as the 
overall availability of H decreases. Besides, \ch{CH4} increases with metallicity at low temperature (where it is 
stable). Similar findings have been reported by \cite{Line2011, Zahnle2014, Madhusudhan2011, Moses2013a, Venot2014}. 
In another study \cite{Charnay2015} found that for the H-rich atmospheres, the verticle mixing will increase as metallicity 
increases. Therefore, higher metallicity is required for a planet like GJ 1214 b to form condensate clouds in the upper atmosphere. 
Similarly, \cite{Drummond2018} used a coupled 3D hydrodynamic model to solve for the chemical equilibrium abundances 
and found that as the metallicity is increased, significant changes in the dynamical and thermal structure affect the phase curve. Also, the opacity effect is the dominant mechanism in altering the circulation and the thermal 
structure compared to the mean molecular weight and heat chemistry. 

These general studies on the metallicity effect are limited to the equilibrium abundance or targeted for 
particular exoplanets where the disequilibrium processes are also considered. For example, the effect of metallicity 
on the disequilibrium abundance has been studied for the exoplanets GJ 436 b \citep{Line2011, Madhusudhan2011, Moses2013, Venot2014}; 
HD 18733 b and HD 209458 b \citep{Moses2011, Dash2022}; WASP-12 b, CoRoT-2 b, XO-1 b, and HD 189733 b \citep{Moses2013a}. However, 
the general behaviour of the metallicity on the nonequilibrium abundance is mostly unexplored.

In the presence of transport, the chemical species mix in different parts of the atmosphere. The abundance of species 
deviates from the equilibrium abundance if the timescale of the chemical reaction becomes larger than the timescale 
of the transport. The disequilibrium abundance due to transport is frozen at the equilibrium abundance at the quenched 
level. The quenching approximation is the most straightforward approach to estimate the disequilibrium abundance. In the 
quenching approximation, the quench level is defined at a pressure level where the chemical conversion timescale is equal 
to the vertical mixing timescale of the atmosphere. Below the quench level, the chemical timescale dominates over the transport 
timescale, and the atmospheric composition remains in the chemical equilibrium. Above the quench level, the abundance of 
the species is frozen with the equilibrium abundance at the quench level. However, one should apply the quenching 
approximation with caution, as the mixing length of the atmosphere cannot be calculated using the first principle approach. 
Also, the quenched molecules control the abundance of other molecules \citep{Tsai2017, Moses2011, Smith1998, Cooper2006}.

Thus the quenching approximation can be used as a powerful tool to understand the general behaviour of the effect of metallicities 
on non-equilibrium abundances. In this work, we aim to find the verticle mixing and the chemical timescales over the temperature range between 
500 and 2500 K, the pressure range between 0.1 mbar and 1 kbar, and metallicity between 0.1 - 1000 .

$\times$ solar metallicity (-1 to 3 [Fe/H]). 
Then we use the timescale data to find the temperature and pressure space curve where these two timescales are equal 
(quenched curve) and study the quenched curve for various mixing strengths in the metallicity parameter space. In \S 2, 
the photochemistry-transport model is discussed, while in \S \ref{S-DisEq}, the disequilibrium chemistry and the quenching 
approximation are discussed. The model results for both the equilibrium and disequilibrium chemistry are given in \S \ref{Results}. 
A comparison between the quenching approximation 
results with the photochemistry-transport model is made in \S 5. Discussions are presented in \S 6 and the concluding remarks are made in \S 7.

\section{1D Atmospheric Model} \label{sec:model}
We have developed a 1D photochemistry-transport model to study the atmospheric composition of exoplanets.  In this section the main features of the model are briefly mentioned while a detailed description with benchmarking is provided in the Appendix. The model 
solves the mass continuity equation for each species as follows:

\begin{equation}
\frac{\partial n_i}{\partial t} = P_i - n_i L_i - \frac{\partial \phi_i}{\partial z}, \label{Eq:1}
\end{equation}
where $n_i$ is the number density of the $i$-th species, $P_i$ and $n_iL_i$ are the species production and loss rates due to
chemical kinetics and photochemical reactions, and $\phi_i$ is the transport flux.
The model couples the chemical kinetics and photochemistry with the atmospheric transport and solves equation \ref{Eq:1} numerically 
for each layer and each species in the network till the convergence criteria take place. The transport processes
include Eddy diffusion and molecular diffusion. The Eddy diffusion coefficient is taken as a parameter, 
and the molecular diffusion coefficient is calculated by the description given in \cite{Chapman1991}. The upper 
atmosphere of exoplanets is exposed to stellar radiation, which produces free radicals that react with 
the other species and change the chemical composition. To find the flux in each atmospheric layer, we use the 
two-stream approximation of radiative transfer following \cite{Heng2014}. In this method, we find the radiation flux, 
which is going out and coming in from both the boundaries of an atmospheric layer. This flux is converted into photon 
density and used to find the photochemical rates. We use the scattering and absorption cross-section data for active 
photochemical species in the network to find the optical depth.

The chemical network contains all the important species up to six hydrogen, two carbon, two nitrogen, and three oxygen 
atoms, and single atoms for He, Na, Mg, Si, Cl, Ar, K, Ti, and Fe. The chemical network is constructed by including chemical 
reactions involving H, C, N, and O from \cite{Tsai2017, Tsai2018}, while for the other elements, we have taken chemical reactions 
from \cite{Rimmer2016}.  For this work we have used a reduced network of 52 species involving H-C-N-O which are connected 
by 600 chemical reactions.

To study the effect of metallicity on the nonequilibrium abundance of a hydrogen-dominated atmosphere, we make a grid in 
the metallicity, temperature, and pressure space. The pressure range is $10^{-4}$ to $10^{3}$ bar, the temperature range is 500 
to 2500 K, and the metallicity range is 0.1 to 1000 $\times$ solar metallicity. In this study, we change the metallicity 
relative to the solar photospheric elemental abundance. The change in metallicity corresponds to an increase or decrease in the 
heavy elemental abundance (elements other than H and He) with respect to the solar metallicity by a common factor. The solar 
photospheric metallicity is taken from \cite{Lodders2009}. The range of bulk abundance of elements in the present 
study are C/H = $2.77\times 10^{-5} - 2.77\times 10^{-1}$, N/H = $8.18\times 10^{-6} - 8.18\times 10^{-2}$ , and
O/H = $6.06 \times 10^{-5} - 6.06\times 10^{-1}$.
 
We ran two sets of models. In the first set, we found the disequilibrium abundances in the presence of transport using quenching 
approximation, for which we followed the method given in \cite{Tsai2018}. Then we used the method to discuss the effect 
of metallicity on the disequilibrium abundance and demonstrated how it can be used to constrain the metallicity and transport 
parameters, for which we used four test exoplanets, namely HR 8799 b, HD 189733 b, GJ 436 b, and WASP-39 b. In the second set, we ran the 
photochemistry-transport model for HD 189733 b and GJ 1214 b. We compared the output of the photochemistry-transport model with 
the quenching approximation which demonstrates the effectiveness of the quenching approximation.

\section {Disequilibrium abundance and quenching approximation}\label{S-DisEq}

\subsection{Disequilibrium Abundance}
The presence of disequilibrium processes such as transport alter the atmospheric composition from its equilibrium 
abundance. In the transport-dominated region, the timescale for a chemical reaction becomes smaller than 
the timescale for atmospheric mixing; therefore, species transported from other layers cannot follow the equilibrium 
composition and the abundance deviates from equilibrium abundance, which has a large impact on the atmospheric composition. 
\cite{Prinn1977} used transport mixing to explain the high abundance of \ch{CO} in the troposphere of Jupiter. Subsequently, 
a few different methods have been used to find the atmospheric composition in the presence of transport, and a comparison of these 
methods is given in \cite{Smith1998}. Among these methods, the quenching approximation is the most straightforward 
approach to constraining the disequilibrium abundance. It provides a way to determine disequilibrium abundance without 
running a full kinetics/transport model.

\subsection{Quenching Approximation}\label{S-Qu}
The quench level is defined at a pressure level where the chemical conversion timescale is equal to the vertical mixing 
timescale of the atmosphere. Below the quench level, the chemical timescale dominates over the transport timescale, 
and the atmospheric composition remains in the chemical equilibrium. Above the quench level, the abundance of species 
is frozen with the equilibrium abundance at the quench level. The quenching approximation as described by \cite{Smith1998} 
can give an accurate result within 15\% of the kinetics/transport model \citep{Moses2011}. This approximation is routinely used 
in various studies to constrain the disequilibrium abundance of exoplanets \citep{Madhusudhan2011, Line2010, Moses2011, 
Visscher2012, Zahnle2014, Tsai2017, Tsai2018, Fortney2020}. We have used the quenching approximation to determine the abundance of 
\ch{CO}, \ch{CH4}, \ch{H2O}, and \ch{CO2} in order to study the effect of metallicity.

\subsubsection{Vertical Mixing Timescale} \label{sec:tau_mix}
The vertical mixing timescale $\tau_{mix}$ can be computed using the mixing length theory, and is given by the 
following equation:

\begin{equation} 
\tau_{mix} = L^2 / K_{zz},
\end{equation}
where $L$ is the mixing length scale of the atmosphere and $K_{zz}$ is the Eddy diffusion coefficient 
\citep{Visscher2011, Heng2017}. Since the Eddy diffusion 
coefficient has a large uncertainty, it is treated as a free parameter. The mixing length scale cannot be computed from the 
first principle, and a simple approximation is to take the pressure scale height as the mixing length. However, \cite{Smith1998} 
found that the mixing length can be $L \approx 0.1-1 \times \text{pressure scale height}$, which leads to $\tau_{mix} = 
(\eta H)^2 / K_{zz}$, where $\eta \in[0.1,1]$ and the exact value of $\eta$ depends upon the rate of change of chemical 
timescale with height. The pressure scale height $H = \frac{K_b  T}{\mu g}$, where $T$, $g$ and $\mu$ are temperature, 
surface gravity, and mean molecular mass of the atmosphere, respectively. It is to be noted that metallicity changes the 
elemental composition, thereby changing the value of $\mu$. When metallicity increases from 0.1 to 1000 $\times$ solar metallicity, $\mu$ 
changes by one order of magnitude.

\subsubsection{Chemical Timescale}\label{sec:CTS}
In chemical equilibrium, the abundance of the chemical species does not change with time and 
the chemical reactions take place in a way that the production and loss rate of any species are balanced, such that
 
\begin{equation}
\frac{dn_{\text{EQ}}}{dt} = P_{EQ} - n_{\text{EQ}} L_{EQ}= 0.
\end{equation} 
Here, $n_{\text{EQ}}$, $P_{\text{EQ}}$ and $n_{\text{EQ}} L_{\text{EQ}}$ are the number density, 
production and loss rate respectively and $L_{\text{EQ}}$ is independent of $n_{\text{EQ}}$. 
When the physical conditions such as temperature and pressure of the system change, or the species are transported 
into other regions of the atmosphere, the chemical abundance deviates from chemical equilibrium, and the species 
are converted among themselves to restore chemical equilibrium. This conversion of species takes place through 
several chains of chemical reactions, which are called conversion schemes. The timescale of reactions in 
a conversion scheme can vary significantly; therefore, the conversion timescale is computed from the slowest reaction in 
the fastest conversion scheme, known as the rate-limiting step (RLS). The following relation gives the timescale of the 
conversion of species $a$ into $b$:  

\begin{equation}
\tau_{a\rightarrow b} = \frac{[a]}{\text{Rate of RLS}_{a\rightarrow b}}.
\end{equation}
Here, [a] is the abundance of species $a$, and $\text{RLS}_{a\rightarrow b}$ is the rate-limiting step in the conversion 
of $a$ into $b$. In a chemical network, a particular species is involved in several reactions; as a result, there are many conversion 
pathways between two species. The number of these pathways increases exponentially as the number of reactions in the 
network increases. However, in a chemical network, only a few conversion schemes are important, as most of the conversion 
schemes are significantly slower than the fastest conversion scheme. 

\begin{figure}[h]
	\centering
	\includegraphics[trim={5cm 2cm 7cm 6cm},clip,width=1\textwidth]{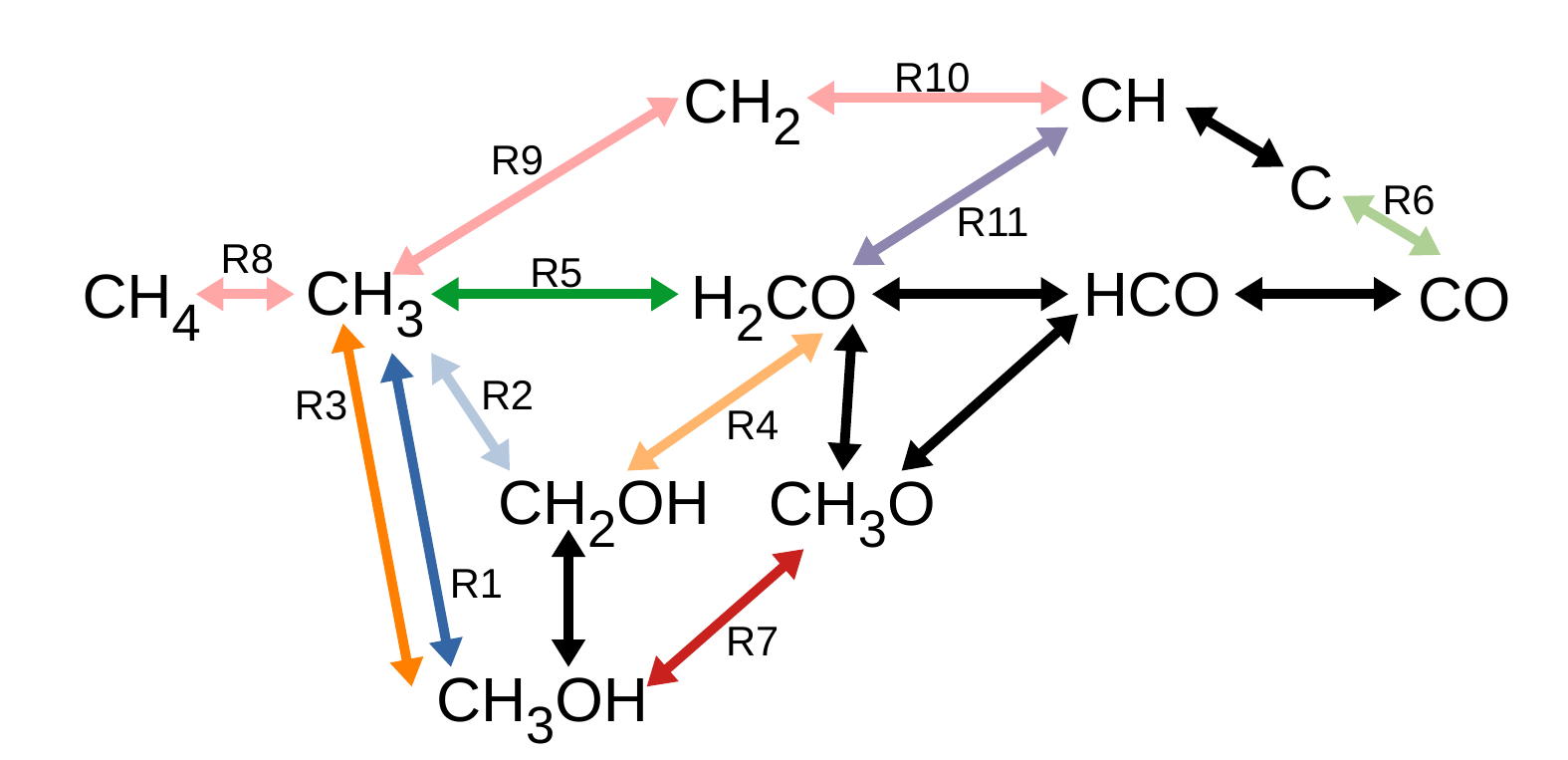}
	\caption{Major chemical pathways between \ch{CH4} and \ch{CO} for a hydrogen dominated atmosphere. The reactant is either \ch{H} 
or \ch{H2} from right to left, but the reactant is \ch{H2O} in the \ch{CH} to \ch{H2CO} conversion. The colored arrows other than black are the 
rate limiting steps at different temperature-pressure values (see also Figure \ref{fig:RLS_area_CO}).}\label{fig:conversion_scheme}
\end{figure}

The timescale of the reactions is a function of several parameters, especially temperature and pressure. Therefore, as these 
parameters change, the RLS also changes, i.e., a reaction that is the RLS for a given set of parameters may not be the RLS as 
these parameters get altered. Thus the rate-limiting reaction can change in the parameter space. In order to understand the effect 
of metallicity on the RLS in the conversion between species, a chemical network analyzing tool is built. This tool takes a 
chemical network and finds all the possible conversion schemes between the species. It then finds the fastest conversion scheme 
and its slowest reaction (RLS). We adopted a reduced chemical network following \cite{Tsai2018} for this task and used their methodology. 
In Figure \ref{fig:conversion_scheme}, the conversion path between \ch{CH4} and \ch{CO} for the parameter range is shown. 
The different colored arrows (other than black) represent the RLS reactions. The RLS in the conversion of $\ch{CH4}\rightleftarrows\ch{CO}$ for our parameter range of temperature, pressure and metallicity is listed below. 

\begin{align*}
\ch{CH3} + \ch{H2O} &\rightleftarrows \ch{CH3OH} + \ch{H} &R1\\
\ch{OH} + \ch{CH3} &\rightleftarrows \ch{CH2OH} + \ch{H} &R2\\
\ch{OH} + \ch{CH3} + \ch{M} &\rightleftarrows \ch{CH3OH} + \ch{M} &R3\\
\ch{CH2OH} + \ch{M} &\rightleftarrows \ch{H} + \ch{H2CO} + \ch{M} &R4\\
\ch{CH3} + \ch{O} &\rightleftarrows \ch{H2CO} + \ch{H} &R5\\
\ch{OH} + \ch{C} &\rightleftarrows \ch{CO} + \ch{H} &R6\\
\ch{CH3OH} + \ch{H} &\rightleftarrows \ch{CH3O} + \ch{H2} &R7\\
\ch{H} + \ch{CH4} &\rightleftarrows \ch{CH3} + \ch{H2} &R8\\
\ch{H} + \ch{CH3} &\rightleftarrows \ch{CH2} + \ch{H2} &R9\\
\ch{H} + \ch{CH2} &\rightleftarrows \ch{CH} + \ch{H2} &R10\\
\ch{H2CO} + \ch{H} &\rightarrow \ch{CH} + \ch{H2O} &R11
\end{align*}

In the interconversion of \ch{CH4}$\rightleftarrows$\ch{CO}, the chemical timescales are as follows \citep{Tsai2018}:

\begin{align}
\tau_{\ch{CH4}} &= \frac{[\ch{CH4}]}{\text{Reaction rate of RLS}} + \tau_{\ch{H2}} \times \frac{3[\ch{CO}]}{\ch{H2}} \label{eq:main_3}\\ 
\tau_{\ch{CO}} &= \frac{[\ch{CO}]}{\text{Reaction rate of RLS}} + \tau_{\ch{H2}} \times \frac{3[\ch{CO}]}{\ch{H2}} \label{eq:main_4}.
\end{align}
Here, $\tau_{\ch{CH_4}}$ and $\tau_{\ch{CO}}$ are the timescales of conversion of $\ch{CH4}\rightarrow\ch{CO}$ and 
$\ch{CO}\rightarrow\ch{CH4}$ respectively. $[\ch{CH4}]$, $[\ch{CO}]$  and  $[\ch{H2}]$ are the number densities of \ch{CH4}, 
\ch{CO}, and \ch{H2} respectively. The interconversion timescale of $\ch{H2}\rightleftarrows\ch{H}$ is $\tau_{\ch{H2}}$, and 
`Reaction rate of RLS' is the rate of RLS relevant for the desired temperature-pressure and metallicity values. The first 
term in equations \ref{eq:main_3} and \ref{eq:main_4} is related to the timescale of the RLS. The second term is related to the interconversion of $\ch{H}\rightleftarrows\ch{H2}$, which is 
necessitated because during the conversion of $\ch{CH4}\rightleftarrows\ch{CO}$, the $\ch{H}\rightleftarrows\ch{H2}$ interconversion also 
occurs. Reconversion of $\ch{H2}\rightarrow\ch{H}$ or $\ch{H} \rightarrow\ch{H2}$ is also required to achieve steady-state.

\section{Results}\label{Results}

\subsection{Equilibrium Results}\label{S-Equi}
We first re-visit the effect of metallicity on the chemical equilibrium abundances as 
described in \cite{Moses2013a}. The chemical equilibrium abundance depends on the local temperature, pressure, and elemental 
abundance. It can be calculated by minimizing the Gibbs free energy of the system or by solving the mass continuity equation 
with the relevant chemical network for the parameter range. The chemical equilibrium composition can be used as the atmospheric 
composition when the chemical timescale is faster than the timescale of any other physical process; therefore, at 
the high-pressure and high-temperature region, the atmospheric composition can be approximated as the chemical equilibrium 
composition.

\begin{figure}[htb!]
	\centering
	\includegraphics[width=0.8\textwidth]{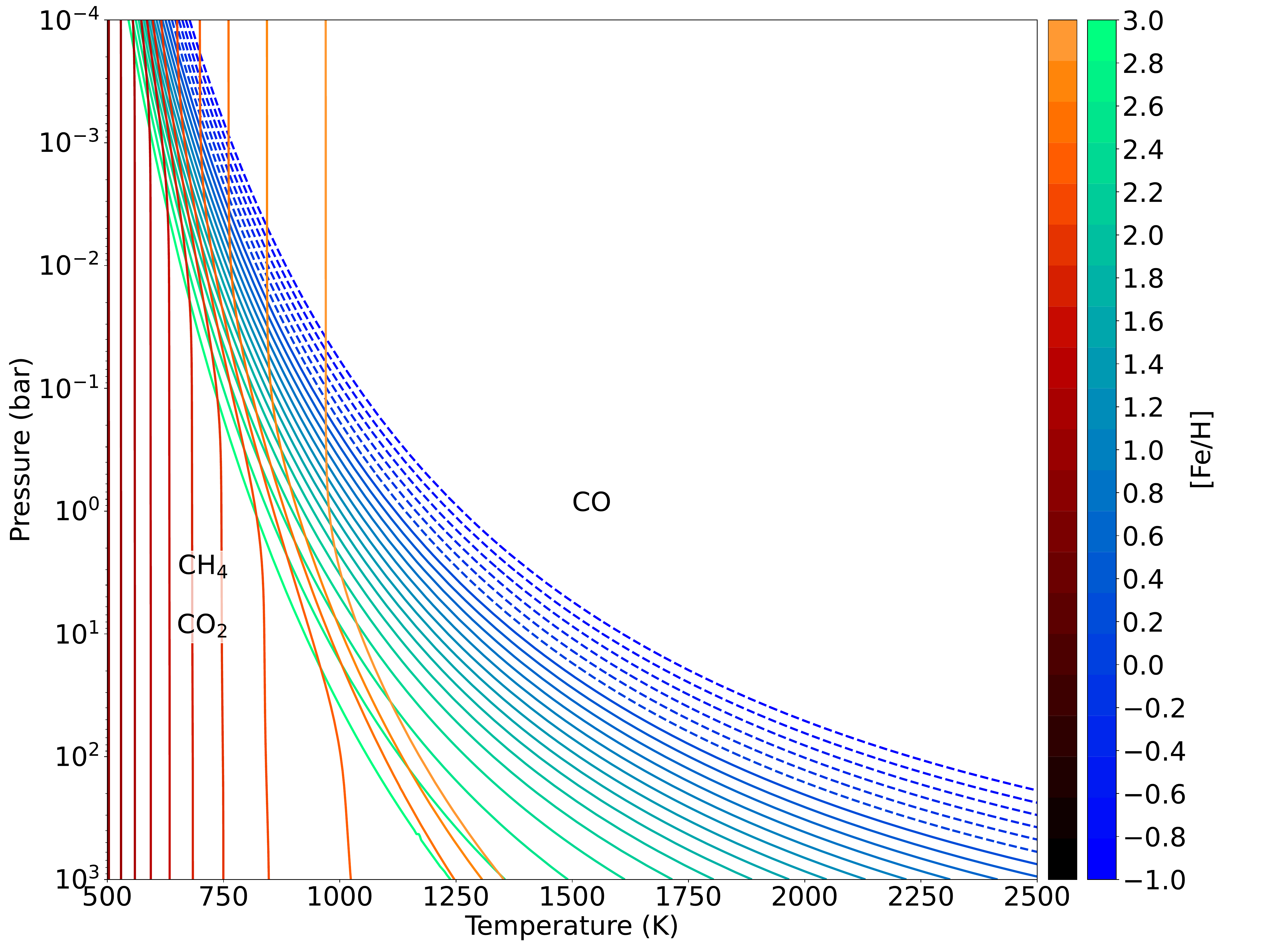}
	\caption{The contours of $\ch{CH4}/\ch{CO}=1$ and $\ch{CO2}/\ch{CO}=1$ for assorted metallicity values are shown. 
The cyan to blue lines delimitate the region where \ch{CH4}(left) or \ch{CO} (right) is the dominant species, and the orange to 
black lines delimitate the region where \ch{CO2} (left) or CO (right) is the dominant species. The metallicity value 
in the colour bar corresponds to the respective colour line in the plot; for example, the rightmost curve for $\ch{CH4}/\ch{CO}=1$  
represents [Fe/H]= -1, the second from right line corresponds to -0.8, and so on. The dashed lines are for subsolar metallicity, and the
solid lines are for supersolar metallicity. The subsolar metallicity lines for $\ch{CO2}/\ch{CO}=1$ are outside the parameter space.
}\label{fig:1}
\end{figure}

An increase in metallicity increases the bulk abundance of \ch{C} and \ch{O}, thereby increasing the abundance of molecules that use them. 
The effect of metallicity on the abundance can vary based on the use of heavy elements by molecules. For example, the metallicity dependence of 
molecules like CO, which consists of only one atom of each element, can be different from molecules like \ch{CO2}, which requires 
two atoms of an element. Similarly, the metallicity dependence of molecules comprising one heavy element and hydrogen atoms, such 
as \ch{CH4} and \ch{H2O}, will 
be different since the bulk abundance of hydrogen decreases at a slow rate with increasing metallicity; therefore, 
these molecules, can have a complex relationship with the metallicity. Although, the general behavior is that the abundance of the 
molecules with heavy elements increases with increasing metallicity, and the increase is more profound when the molecules consist of 
multiple heavy elements 
\citep{Moses2013a}. In Figure \ref{fig:1}, the boundaries of \ch{CH4-CO} and \ch{CO2-CO} for 20 different metallicities are plotted. 
The boundary of \ch{CH4-CO} shifts towards low temperature with metallicity, whereas the boundary of \ch{CO2-CO} shifts towards high 
temperature. At 100 mbar pressure, the \ch{CH4-CO} boundary shifts from 800 K to 600 K as the metallicity increases from 0.1 to 1000 $\times$ solar 
metallicity. This behavior is similar as that obtained by \cite{Moses2013}. 

\begin{figure}[htb!]
	\centering
	\includegraphics[width=1\textwidth]{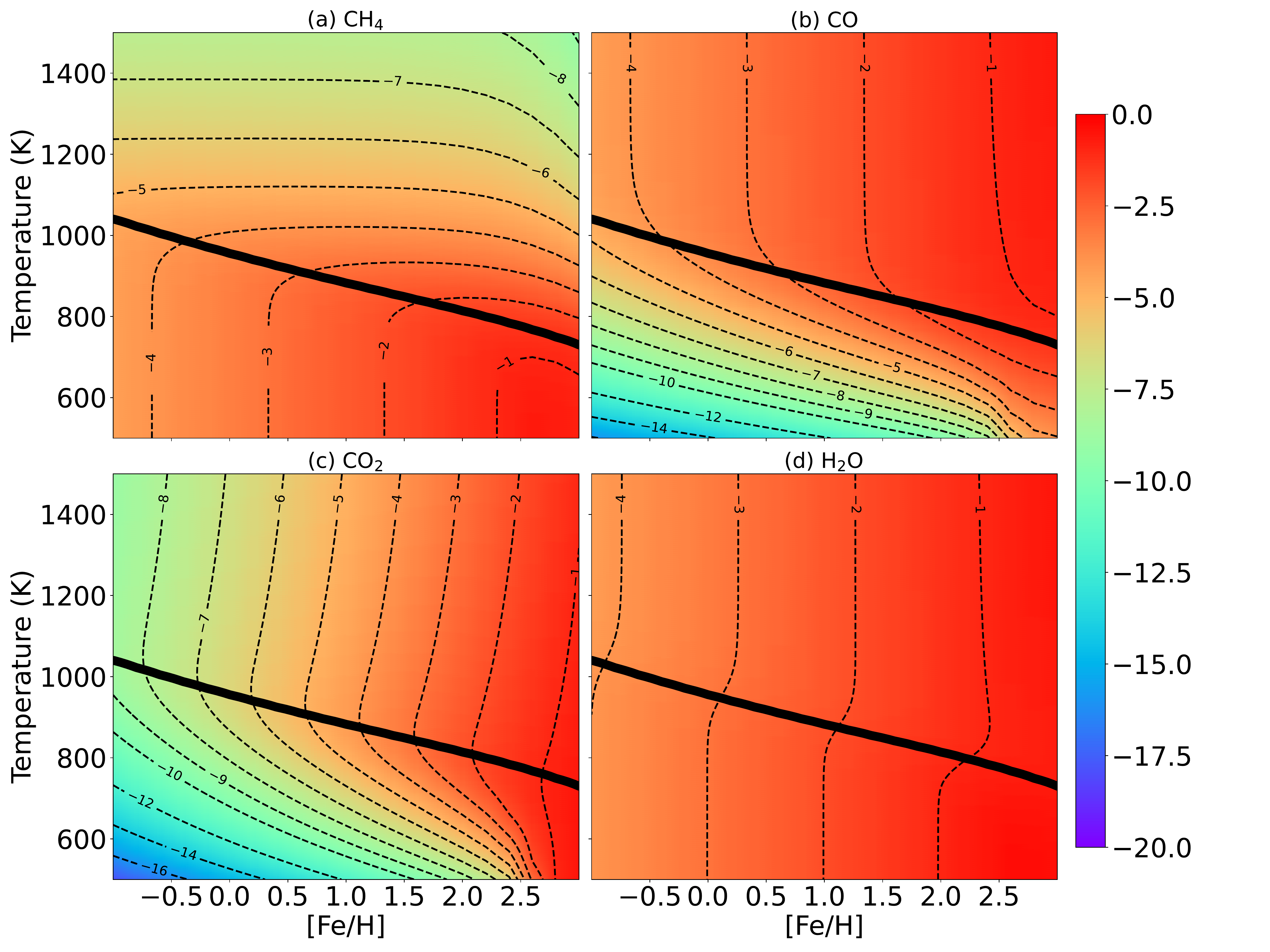}	
	\caption{Variation of the equilibrium mole fraction for (a) \ch{CH4}, (b) \ch{CO}, (c) \ch{CO2},  
and (d) \ch{H2O} with metallicity and temperature at 100 mbar pressure is shown. The solid black line is the 
equal-abundance curve of \ch{CH4} and \ch{CO}, and the black dashed lines are the constant contours of the mole fraction (in log10 scale) of the
respective gas-phase species.}\label{fig:equi}	
\end{figure}

In Figure \ref{fig:equi}, we have plotted the constant contour of the mixing ratio of \ch{CH4}, \ch{CO}, \ch{CO2}, and \ch{H2O} 
in the metallicity and temperature parameter space at 100 mbar pressure. We have also over-plotted an equal-abundance curve of \ch{CH4-CO}, 
which divides Figure \ref{fig:equi} into two regions. In region A, \ch{CH4} dominates over \ch{CO} (below the solid black line), while in region B, 
\ch{CO} dominates over \ch{CH4} (above the solid black line). The \ch{CH4} abundance is increased in region A, whereas in 
region B, the \ch{CH4} abundance is constant with metallicity. However, the decrease of bulk H at significantly high metallicity 
leads to a decrease in the \ch{CH4} abundance. The \ch{CO} abundance increases with metallicity for all temperature-pressure ranges. 
In region B, the \ch{CO} abundance increases linearly with metallicity, and in region A, the increment rate increases as the temperature 
decreases. Available oxygen to form \ch{H2O} and \ch{CO} is high in region A as compared to region B because in region B, atomic oxygen is locked in \ch{CO}. For the solar \ch{C/O} ratio, the \ch{O} reservoir 
is almost twice that of the C reservoir; as a result, the \ch{H2O} abundance is linearly increased with metallicity \citep{Moses2013}, 
since \ch{H2O} is among the most stable molecules in most of the temperature and pressure ranges. At [Fe/H] $>$ 3, the increment of \ch{H2O} is 
restricted by the availability of the hydrogen reservoir. Also, in region B, the contours of the constant mole fraction of \ch{H2O} 
are at a relatively higher metallicity than in region A. The \ch{CO2} abundance mostly follows the \ch{CO} abundance in region A. 
However, in region B, the \ch{CO2} abundance increases with the square of metallicity. 

\begin{figure}[htb!]
	\centering
	\includegraphics[trim={0cm 0cm 0cm 0cm},clip,width=1\textwidth]{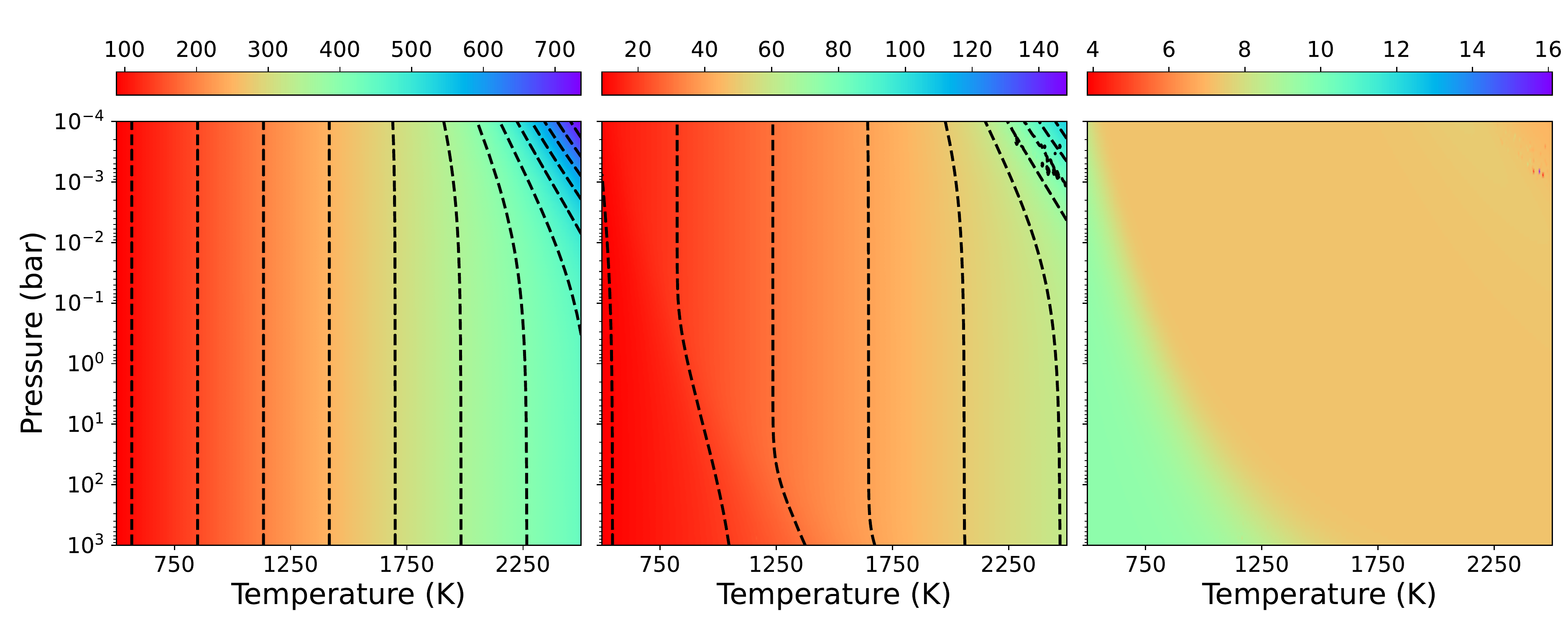}
	\caption{The constant contour lines for the scale height for two different metallicity values are plotted. 
		The left panel is for 0.1 $\times$ solar metallicity and the middle panel is for 1000 $\times$ solar metallicity. The right panel is the ratio of 
		0.1 and 1000 $\times$ solar metallicity.}
	\label{fig:scale_height}
\end{figure}

To understand the effect of metallicity on the pressure scale height of the atmosphere, contours 
of constant scale height in the temperature and pressure parameter space for 0.1 and 1000 $\times$ solar metallicity and g = 2000 cm s$^{-2}$ are shown in figure \ref{fig:scale_height}. It can be seen that the scale height has decreased only by one order of magnitude as the metallicity increases by four orders of magnitude. The change in the scale height is corespondes to the increase of the mean molecular mass of the atmosphere with increasing metallicity.  The effect of metallicity on $\tau_{mix}$ comes from the dependence of mixing length on the pressure scale height (see section \ref{sec:tau_mix}). $\tau_{mix}$ has a squared dependence on the pressure scale height and the constant $\eta$ can have a range of one order of magnitude; it makes $\tau_{mix}$ vary by four orders of magnitude as the metallicity increased from 0.1 to 1000 $\times$ solar.

\subsection{RLS and Disequilibrium Abundance}\label{sec:RLS}
\begin{figure}[b!]
	\centering
	\includegraphics[trim={0cm 0cm 8cm 0cm},clip,width=1\textwidth]{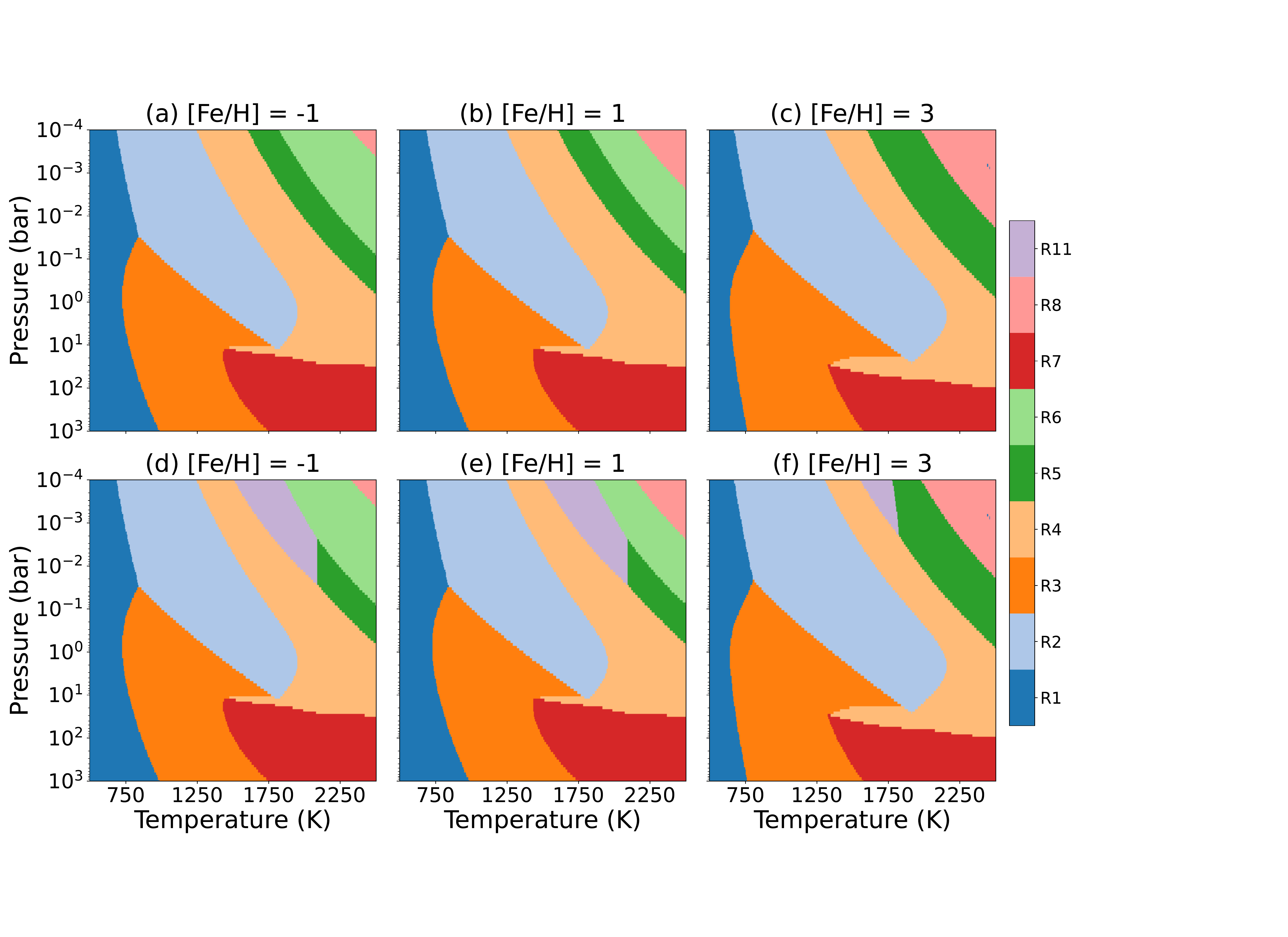}
	\caption{The temperature-pressure range of different rate-limiting steps is shown for three different metallicities. Each 
		color corresponds to the different RLS in Figure \ref{fig:conversion_scheme}. Panels (a)-(c) and (d)-(f) respectively represent the RLS 
		parameter space for the conversion of $\ch{CH4}\rightarrow\ch{CO}$ ($\tau_{\ch{CH4}}$) and $\ch{CO}\rightarrow\ch{CH4}$ ($\tau_{\ch{CO}}$).}\label{fig:RLS_area_CO}	
\end{figure}

Figure \ref{fig:RLS_area_CO} shows the effect of metallicity on the parameter space of the RLS reactions. Each coloured area 
represents the RLS reaction as shown in Figure \ref{fig:conversion_scheme}. Maximum change occurs in the high-temperature and 
low-pressure region, where H is more stable than \ch{H2}. In this region, \ch{CH4} is converted into \ch{CO} via 
progressive dehydrogenation and subsequent oxidation. The different steps of dehydrogenation become RLS as the metallicity increases. 
The pressure-temperature range for most of the RLS shift towards low temperature as the metallicity increases. It can be seen that R11 is only present 
in the $\ch{CO}\rightarrow\ch{CH4}$ conversion but not in the $\ch{CH4}\rightarrow\ch{CO}$ conversion. Barring R11, all the RLS are common in both $\ch{CO}\rightarrow\ch{CH4}$ and $\ch{CH4}\rightarrow\ch{CO}$ conversions. We find that 
the conversion pathways are the same as \cite{Tsai2018}, though the RLS in the high metallicity region are slightly different. 

\cite{Tsai2018} found three distinct pathways for the $\ch{CH4}\rightarrow\ch{CO}$ conversion. First, \ch{CH4} is converted into \ch{CO} through progressive 
dehydrogenation of \ch{CH4} into C and further oxidization of C to form \ch{CO}. Second, \ch{H2CO} forms by oxidation of \ch{CH3} 
and converts into \ch{CO} through \ch{HCO}. Finally, third, the conversion path can differ depending on the temperature and pressure, 
and molecules like \ch{CH2OH}, \ch{CH3OH}, and \ch{CH3O} are intermediate molecules. We divided these RLS into two types based on 
the effect of metallicity on the RLS. In the first type, \ch{H} or \ch{H2} come as the reactants of the RLS. In the second, both 
the reactants of the RLS are heavy molecules. The abundance of \ch{H} decreases with metallicity, but the relative change is slow. 
Thus the timescale of RLS in the first type increase slowly and the average abundance of heavy molecules increases linearly with metallicity \citep{Zahnle2014}; in the second type, it is inversely proportional to metallicity.

\subsection{Effect of Metallicity on the Conversion Timescale}

\subsubsection{\ch{CH4} and \ch{CO} Conversion Timescale}
\begin{figure}[h!]
	\centering
	\includegraphics[width=1\textwidth]{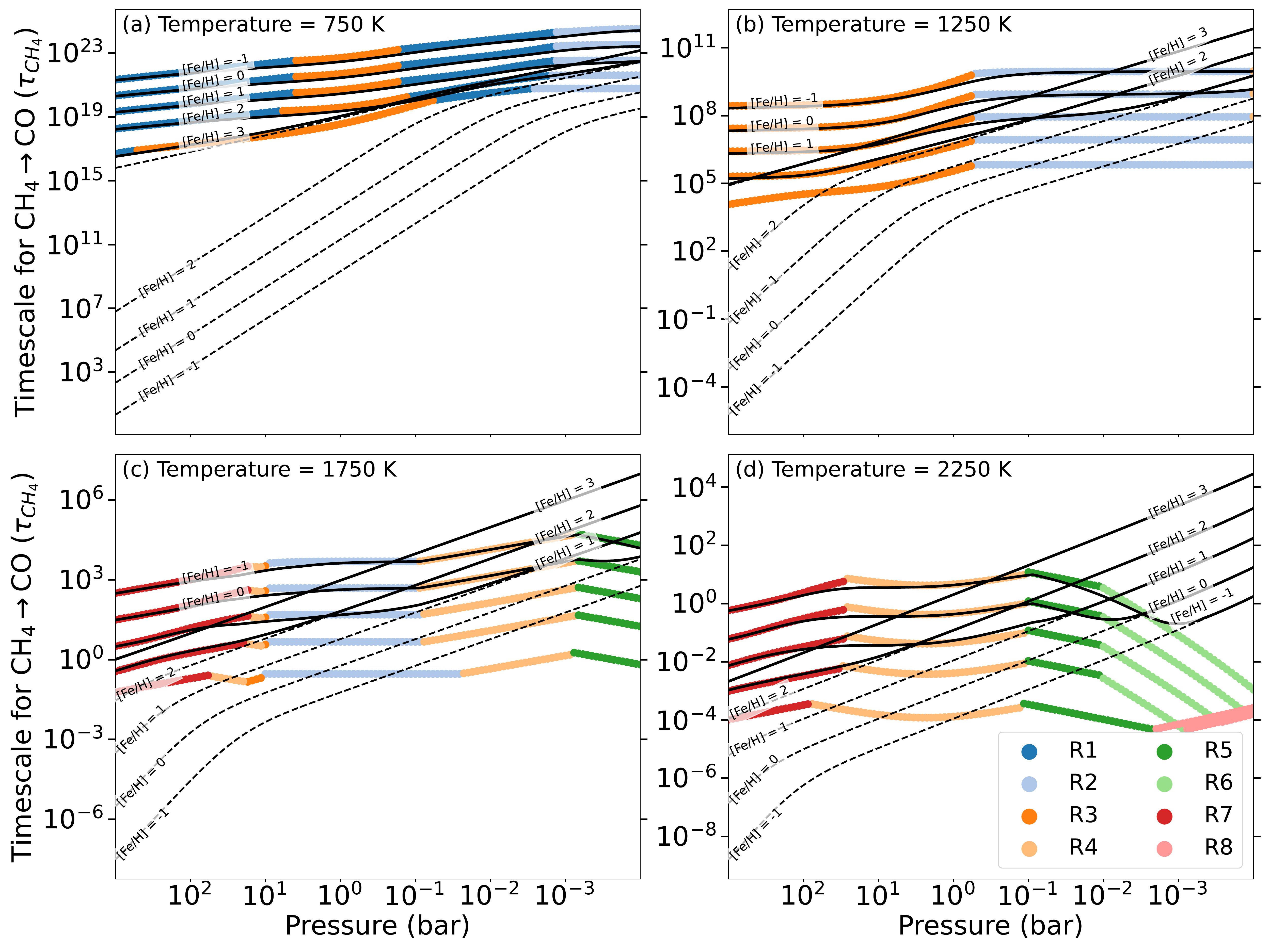}
	\caption{The timescale of conversion of $\ch{CH4}\rightarrow\ch{CO}$ ($\tau_{\ch{CH4}}$) for four different temperatures 
		(750 K, 1250 K, 1750 K, and 2250 K) with five different metallicities (0.1, 1, 10, 100, 1000 $\times$ solar metallicity are shown). 
		The colored lines represent the timescale of RLS in the corresponding temperature-pressure value, and the black dashed lines represent  $\tau_{\ch{H2}} \times \frac{3[\ch{CO}]}{\ch{H2}}$. The solid black line is $\tau_{\ch{CH4}}$, labeled with the respective 
		metallicity. }	\label{fig:time_scale_CH4}
\end{figure}

\begin{figure}[h!]
	\centering
	\includegraphics[width=1\textwidth]{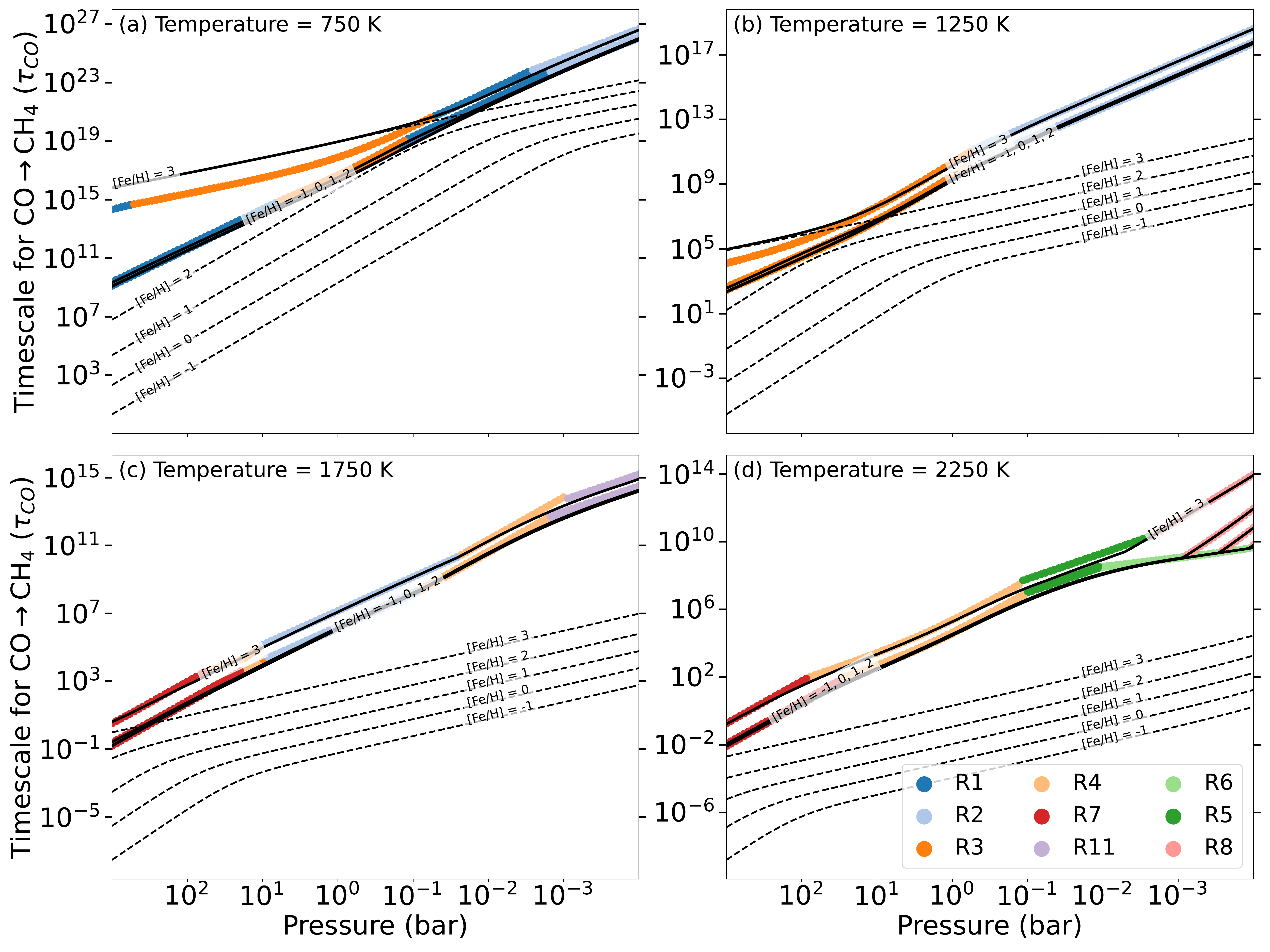}
	\caption{Same as Figure \ref{fig:time_scale_CH4}, but for $\ch{CO}\rightarrow\ch{CH4}$ conversion. } 
	\label{fig:time_scale_CO}
\end{figure}

In Figures \ref{fig:time_scale_CH4} and \ref{fig:time_scale_CO}, we have shown $\tau_{\ch{CH_4}}$ (Equation \ref{eq:main_3}) 
and $\tau_{\ch{CO}}$ (Equation \ref{eq:main_4}) for four different temperatures (750 K, 1250 K, 1750 K, and 2250 K) and five different 
metallicities (0.1, 1, 10, 100, and 1000 [Fe/H]). The contributions from each term in Equations \ref{eq:main_3} and \ref{eq:main_4} 
i.e., the RLS (the first term in Equations \ref{eq:main_3} and \ref{eq:main_4}) and the $\ch{H}\rightleftarrows\ch{H2}$ conversion term 
(the second term in Equations \ref{eq:main_3} and \ref{eq:main_4}) are shown by the colored lines and the black dashed lines, respectively. 
The rates of increase of these two terms in Equations \ref{eq:main_3} and \ref{eq:main_4} are different, as is evident from Figures 
\ref{fig:time_scale_CH4} and  \ref{fig:time_scale_CO}. In addition, their relative strengths are a strong function of temperature.

It can be seen that for the conversion of $\ch{CH4}\rightarrow\ch{CO}$, the relative importance of the $\ch{H}\rightleftarrows\ch{H2}$ 
conversion term changes profoundly over the parameter space. The magnitude of $\ch{H}\rightleftarrows\ch{H2}$ 
conversion term in equation \ref{eq:main_4} and \ref{eq:main_3} increases with increasing metallicity and decreases with 
increasing temperature and pressure. Although, the contribution for this term is lesser for the $\ch{CO}\rightarrow\ch{CH4}$ conversion. It can be seen that 
with an increase in metallicity and temperature, the point where the slope of $\ch{H}\rightleftarrows\ch{H2}$ conversion term changes in Figures \ref{fig:time_scale_CH4} and 
\ref{fig:time_scale_CO} moves towards higher pressure region, indicating that it is moving from \ch{CH4} to \ch{CO} dominated region.

For 750 K (panel (a) in Figures \ref{fig:time_scale_CH4} and \ref{fig:time_scale_CO}), as the metallicity increases from 0.1 to 
1000 $\times$ solar, the contribution of the first term in Equation \ref{eq:main_3} is decreased by five orders of magnitude. 
In contrast, in Equation \ref{eq:main_4}, the first term is increased by more than five orders of magnitude. The second 
term in Equation \ref{eq:main_3} dominates in the low-pressure region; for a given pressure, it increases with metallicity. 
The extent of increase is more in the high-pressure region when compared to the low-pressure region; in the high-pressure region, it increases by 15 orders of magnitude (Figure~\ref{fig:time_scale_CO}) when metallicity increases from 0.1 to 
1000 $\times$ solar value. Also, in this region, \ch{CH4} dominates over CO; with the gradual reduction in pressure, we move 
from the \ch{CH4} to \ch{CO} dominated region.

Panels (b), (c), and (d) in Figures \ref{fig:time_scale_CH4} and \ref{fig:time_scale_CO} show the pressure variation of timescales for the 
temperatures 1250 K, 1750 K, and 2250 K, respectively. It is evident that as the temperature increases, the magnitude of these two 
terms starts to decrease. The metallicity dependence of $\tau_{\ch{CH_4}}$ at a particular temperature remains qualitatively similar. The contribution of the 
first term in Equation \ref{eq:main_3} decreases by nearly five orders of magnitude with increasing metallicity for a given 
pressure. However, for $\tau_{\ch{CO}}$ (Equation \ref{eq:main_4}), the first term increases only slightly with an increase in 
metallicity (Figure~\ref{fig:time_scale_CO}) because for $\tau_{\ch{CO}}$ the RLS belongs to the first type (see section \ref{sec:RLS}). For $\tau_{\ch{CH_4}}$, with the increase of temperature, the contribution of 
the first term initially flattens out (1250 K) and then starts to decrease in the low-pressure region. Also, for $\tau_{\ch{CH_4}}$, 
the second term increases with metallicity by around 7 to 10 orders of magnitude for a given pressure, and it starts to dominate 
for a pressure of $\approx$ 10$^{-3}$ bar but at different metallicities: [Fe/H] $> 2$, [Fe/H] $> 1$ and [Fe/H] $> -1$ for temperatures 1250 K, 
1750 K and 2250 K, respectively. Thus, the $\ch{H2}\rightleftarrows\ch{H}$ conversion term plays a more prominent role, especially in the low-pressure region for 
$\tau_{\ch{CH_4}}$, which makes the metallicity dependence complex. In comparison, the RLS term dominates most parts of the parameter 
space for $\tau_{\ch{CO}}$, thereby diminishing the contribution of the second term greatly.

\begin{figure}[t!]
	\centering
	\includegraphics[trim={1.5cm 1cm 3cm 2cm},clip, width=0.4\textwidth]{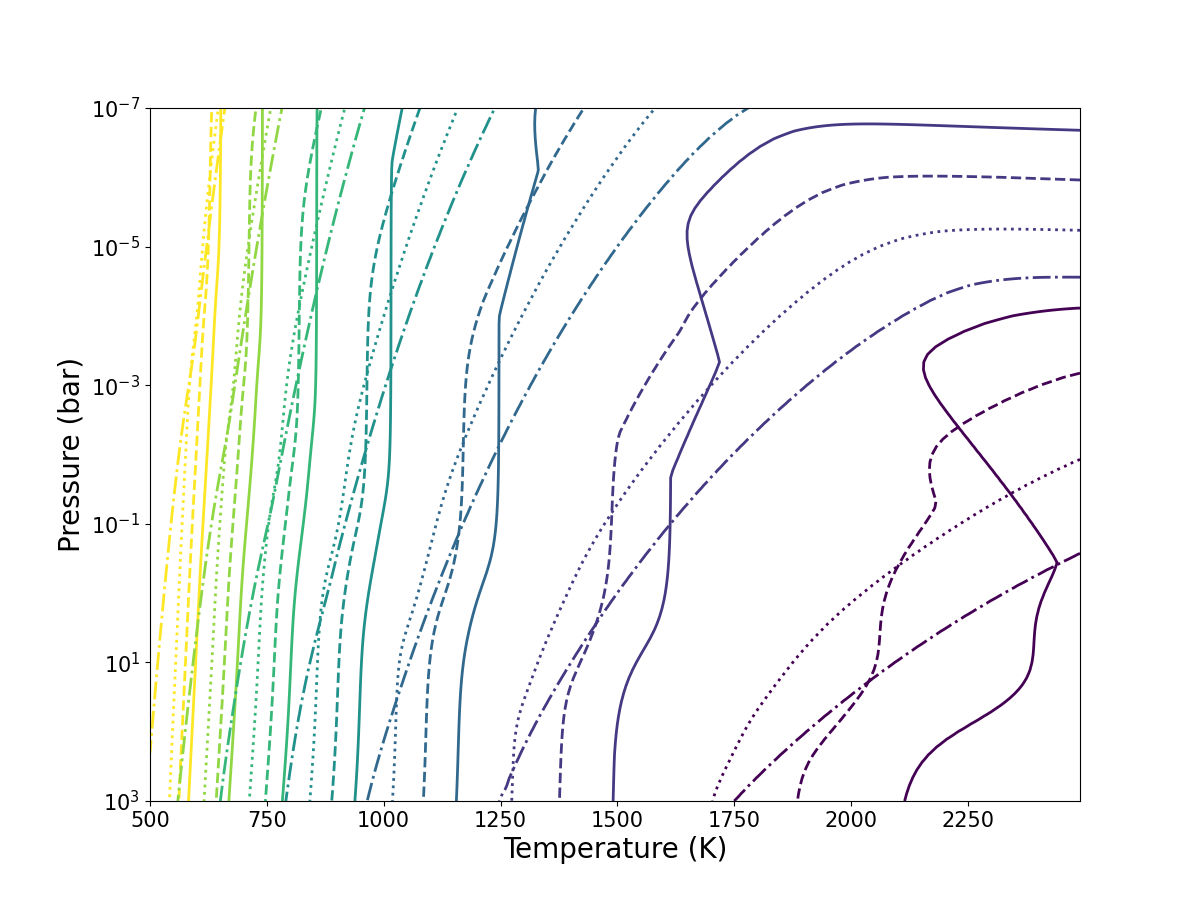}
	\includegraphics[trim={1.5cm 1cm 3cm 2cm},clip,width=0.4\textwidth]{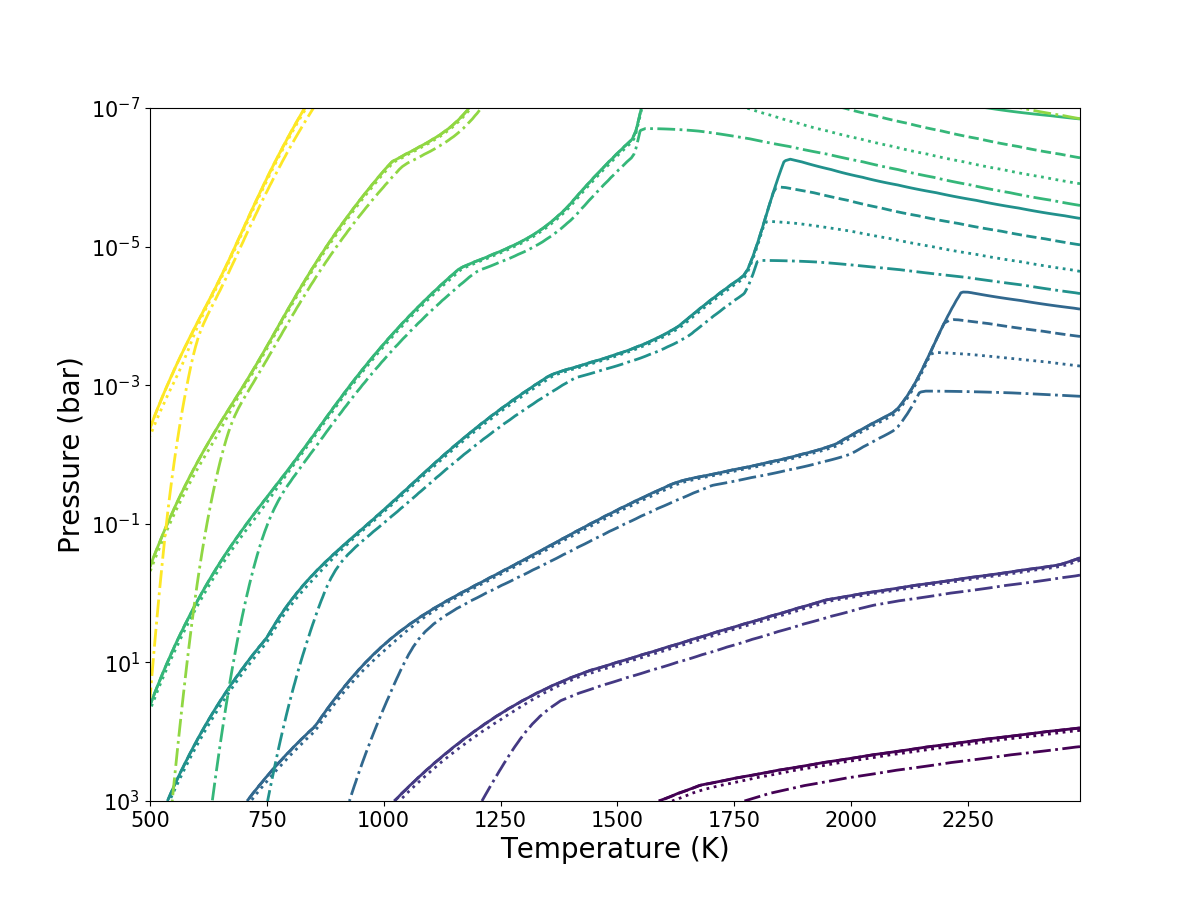}
	\caption{The constant contour line of $\tau_{\ch{CH4}}$ (left) and $\tau_{\ch{CO}}$ (right) are plotted for four different 
metallicities (0.1, 1, 10, 1000 $\times$ solar metallicity). The solid, dashed, dotted, and dotted-dashed lines show 0.1, 1, 10, and  1000 $\times$ solar 
metallicity respectively. The coloured lines from blue to yellow represent the constant contour of timescales in log10 scale for 0, 5, 10, 15, 20, 25, 
and 30.}	\label{fig:contour_chemical_time_scale}
\end{figure}

In Figure \ref{fig:contour_chemical_time_scale}, we have shown the constant contour lines of $\tau_{\ch{CH4}}$ and $\tau_{\ch{CO}}$ in the
temperature and pressure parameter space for four different metallicities. The solid, dashed, dotted, and dotted-dashed lines are for 
1000, 10, 1, and 0.1 $\times$ solar metallicity, respectively. The lines from blue to yellow are the constant contour lines of $10^{0}$ to 
$10^{30}$ s. Both the conversion timescales decrease with increasing temperature and pressure. However, the dependence of 
the timescale on the metallicity is complex and has different dependencies at different pressure levels. The contours of constant $\tau_
{\ch{CH4}}$ move towards the high-pressure region with an increment in metallicity in the region where the second term in Equation \ref{eq:main_3} 
dominates, since this term increases with metallicity. In the region where RLS is dominant, $\tau_
{\ch{CH4}}$ is proportional to the metallicity and the contours of constant $\tau_{\ch{CH4}}$ move towards low temperature as the 
metallicity increases (see also Figures \ref{fig:time_scale_CH4} and \ref{fig:time_scale_CO}). In the region where R11 acts as the RLS, the contour of constant $\tau_{\ch{CO}}$ shifts 
towards the high-pressure 
region with an increment in metallicity. For the 
intermediate pressure, temperature, and metallicity regions, $\tau_{\ch{CO}}$ is unaffected by the metallicity. For [Fe/H] $<2$, 
$T <1250$ K and $P>10^{-3}$ bar, the contour of constant $\tau_{\ch{CO}}$ shifts towards the high-pressure region.

\begin{figure}[b!]
	\centering
	\includegraphics[trim={0cm 0cm 9cm 1.5cm},clip, width=1\textwidth]{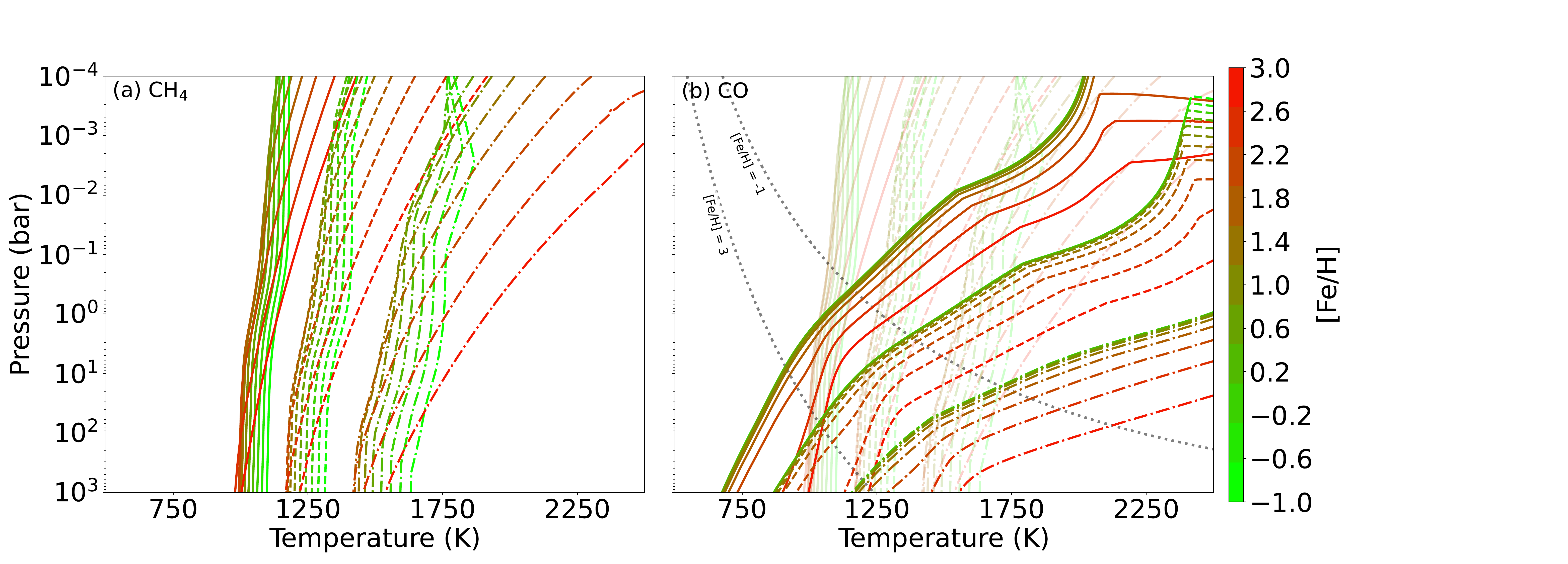}
	\caption{
The contour of (a) $\tau_{\ch{CH4}} /\tau_{mix} = 1$ and (b) $\tau_{\ch{CO}} /\tau_{mix} = 1$ for assorted parameters are shown. 
The colour bar (each value represents one metallicity contour) represents the metallicity value for the plotted quenched 
curve. The three sets of lines are for different values of $K_{zz}$ coefficients; solid, dashed, and dashed-dotted lines are for  
$K_{zz}$ = $10^4$, $10^8$, and $10^{12}$ cm$^2$ s$^{-1}$ respectively. In panel (b), the quenched curves of \ch{CH4} are shown 
in faded lines with the same colour and the equal-abundance curves for 0.1 ([Fe/H] = -1) and 1000 
([Fe/H] = 3) times solar metallicity are also shown.
}
	\label{quenching_contiur_plot}
\end{figure}

\subsection{Effect of Metallicity on the Quench Level}
To find the transport-induced disequilibrium abundances, we have used vertical mixing and chemical timescales and found the 
atmosphere's quench level (see section \ref{S-Qu}). At the quench level, the vertical mixing timescale is comparable to the 
chemical timescale. Since the chemical timescale is different for different species, the species quench at different atmospheric 
heights. Figures \ref{quenching_contiur_plot} (a) and (b) show the contour lines on which the vertical mixing and 
chemical conversion timescales are equal for \ch{CH4} and CO, respectively. The three sets of lines are for different values 
of $K_{zz}$ coefficients (solid: $10^4$ cm$^2$ s$^{-1}$, dotted: $10^8$ cm$^2$ s$^{-1}$, and dashed-dotted: $10^{12}$ cm$^2$ s$^{-1}$). 
Each set of lines runs from green to red (shown in the color bar), representing eleven different metallicity values. The figure shows 
both species' general behavior of vertical mixing and chemical timescales. As the temperature and pressure increase, 
the chemical timescale decreases; therefore, for quenching to happen, the corresponding vertical mixing timescale also has to reduce, 
which will require a higher $K_{zz}$ value. It is evident from Figure \ref{quenching_contiur_plot} that as we move towards the high pressure 
and temperature regions, $K_{zz}$ increases. For each $K_{zz}$ value, the quenched curve for eleven different metallicities are shown. 
The vertical mixing timescale decreases with increasing metallicity, leading to 
shifting of the quenched curve in the high temperature and high-pressure region with increasing metallicity. However, the effect 
of metallicity on the chemical timescale is complex, making the quenched curve a complex function of metallicity. For a fixed $K_{zz}$ 
value, the \ch{CO} quenched curve shifts towards the lower atmosphere with increasing metallicity and follows $\tau_{\ch{CO}}$ 
for 10$^3$ - 10$^{-4}$ bar and 500 - 2500 K. Increasing the metallicity decreases the vertical mixing timescale, but $\tau_{\ch{CO}}$ 
mostly remains constant. As a result, the quenched curve of \ch{CO} moves from the low-pressure to the high-pressure region in the 
parameter space. In the $\ch{CH4}\rightarrow\ch{CO}$ conversion, $\tau_{\ch{CH4}}$ shows complex behavior with metallicity, which 
is also reflected in Figure \ref{quenching_contiur_plot} (a). The quenched curve shifts towards the low-temperature and low-pressure 
region for the parameter region in which the first term dominates in Equation \ref{eq:main_3}. However, when the second term dominates, 
the quenched curve moves to the higher-pressure region with increasing metallicity.

\begin{figure}[h]
	\centering
    \includegraphics[trim={0cm 3cm 10cm 4.5cm},clip,width=1\textwidth]{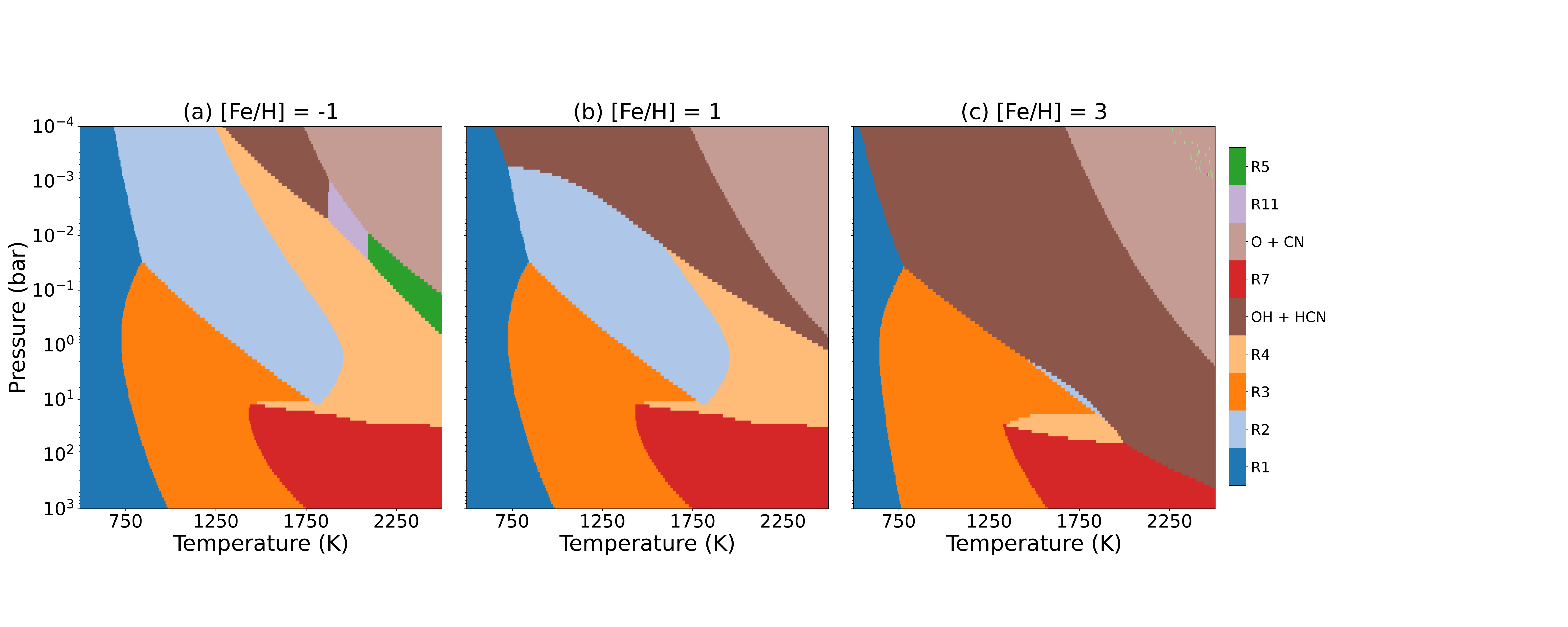}
	\includegraphics[trim={0cm 0cm 7.8cm 6cm},clip,width=0.7\textwidth]{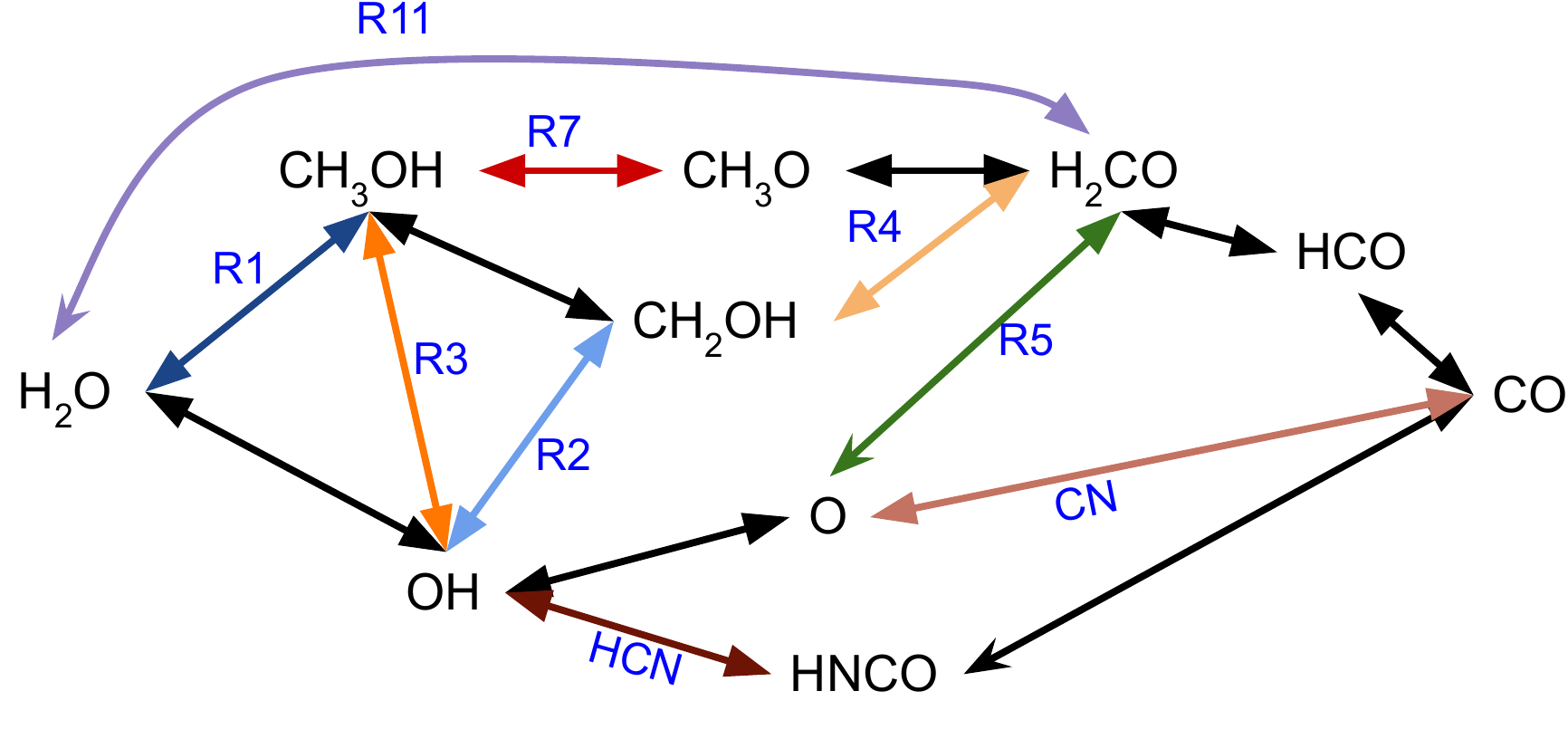}
	\caption{The top row shows the temperature-pressure range of different rate-limiting steps for the conversion of 
$\ch{H2O}\rightleftarrows\ch{CO}$ 
for three different metallicities. The colored regions correspond to the RLS in the bottom plot which shows the major chemical pathways for
$\ch{H2O}\rightleftarrows\ch{CO}$ conversion in hydrogen dominated atmosphere.}\label{fig:RLS_H2O}
\end{figure}

\subsubsection{\ch{H2O}}
For solar C/O ratio, atomic oxygen is more than twice as abundant as C \citep{Lodders2009}. Therefore even if all 
the C is locked into \ch{CO}, the available atomic oxygen to form \ch{H2O} does not decrease significantly. Thus, water is the most abundant 
O-bearing species for the parameter space, as pointed out by \cite{Moses2013}. Quenching will not affect the \ch{H2O} abundance, as it 
remains dominant at the quench level \citep{Moses2011} and follows its chemical equilibrium abundance. In Figure \ref{fig:RLS_H2O}, we 
have shown the different regions of the RLS and the pathways for the $\ch{H2O}\rightarrow\ch{CO}$ conversion. The conversion of 
$\ch{H2O}\rightarrow\ch{CO}$ has mainly two pathways. In the first pathway, the atomic carbon to form 
\ch{CO} comes from \ch{CH4}, and the RLS in this conversion is the same as in the  $\ch{CH4}\rightarrow\ch{CO}$ conversion and reactions R1 to R5 
become the RLS. In the second pathway, \ch{C} comes from \ch{CN} or \ch{HCN} to form \ch{CO}. The increment of metallicity significantly 
increases the second pathway. This is because increasing the metallicity decreases the relative abundance of \ch{CH4}, resulting in the relatively 
lower reaction rate of RLS in the conversion of $\ch{CH4}\rightarrow\ch{CO}$. The timescale of \ch{H2O} or $\tau_{\ch{H2O}}$ is given as follows:

\begin{align}
\tau_{\ch{H2O}} = \frac{[\ch{H2O}]}{\text{Reaction rate of RLS}} +\bigg( \tau_{\ch{H2}} \times \frac{3[\ch{CO}]}{\ch{H2}}\text{ or }
\tau_{\ch{NH3}} \times \frac{[\ch{CO}]}{\ch{NH3}}\bigg) \label{eq:main_5},
\end{align}
where the $\tau_{\ch{H2O}}$ is the timescale of conversion of $\ch{H2O}\rightarrow\ch{CO}$, and $[\ch{H2O}]$, $[\ch{CO}]$, 
$[\ch{NH3}]$  and  $[\ch{H2}]$ are the number densities of \ch{H2O}, \ch{CO}, \ch{NH3} and \ch{H2} respectively. $\tau_{\ch{H2}}$ is 
the interconversion timescale of $\ch{H2}\rightleftarrows\ch{H}$ and $\tau_{\ch{NH3}}$ is the timescale of $\ch{ NH3 + H}\rightarrow\ch{NH2 + H2}$. 
The second term is either $\tau_{\ch{H2}} \times \frac{3[\ch{CO}]}{\ch{H2}}$ or $\tau_{\ch{NH3}} \times \frac{[\ch{CO}]}{\ch{NH3}}$, which is 
decided by the temperature-pressure range. When atomic carbon for \ch{CO} comes from \ch{CH4}, then $\tau_{\ch{H2}} 
\times \frac{3[\ch{CO}]}{\ch{H2}}$ is used whereas, if it comes from \ch{HCN} or \ch{CN}, then $\tau_{\ch{NH3}} \times \frac{[\ch{CO}]}{\ch{NH3}}$ 
is used.

\begin{figure}[h]
	\centering
	\includegraphics[trim={0cm 0cm 0cm 0cm},clip,width=1\textwidth]{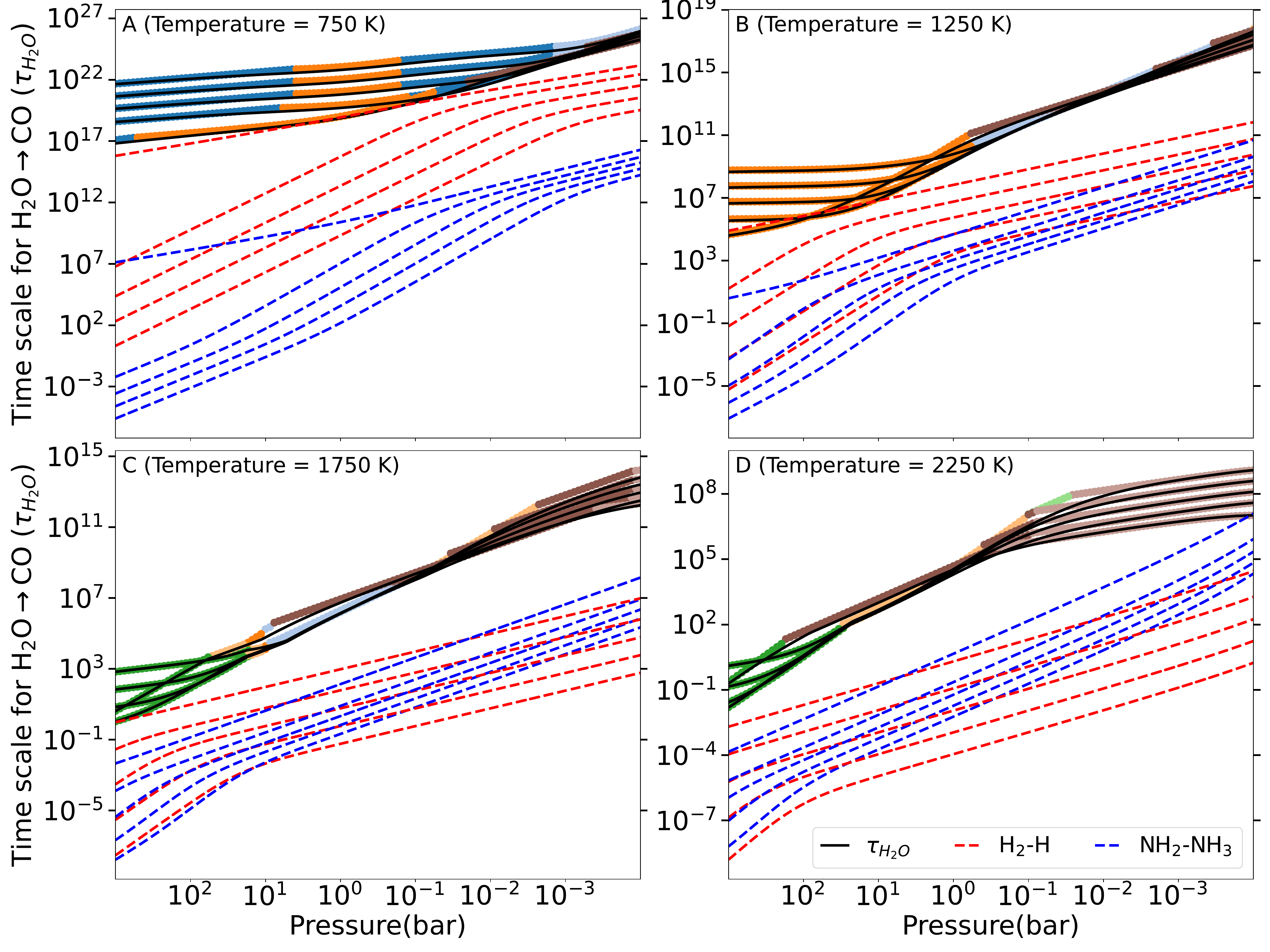}
	\caption{
		The timescale of conversion of $\ch{H2O}\rightarrow\ch{CO}$ ($\tau_{\ch{H2O}}$) for four different temperatures 
		(750 K, 1250 K, 1750 K, and 2250 K) with five different metallicities (0.1, 1, 10, 100, 1000 $\times$ solar metallicity). The colored lines represent the timescale of RLS in corresponding temperature-pressure value from Figure \ref{fig:RLS_H2O}, 
		the red dashed lines represent $\tau_{\ch{H2}}	\times \frac{3[\ch{CO}]}{\ch{H2}}$ and the blue dashed represents $\tau_{\ch{NH3}}	
		\times \frac{[\ch{CO}]}{\ch{NH3}}$. The solid black line represents $\tau_{\ch{H2O}}$.}\label{fig:Time_scale_H2O}
\end{figure}

In Figure \ref{fig:Time_scale_H2O}, we have shown $\tau_{\ch{H2O}}$ (Equation \ref{eq:main_5}) for four different temperatures (750 K, 1250 K, 
1750 K, and 2250 K) and five different metallicities (0.1, 1, 10, 100, and 1000 $\times$ solar). The RLS term, i.e., the first term in Equation \ref{eq:main_5} dominates over the second term for all the temperature and pressure regions. $\tau_{\ch{H2O}}$ is slower than $\tau_{\ch{CO}}$ and $\tau_{\ch{CH4}}$ 
for all the temperature-pressure ranges, which makes the quench level of \ch{H2O} lie deep in the atmosphere. However, it is to be 
noted that if C/O increases from the solar value, the relative abundance of atomic carbon will increase. For \ch{C/O} $>$ 0.5, 
the available atomic oxygen to form \ch{H2O} depends upon the \ch{CO/CH4} ratio \citep{Moses2013}, as most of the oxygen is locked in \ch{CO} 
at high temperature, which is mostly the CO-dominated region, and the remaining 
available \ch{O} forms \ch{H2O}. In that case, the \ch{H2O} abundance can deviate from the equilibrium abundance, and the quench level 
of \ch{H2O} can be used to determine the composition of the atmosphere.

\begin{figure}[h]
	\centering
	\includegraphics[width=1\textwidth]{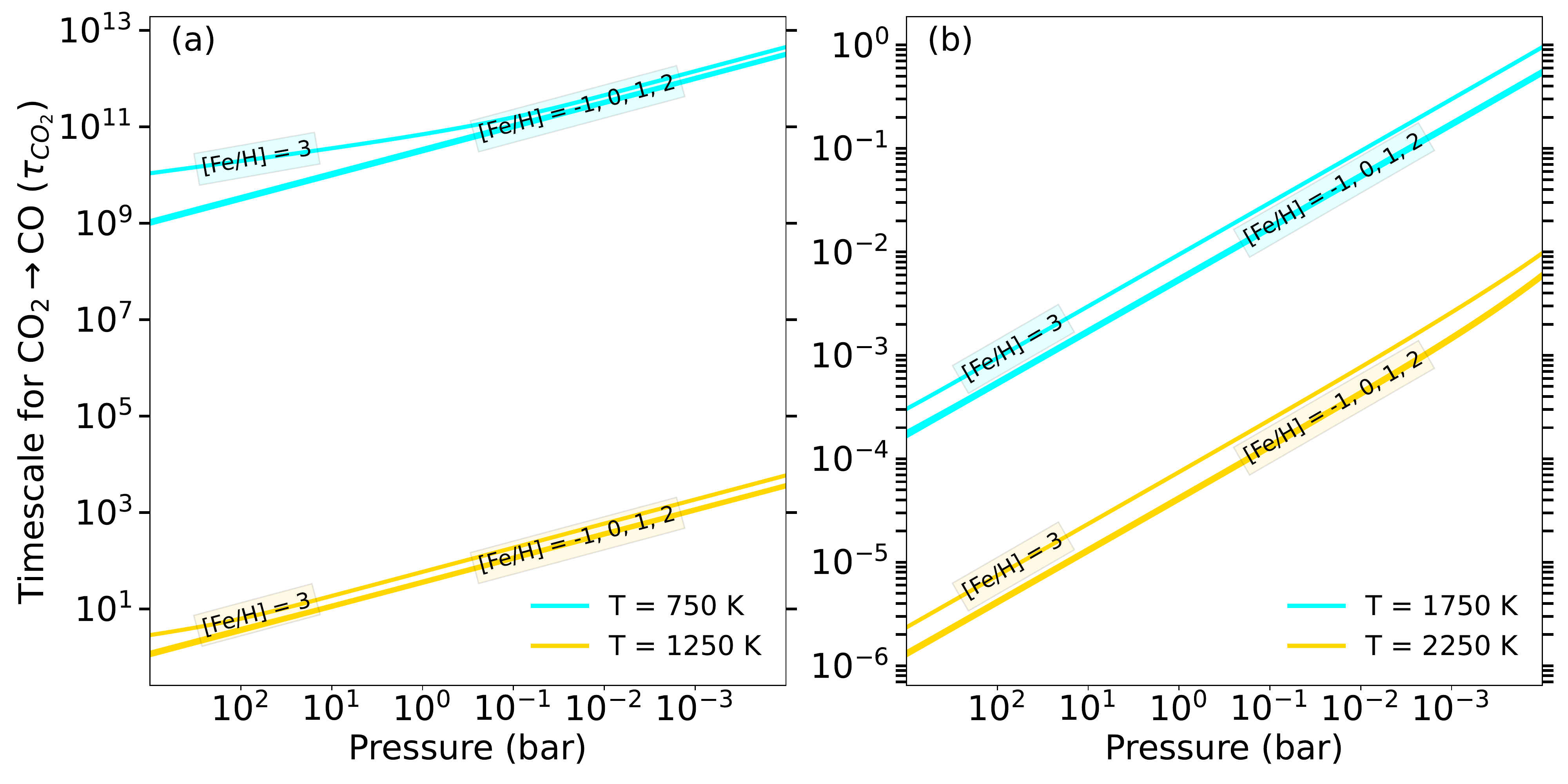}
	\caption{The timescale of conversion of $\ch{CO2}\rightarrow\ch{CO}$ ($\tau_{\ch{H2O}}$) for four different temperatures 
		(750 K, 1250 K, 1750 K, and 2250 K) with five different metallicities (0.1, 1, 10, 100, 1000 $\times$ solar metallicity). The colored lines 
		represent the timescale of $\ch{CO}\rightarrow\ch{CO2}$ for the labeled parameter.}
	\label{fig:CO2_time_scale}
\end{figure}

\subsubsection{\ch{CO2}}
\ch{CO2} recycles through \ch{H2O} and \ch{CO} by a relatively fast conversion scheme \citep{Line2010, Moses2011, Tsai2018}.

\begin{align*}
\ch{H2O} + \ch{H} & \rightleftarrows \ch{OH} + \ch{H2} \\
\ch{CO + OH} & \rightleftarrows \ch{CO2 + H} \\
\cline{1-2}
\ch{CO + H2O} &\rightleftarrows \ch{CO2} + \ch{H2}
\end{align*}
The RLS for this conversion is the $\ch{CO + OH} \leftrightarrows \ch{CO2 + H}$  reaction. The $\ch{CO}\leftrightarrows\ch{CO2}$ 
conversion timescale is relatively fast compared to the $\ch{CH4}\rightleftarrows\ch{CO}$ timescale. The RLS of $\ch{CO}\rightarrow\ch{CO2}$ ($\ch{CO + OH} 
\rightarrow \ch{CO2 + H}$) comes into the second type (both the reactants are heavy molecules), and the RLS of $\ch{CO2}\rightarrow\ch{CO}$ 
($\ch{CO2 + H}\rightarrow\ch{CO + OH}$) comes into the first type of RLS (H is one of the reactants). Thus, $\tau_{\ch{CO2}}$ increases with 
metallicity at a slow rate. In Figure \ref{fig:CO2_time_scale}, we have plotted $\tau_{\ch{CO2}}$ with pressure at different temperatures
and metallicities. $\tau_{\ch{CO2}}$ is more than four orders of magnitude faster than $\tau_{\ch{CO}}$ for 1750 K and 2250 K temperatures, and this 
difference drastically increases with the decreasing pressure. For 750 K, 1000 bar, and 0.1-100 $\times$ solar metallicity, $\tau_{\ch{CO2}}$ 
is comparable to $\tau_{\ch{CO}}$, though for all the other parameter ranges,  $\tau_{\ch{CO2}}$ is faster then  $\tau_{\ch{CO}}$.  
The fast timescale of  $\tau_{\ch{CO2}}$  makes the  \ch{CO2} abundance remain in equilibrium  with \ch{H2O} and \ch{CO}. Thus, for 
the pressure level, where \ch{CO} follows its quenching abundance, \ch{CO2} also deviates from its chemical equilibrium abundance. The 
chemical timescale of \ch{CO2} increases with height. For a  sufficiently high chemical timescale, the \ch{CO2} abundance can be quenched at 
a high altitude where the \ch{CO2} chemical timescale becomes comparable to the vertical mixing timescale. However, the quenched \ch{CO2} abundance does not contribute to the infrared part of the transmission or emission spectra of the planets because the infrared spectrum is shaped well below 
the \ch{CO2} quench level (P  $>10^{-4}$ bar) \citep{Moses2011}. 

\section{Applying on the Test Exoplanets}

\begin{table}
	\caption{Parameters for the exoplanets studied}
	\begin{center}
		\resizebox{\textwidth}{!}{
			
			\begin{tabular}{||c c c c c c||} 
				\hline
				Name & [Fe/H] & C/O &  Intrinsic temperature & Heat redistribution & Reference\\ [0.5ex] 
				\hline\hline
				HD 189733 b & 0 & Solar & - & No redistribution  & \cite{Moses2011} \\ 
				\hline
				GJ 1214 b  &  2 & Solar & 60 K & Equal redistribution & \cite{Charnay2015}\\
				\hline
				HR 8799 b & 0 & 0.66 & - & - &\cite{Moses2016}\\
				\hline
				GJ 436 b & - & 1 & - & - & \cite{Line2014} \\
				\hline
				WASP-39 b & 1 & 0.35 & - & Equal redistribution & \cite{Tsai2022}; \\ [1ex] 
				&  &  &  &  & \cite{JWSTTECERST2022} \\ [1ex] 
				\hline
			\end{tabular}
		}
	\end{center}
\end{table}

The quenching approximation has the potential to constrain the disequilibrium abundance. Here, we have benchmarked the 
quenching approximation results with the photochemistry-transport model (described in Section \ref{sec:model} and Appendix). 
For the benchmarking, we used two exoplanets, namely HD 189733 b and GJ 1214 b. HD 189733 b is a gas giant with 
an orbital period of 2.22 days \citep{Moutou2006}, and GJ 1214 b is a Neptune-sized planet with an orbital period 
of 1.58 days \citep{Charbonneau2009}. These two exoplanets are excellent candidates for this study since their 
equilibrium temperature is such that it intersects the \ch{CH4-CO} boundary at 10 bar and 10$^{-5}$ bar pressure which 
leads to having a more pronounced effect of the transport on \ch{CH4} and \ch{CO} abundances. HD 189733 b and GJ 1214 b 
have T$_{eq}\approx $ 1200 K and 600 K, and surface gravity  g$_{surface}\approx $ 21.5 $\text{m s}^{-2}$ and 8.9 
$\text{m s}^{-2}$ respectively. To find the quench level of \ch{CH4} and \ch{CO}, we overplotted the T-P profile 
of HD 189733 b and GJ 1214 b with the constant contour of the quench level (given in Figure~\ref{quenching_contiur_plot}) 
for assorted metallicities. Using the quench level, we found the quenched abundance and benchmarked it with the 
photochemistry-transport model. For the benchmarking, we turned off the photochemistry and molecular diffusion 
for the photochemistry-transport model since the quenching approximation only deals with the Eddy diffusion.  
However, molecular diffusion and photochemistry can affect our results, depending upon the strength of the stellar 
radiation at the top of the atmosphere and the strength of transport. For moderate values of the Eddy diffusion 
coefficient (10$^6$-10$^9$ cm$^{2}$ s$^{-1}$), Eddy diffusion dominates over molecular diffusion in the parameter 
range we considered. As described in \cite{Hu2014}, the homopause (the pressure level where both the coefficients become equal) 
is at 10$^{-5}$ bar for $K_{zz} = 10^6$ cm$^{2}$ s$^{-1}$ for GJ 1214 b. The photochemical products formed at high 
altitudes can diffuse in the deep atmosphere and change the composition. As discussed in \cite{Moses2011} and 
\cite{Madhusudhan2016}, the photochemical products can reach up to 10$^{-4}$ bar for high radiated hot Jupiter. 
However, photochemical products can affect the atmospheric composition of a lower-temperature atmosphere.

\subsection{HD 189733 b}
\begin{figure}[h]
	\centering
	\includegraphics[width=1\textwidth]{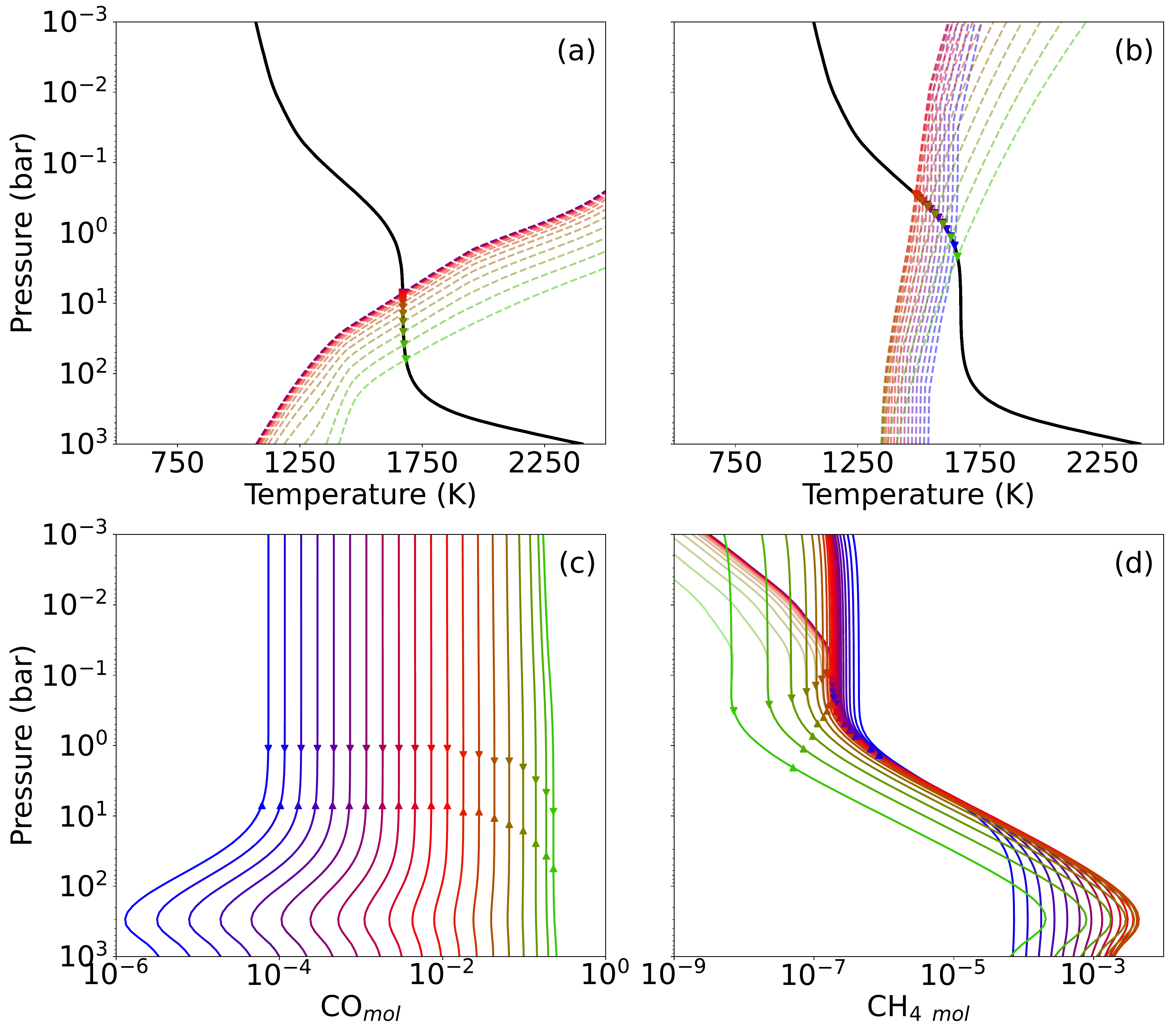}

	\caption{The top panel shows the T-P profile overplotted with the quenched curve of (a) \ch{CO} and (b) \ch{CH4} 
for 20 different metallicities (blue-red-green lines are from 0.1-10-1000 $\times$ solar metallicity). The quench level contour 
is calculated for the $K_{zz}$ = 10$^9$ cm$^2\text{ s}^{-1}$ 
and assuming the mixing length is equal to 0.1 $\times$ atmospheric scale height. The bottom panel shows the mixing ratio of (c) \ch{CO} 
and (d) \ch{CH4}, for the same set of metallicities. The colored lines are the output of the photochemistry-transport model and the 
corresponding faded 
colored lines are the equilibrium abundances. The upper and lower triangles are the quench levels calculated using $0.1 H$ and $1 H$ as the 
mixing length.}\label{fig:10}
\end{figure}

The T-P profile for HD 189733 b is adopted from \cite{Moses2011} (see Table 1 for parameters) and overplotted with the contours of the quench level of \ch{CO} 
and \ch{CH4} in Figures \ref{fig:10} (a) and (b) respectively. The T-P profile crosses the \ch{CH4-CO} boundary at the 10 
bar pressure level for solar metallicity.  The temperature is warm enough to make \ch{CO} the dominant species. Below 10 bar, 
\ch{CH4} is the dominant C-bearing species, and above 10 bar, \ch{CO} becomes the primary C-bearing species. Figures \ref{fig:10} (c) 
and (d) show the mixing ratio of \ch{CO} and \ch{CH4} respectively for 20 different metallicities. With increasing metallicity, 
the \ch{CH4-CO} boundary moves towards the high-pressure and low-temperature region, shifting a greater part of the thermal profile 
into the \ch{CO} region. The \ch{CO} equilibrium abundance increases for the entire pressure range as the metallicity increases. However, 
the \ch{CH4} equilibrium abundance first increases and then decreases with increasing metallicity in the high-pressure region, and in the 
low-pressure region, the abundance decreases with the increasing metallicity. The quench level of \ch{CO} lies well below the \ch{CH4} 
quench level. \ch{CO} is the dominant species at the \ch{CO} quench level, making \ch{CO} the major source of carbon in the 
transport-dominated atmosphere. Increasing metallicity does not alter the dominant C-bearing species; as the metallicity 
increases, \ch{CO} becomes more abundant. The \ch{CO} quench level shifts towards the high-pressure region, increasing the mixing 
ratio of \ch{CO} in the transport-dominated region with metallicity. Also, \ch{CO2} and \ch{H2O} follow the equilibrium abundance.

\subsection{GJ 1214 b}
\begin{figure}[h]
	\centering
	\includegraphics[width=1\textwidth]{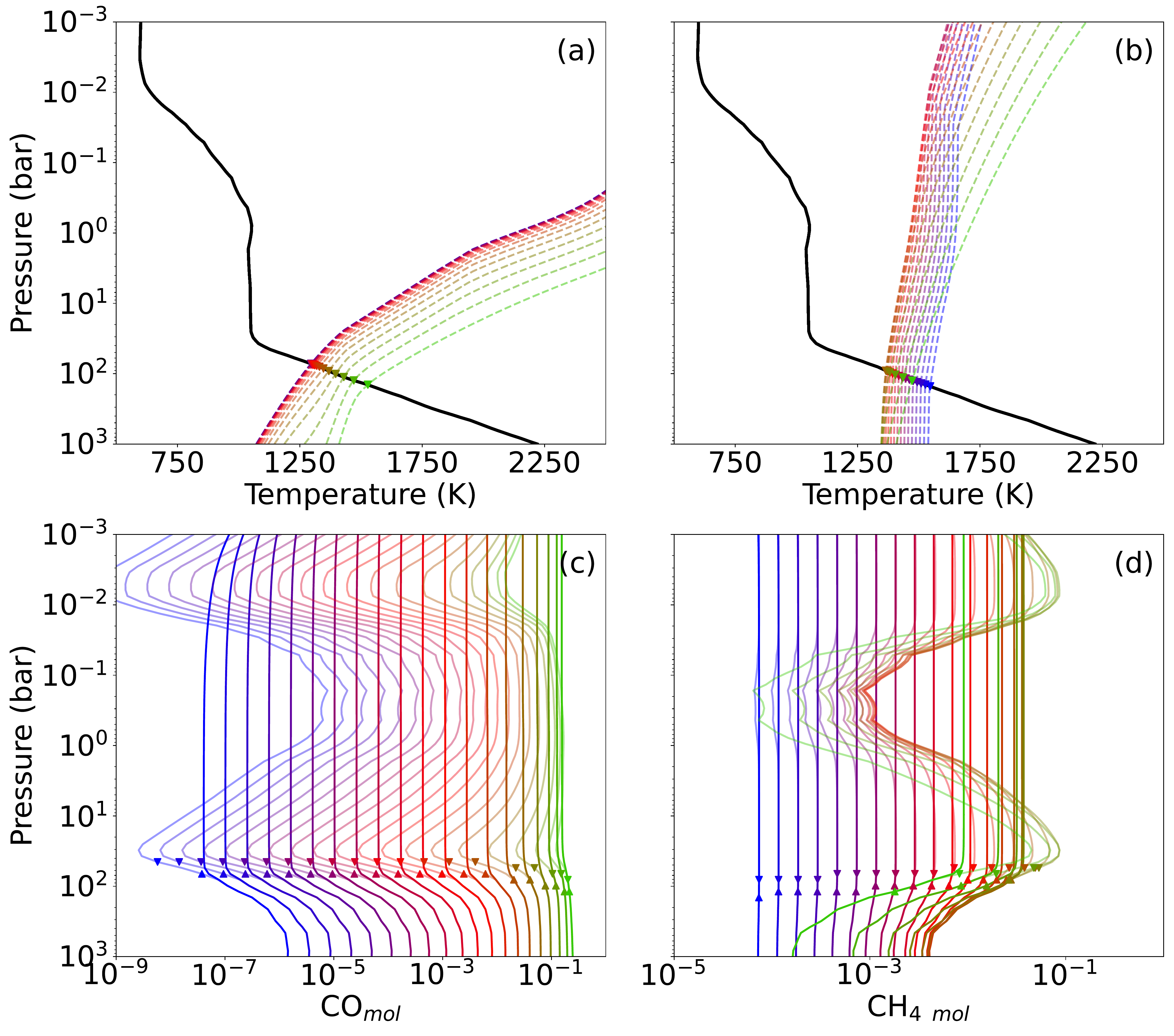}
	\caption{Same as Figure \ref{fig:10}, but for GJ 1214 b}\label{fig:11}
\end{figure}

Figure \ref{fig:11}  shows the T-P profile of GJ 1214 b (see Table 1 for parameters), overplotted with the contours of the quench level of (a) \ch{CO} and (b) \ch{CH4}. 
The T-P profile is adopted from \cite{Charnay2015}, and for solar [Fe/H], it crosses the \ch{CH4-CO} boundary at $10^{-5}$ bar pressure level. 
The temperature is sufficiently low to make \ch{CH4} the dominant C-bearing species at solar metallicity. However, as the metallicity increases, the 
equal-abundance curve of \ch{CH4-CO} moves towards low temperature, and as a result, the region where the T-P profile lies changes from \ch{CH4} 
to \ch{CO} dominant region. In the transport-dominated region and $P<50$ bar, \ch{CH4} remains the dominant C-bearing species
for [Fe/H] $ < 100 \times$ solar metallicity. For $P>50$ bar, \ch{CH4} dominates over CO for [Fe/H] $ < 30 \times$ solar metallicity.
In Figures \ref{fig:11} (c) and (d), the mixing ratio of \ch{CO} and \ch{CH4} for 20 different metallicities are shown. The quench level of \ch{CO} 
and \ch{CH4} lies at $10^2$ bar pressure level; the temperature determines whether \ch{CO} or \ch{CH4} will dominate in the transport region. 
As the metallicity increases, the \ch{CO} equilibrium abundance increases for the entire pressure range. However, the \ch{CH4} equilibrium abundance 
shows complex behavior with metallicity. For $P<10^2$ bar and metallicity $<$ 600 $\times$ solar metallicity, the \ch{CH4} abundance increases with increasing 
metallicity, and for metallicity $>$ 600 $\times$ solar metallicity, the \ch{CH4} abundance starts to decrease, and \ch{CO} starts to become the major C-bearing source.
\subsection{Quench level analysis on the test exoplanets}
It is apparent that \ch{CH4} and \ch{CO} can quench at different pressure levels depending on the thermal profile. 
In the equilibrium composition, $\tau_{\ch{CH4}}$ and $\tau_{\ch{CO}}$ become equal at the equal-abundance curve. 
If the T-P profile intersects the equal-abundance curve, say at a pressure level $P_i$ and the atmospheric mixing is such 
that the quench level lies close to $P_i$, then \ch{CH4} and CO can quench at the same pressure level. If the quench 
level lies well below/above the $P_i$ pressure level, they can quench at different atmospheric heights.
We can find which species quench first by tracing the equal abundance and thermal profile. 
In Figure 9(b), it is shown that for a fixed metallicity and $K_{zz}$ value, the CO quenched curve is at a higher 
temperature relative to the \ch{CH4} quenched curve in the CO dominated region,  and the \ch{CH4} 
quenched curve is at a higher temperature in the \ch{CH4} dominated region. Along with this, assuming the temperature 
is decreasing with decreasing pressure (no thermal inversion) or remains isothermal, then in the CO dominated region 
(right side of the equal-abundance curve ), CO will quench first, and in the \ch{CH4} dominated the region, \ch{CH4} 
quenches first.

We used two extreme values, i.e., $0.1 \times $H and $1 \times $H as the mixing length and calculated the quenched 
abundances instead of using the method given in \cite{Smith1998} to find the mixing length. As evident from Figures \ref{fig:10} 
and \ref{fig:11}, all the modeled transport abundances (mixing ratio from full chemical kinetics model) lie in between 
the quenched abundances with mixing length $L =0.1 \times H$ and $1 \times H$. We define the error as the difference between 
the maximum and minimum equilibrium abundances between $P_{0.1}$ and $P_{1}$, where $P_{0.1}$ is the quench pressure given by 
$L = 0.1 \times $H and $P_{1}$ is the  quench pressure given by $L = 1 \times$H.  For HD 189733 b, the \ch{CO} abundance is accurate 
within $\approx$ a factor of two for low metallicity and matches with the chemical kinetics model for high metallicity. 
The reason for the good agreement of quenching approximation and chemical kinetics model for \ch{CO} is that 
the equilibrium mixing ratio of \ch{CO} does not change with pressure. As a result, the quenched abundance of \ch{CO} becomes 
independent of transport. However, for \ch{CH4}, the accuracy is within $\approx$ an order of magnitude. For GJ 1214 b, the \ch{CO} 
abundance is accurate within $\approx$ a factor of five for low metallicity and within $\approx$ a factor of two for high metallicity. 
For \ch{CH4}, it is accurate within $\approx$ a factor of two and seven for low and high metallicity, respectively. The quenching accuracy 
depends upon the pressure difference between the quench pressure with $L = 0.1 \times H$ and $1 \times H$ and the change of the equilibrium 
abundance between these pressure levels. As metallicity affects both these quantities, the accuracy of the quenching approximation is changed 
with metallicity. Also, the metallicity effect on the quenching approximation differs for different molecules and T-P profiles. For GJ 1214 b, 
the increment of metallicity improves the quenching accuracy for CO while decreasing it for \ch{CH4}.

\subsection{Constraint on Metallicity and Transport Strength}
\begin{figure}[h]
	\centering
	\includegraphics[width=1\textwidth]{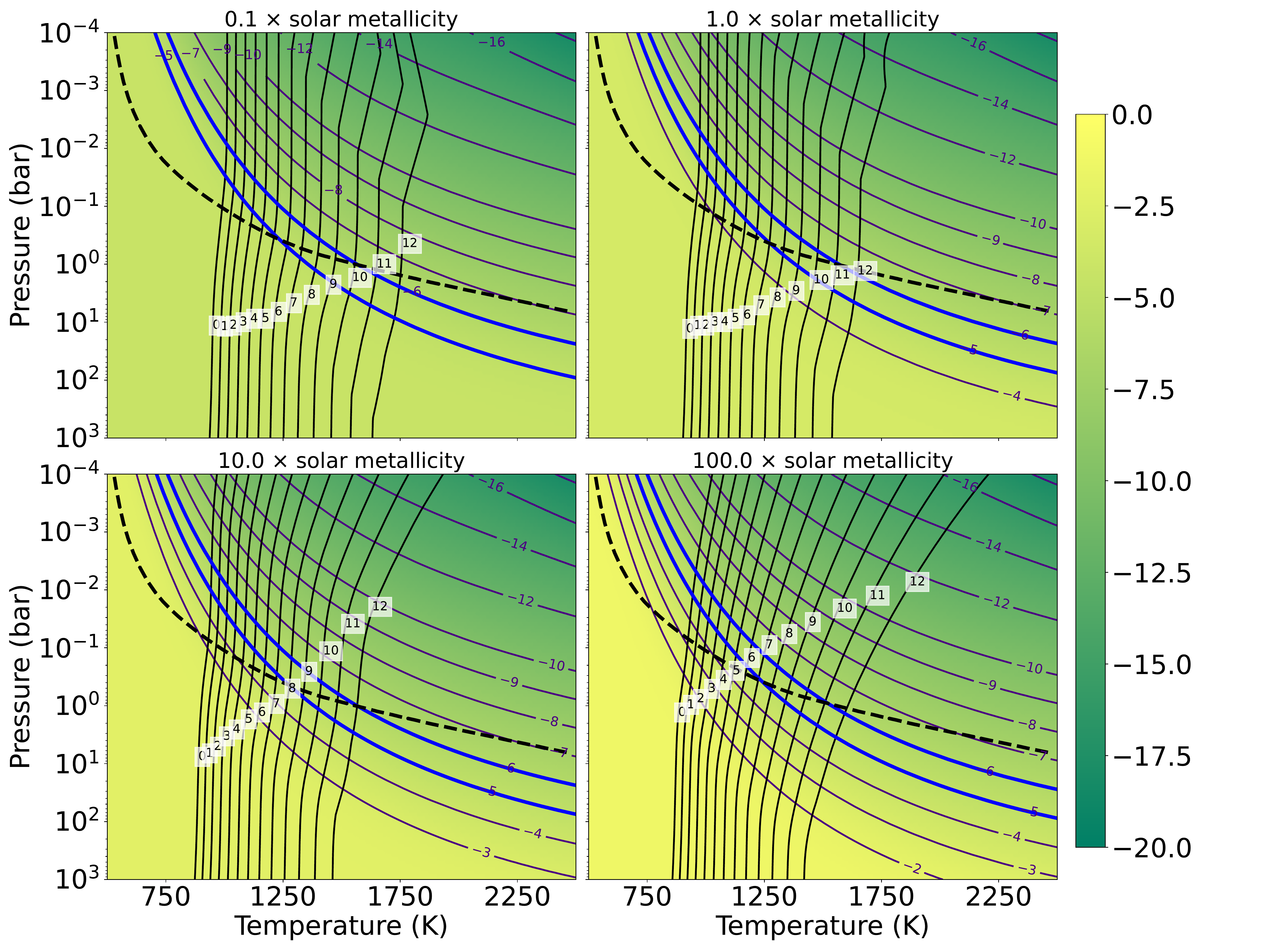}
	\caption{The color bar shows the mole fraction of \ch{CH4} in the log10 scale as a function of pressure and temperature. The region 
between the solid blue lines are the observational constrain of HR 8799 b taken from \cite{Barman2015}, the black dashed line is the 
T-P profile adopted for T$_{eff}$ = 1000 K and $g$ = 3000 cm s$^{-2}$, which is adopted from \cite{Moses2016}. The solid blacks lines are the 
quench lines for different $K_{zz}$ values (label in the plots). The four panels represent the different values of atmospheric metallicity.}\label{fig:13}
\end{figure}

\begin{figure}[h]
	\centering
	\includegraphics[width=1\textwidth]{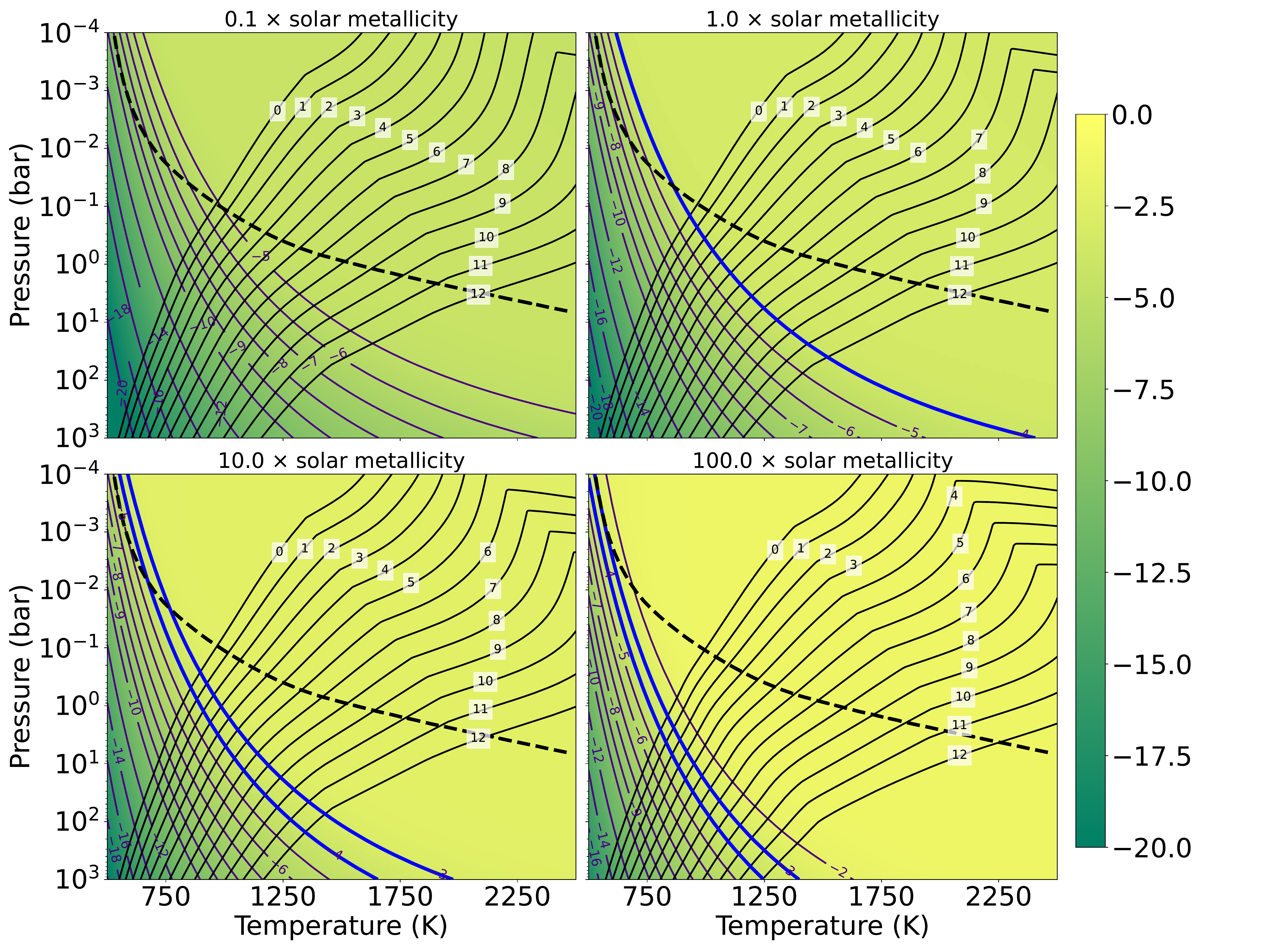}
	\caption{Same as Figure \ref{fig:13}, but for CO (for HR 8799 b). }\label{fig:12}
\end{figure}

\begin{figure}[h]
	\centering
	\includegraphics[width=1\textwidth]{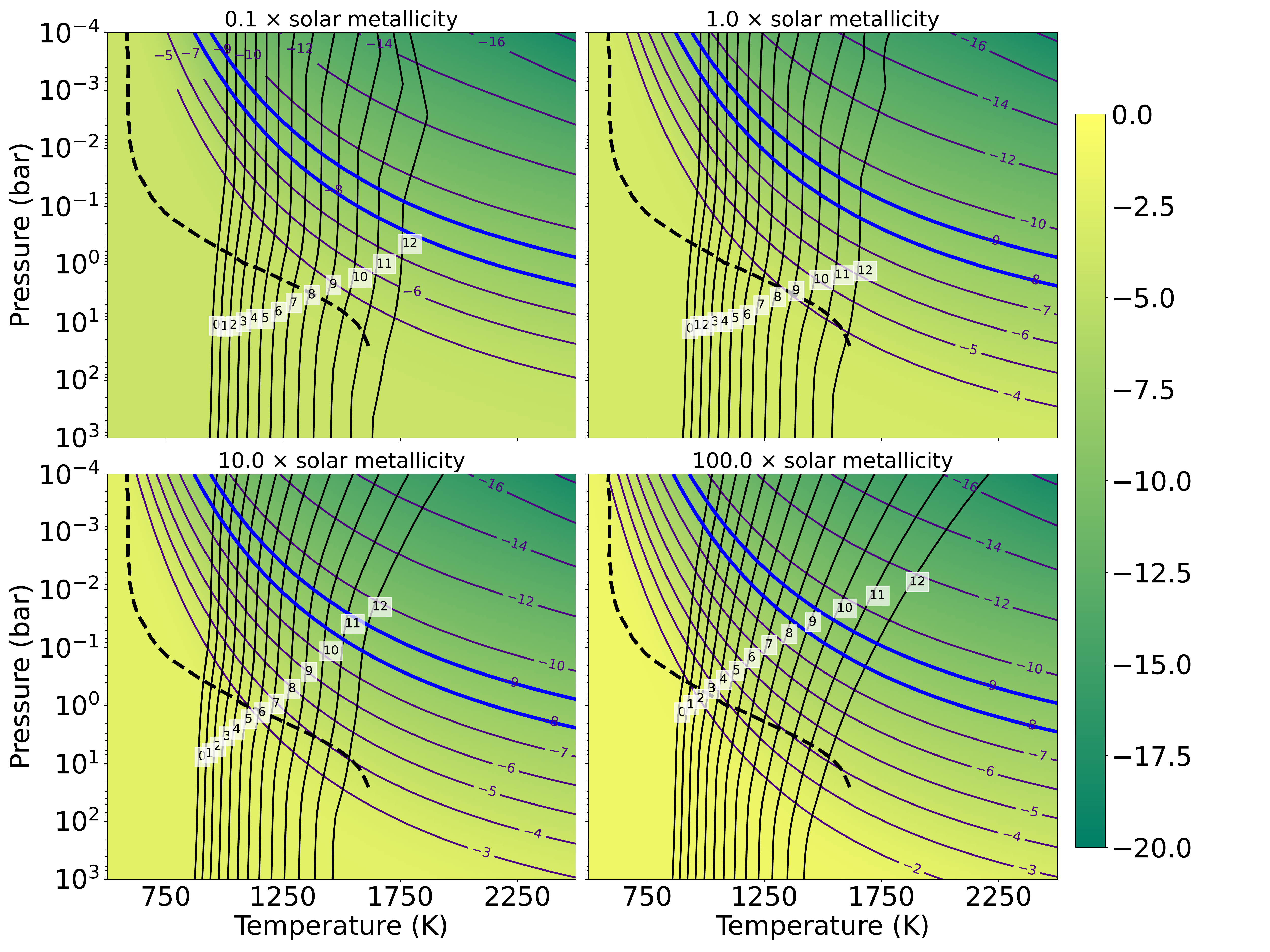}
	\caption{The colour bar shows the equilibrium mole fraction of \ch{CH4} in the log10 scale as a function of pressure and temperature. Along with the color bar, the contour of constant mixing ratio also plotted. The region between the solid royal-blue lines are the observational constrain of the GJ 436 b taken from \cite{Line2014}, the black dashed line is the T-P profile adopted from \cite{Line2014}. The solid black lines are the quench lines for different $K_{zz}$ values (label in the plots). The four panels represent the different values of atmospheric metallicity}\label{fig:14}
\end{figure}

\begin{figure}[h]
	\centering
	\includegraphics[width=1\textwidth]{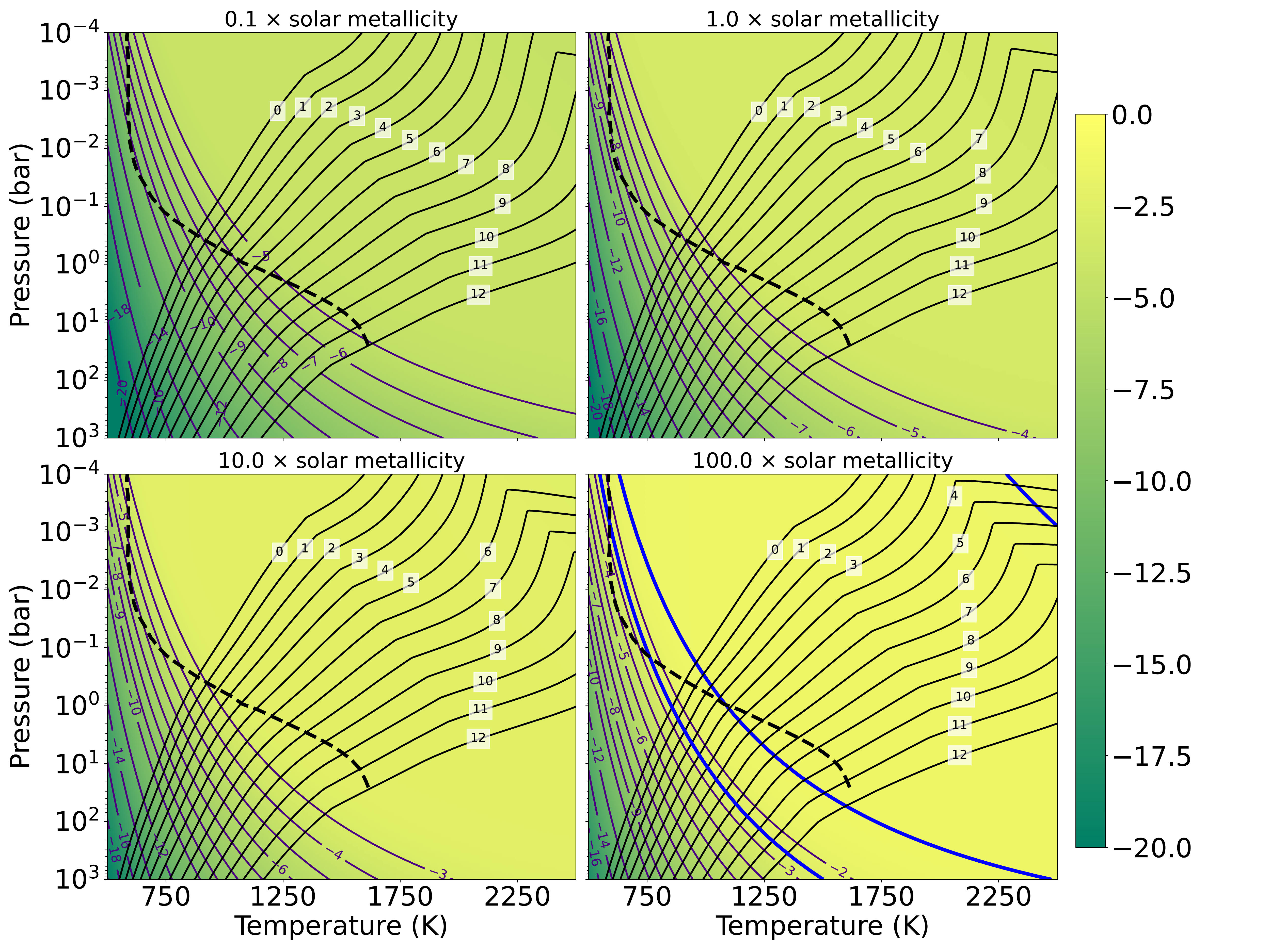}
	\caption{Same as Figure \ref{fig:14}, but for CO (for GJ 436 b). For visualization, the royal-blue curve are 
at $10^{-4}$ and $10^{-3}$ contour of equilibrium mixing ratio.}\label{fig:15}
\end{figure}

The quenching approximation is a powerful tool that can constrain the metallicity and transport strength without 
solving the photochemistry-transport model. To constrain the metallicity and transport strength, the observed abundance of molecules 
and the thermal profiles of the exoplanets 
are overplotted with the color-mesh plot of the equilibrium mixing ratio and quenched curve of the molecules (see Figures \ref{fig:13}-\ref{fig:15}).

To demonstrate this fact, we use four case studies. We chose HR 8799 b as our first case study; it is a 
gas giant with 7 $M_{J}$ mass and 1.2 $R_{J}$ radius, which was discovered by direct imaging \citep{Marois2008}. For the  
second and third cases, we take GJ 436 b and the previously mentioned exoplanet HD 189733 b. GJ 436 b is a Neptune-sized exoplanet with 
a 2.6 days orbital period, discovered in 2004 \citep{Bailey2004} and having a high C/O ratio $\approx 1$, 
whereas HR 8799 b has C/O $\approx 0.66$ \citep{Line2014,Barman2015}.

\begin{figure}[h]
	\centering
	\includegraphics[width=1\textwidth]{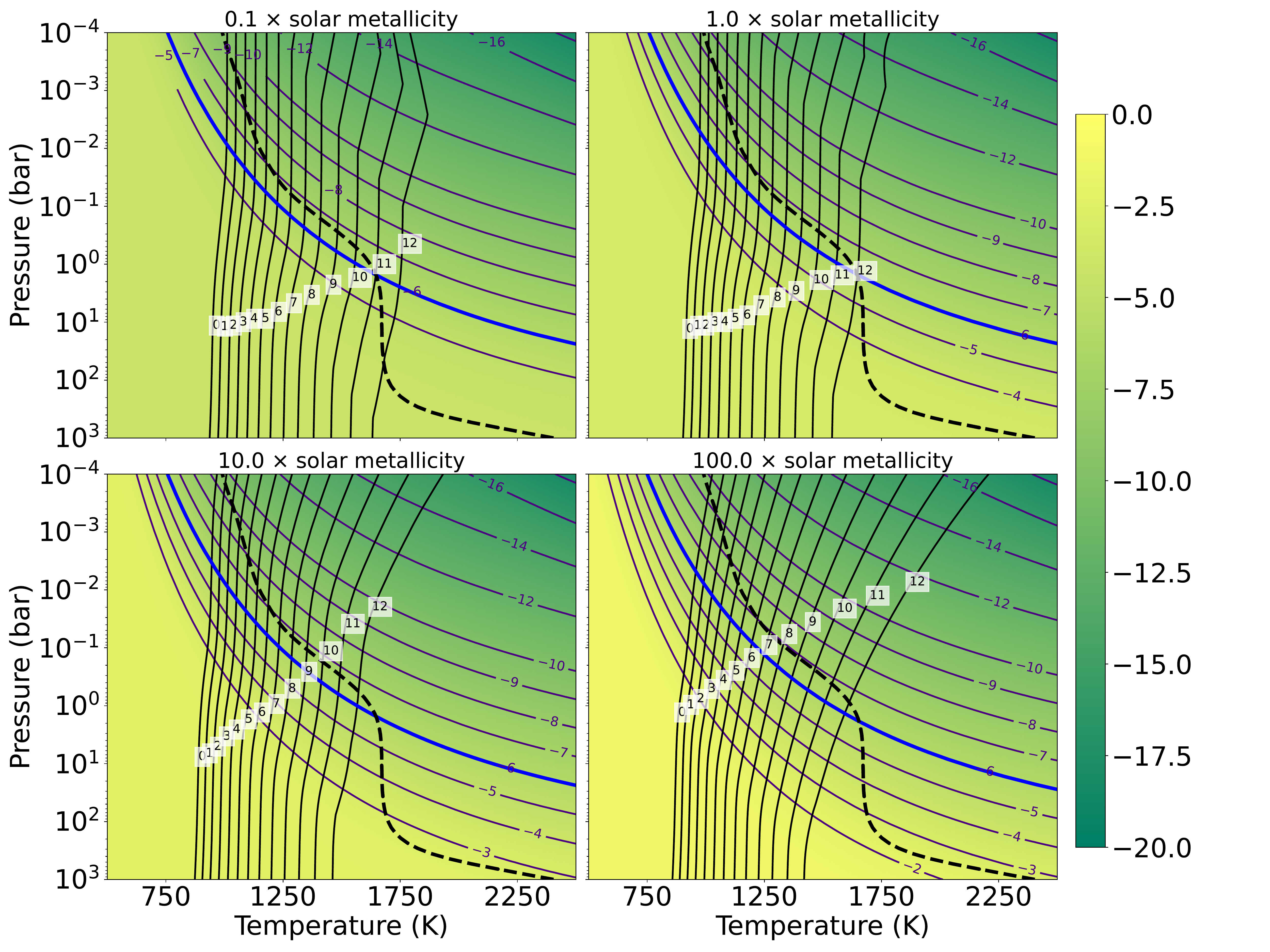}
	\caption{Same as Figure \ref{fig:14}, but for HD 189733 b with \ch{CH4}.}\label{fig:16}
\end{figure}

\begin{figure}[h]
	\centering
	\includegraphics[width=1\textwidth]{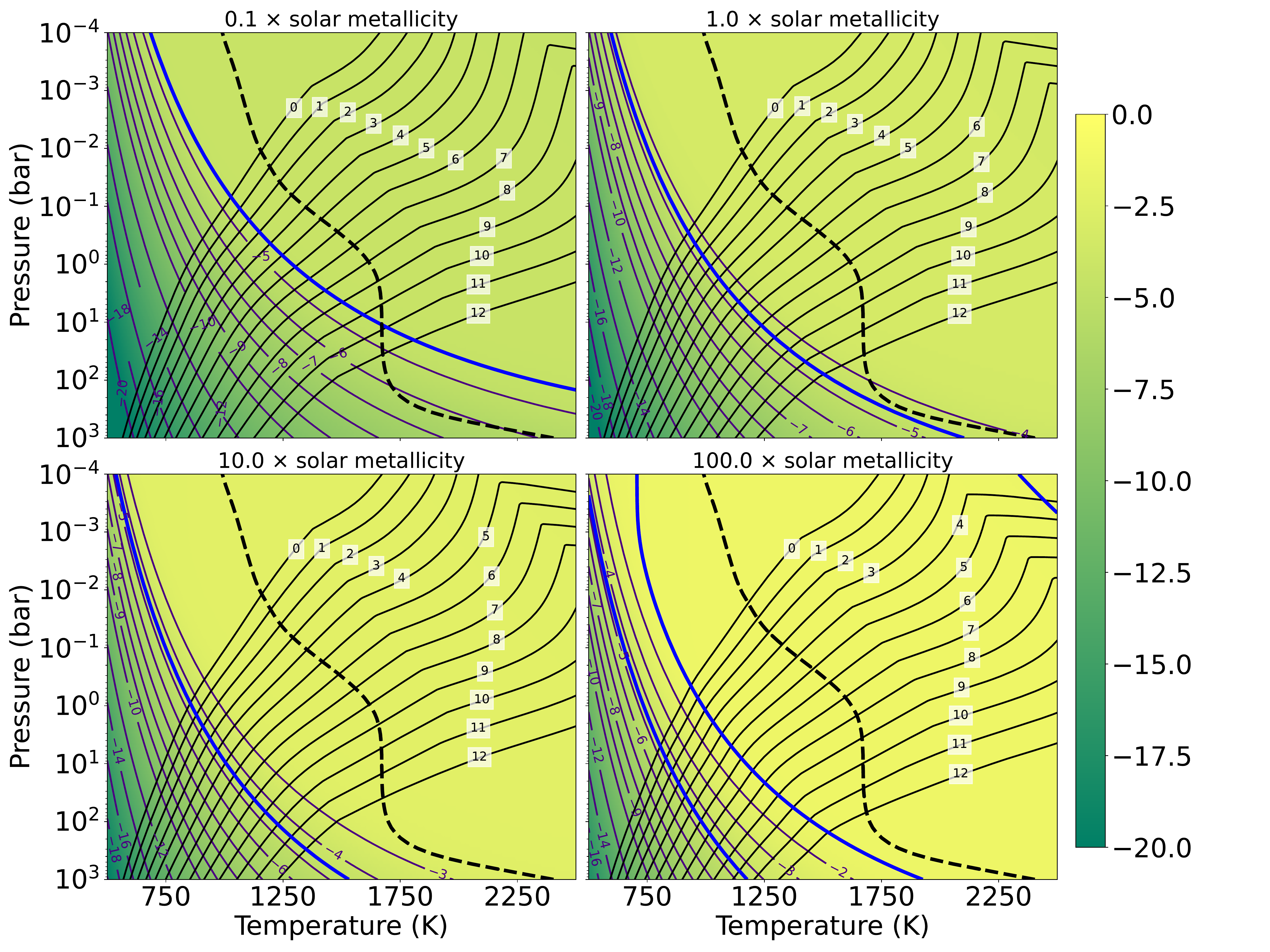}
	\caption{Same as Figure \ref{fig:14}, but for HD 189733 b with \ch{CO}.}\label{fig:17}
\end{figure}

\begin{figure}[hb]
	\centering
	\includegraphics[width=1\textwidth]{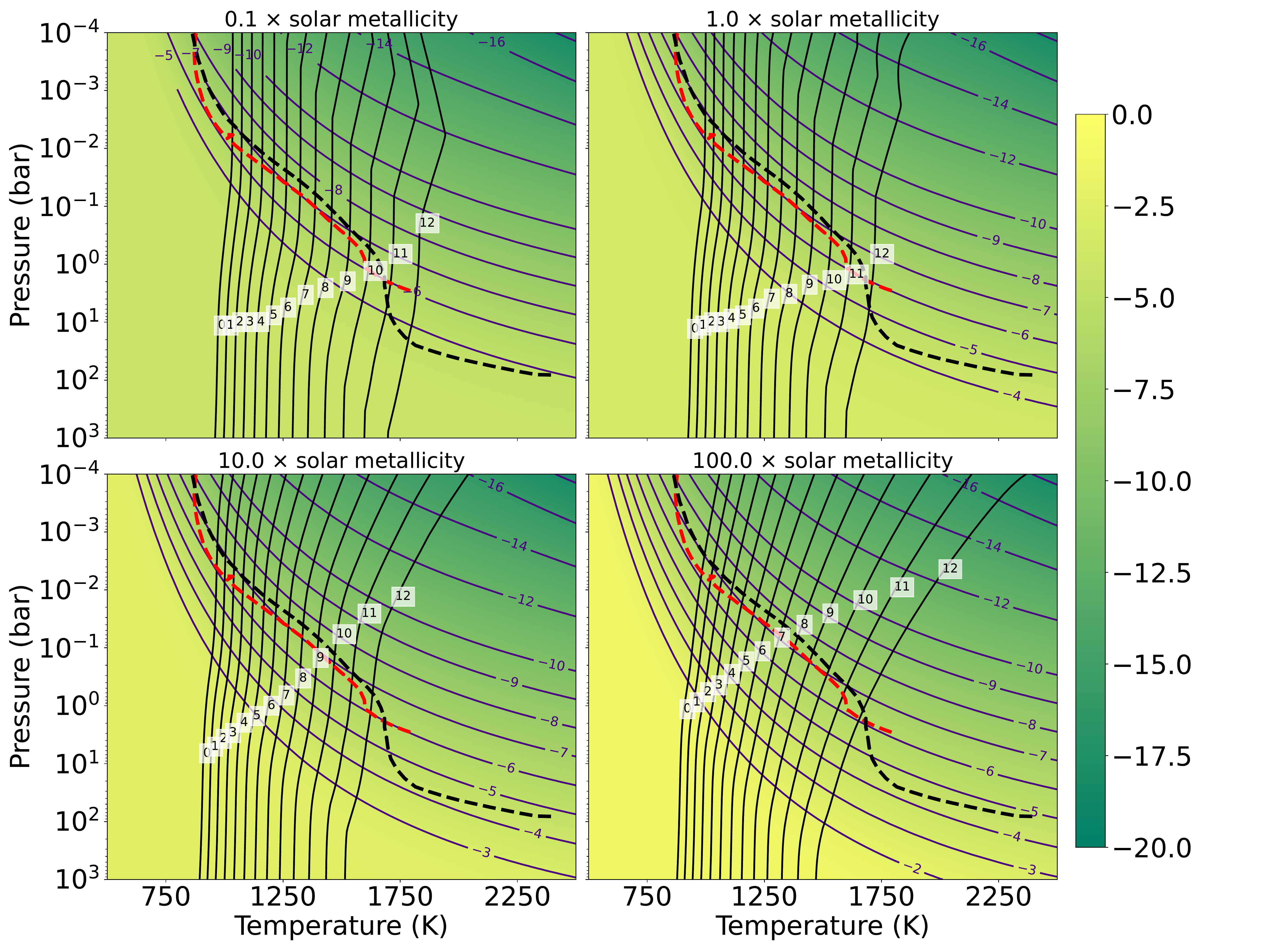}
	\caption{The colour bar shows the equilibrium mole fraction of \ch{CH4} in the log10 scale as a function of pressure and temperature for WASP-39 b. 
Along with the color bar, the contours of constant mixing ratio are also plotted. The T-P profiles are shown in black dashed and red dashed lines, and are taken 
from  \cite{JWSTTECERST2022} and \cite{Tsai2022} respectively. The solid black lines are the quench lines for different $K_{zz}$ values (label in the plots). The 
four panels represent the different values of atmospheric metallicity. }\label{fig:WASP29b_CH4}
\end{figure}

\begin{figure}[h]
	\centering
	\includegraphics[width=1\textwidth]{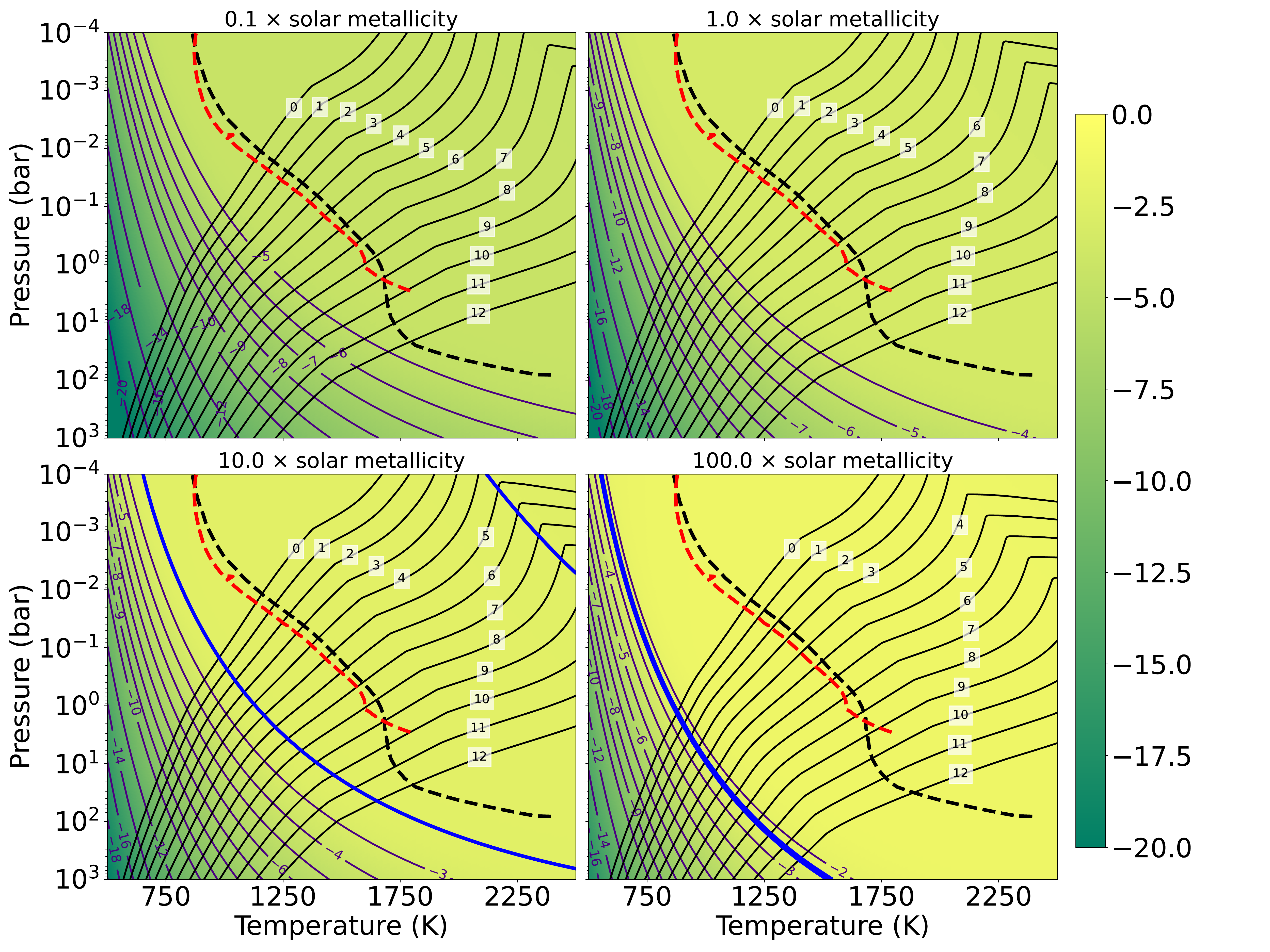}
	\caption{Same as Figure \ref{fig:WASP29b_CH4}, but for \ch{CO} (for WASP-39 b). Here, blue lines represent the equilibrium mixing ratio contours of $4\times 10^{-3}$ and $5\times 10^{-3}$.}\label{fig:WASP29b_CO}
\end{figure}

\begin{figure}[h]
	\centering
	\includegraphics[width=1\textwidth]{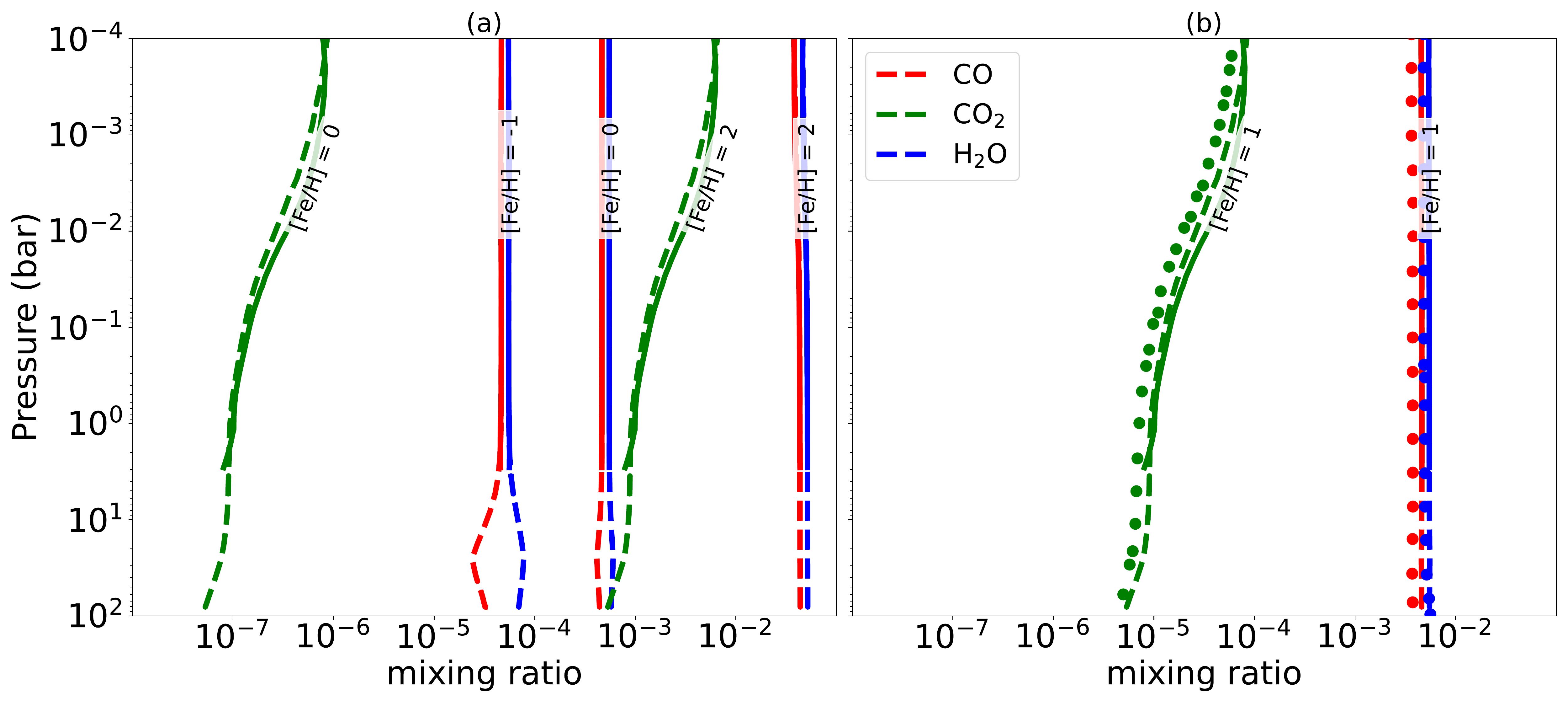}
	\caption{The equilibrium abundances of \ch{CO2}, CO and \ch{H2O} for two different T-P profiles of WASP-39 b (dashed: \cite{JWSTTECERST2022} and solid: \cite{Tsai2022}) are shown for metallicity (a) [Fe/H] = [-1, 0, 2] and (b) [Fe/H] = [1]. The colored circles in panel (b) are the abundances taken from \cite{JWSTTECERST2022}. }\label{fig:WASP_29b}
\end{figure}

\begin{itemize}
\item We overplotted the observed \ch{CH4} (10$^{-6}$ $<$ \ch{CH4} mole fraction $<$ 10$^{-5}$) and CO 
(10$^{-4}$ $<$ \ch{CO} mole fraction $<$ 3 $\times$ 10$^{-4}$) abundance of HR 8799 b (T-P profile taken from \cite{Moses2016}, see Table 1 for parameters) 
with the quenched curve in 
the equilibrium abundance data in Figures \ref{fig:13} and \ref{fig:12}. The observed abundance is taken 
from \cite{Barman2015}. The region between the solid blue lines marks the observed abundance for \ch{CH4}. It can be 
seen that the T-P profile cuts this region for all four metallicities, but the corresponding  log$_{10}$($K_{zz}$) value
lies between 7 and 10. Therefore, \ch{CH4} cannot constrain the metallicity value but provides the $K_{zz}$ value. 
When we examine the observed CO abundance, it can only be constrained by the solar metallicity (see Figure \ref{fig:12}). 
Since a large part of the T-P profile lies in the CO dominated region, it cannot constrain $K_{zz}$; in fact all the $K_{zz}$ 
values are possible. Also, for the subsolar metallicity, the \ch{CO} abundance is very low ($\ch{CO}_{mix} = 10^{-5}$) throughout 
the thermal profile. For the supersolar metallicity, the quench level to achieve observational CO abundance ($\ch{CO}_{mix} \approx 10^{-4}$) 
lies in low-temperature and high-pressure regions, which require low internal temperature. Simultaneous constrain on the observed 
\ch{CH4} and \ch{CO} abundance conclude that the solar metallicity can explain the abundance along with $K_{zz}$ = 10$^7$ to 
10$^{10}$ cm$^2$ s$^{-1}$. The same conclusion has been made by \cite{Moses2016} with a kinetic/transport model. They use $K_
{zz}$ = 4 $\times$ 10$^7$ cm$^2$ s$^{-1}$, the solar metallicity, and C/O = 0.66 to best fit the observed abundance. 

\item 
In the second case, we took GJ 436 b (T-P profile taken from \cite{Line2014}, see Table 1 for parameters), 
and overplotted the observed \ch{CH4} ($10^{-9}<$ \ch{CH4} mole fraction $<10^{-8}$) and 
CO ($3.4 \times 10^{-3}\approx$ \ch{CO} mole fraction) abundance with the quenched curve 
in the equilibrium abundance data in Figures \ref{fig:14} and \ref{fig:15}. The observed \ch{CH4} abundance cannot be achieved 
from the parameter range used in this study. However, the \ch{CO} abundance can be achieved for high metallicity ($>100 \times$ 
solar), which is also concluded in \cite{Line2014}. 
It is to be noted that the current study only focuses on the solar C/O ratio; therefore, it is unable to explain the \ch{CH4} abundance. 
The super-solar C/O ratio is required to reach the observed \ch{CH4} abundance, as increasing the C/O ratio shifts the \ch{CH4-CO} 
boundary towards lower pressure and higher temperature.

\item  In the third case, we took HD 189733 b and overplotted the observed \ch{CH4} (10$^{-5}$ $<$ \ch{CH4} mole fraction 
$<$ 2$\times 10^{-5}$) and CO (2$\times10^{-5}$ $<$ \ch{CO} mole fraction $< 3\times 10^{-2}$) abundances with the quenched curve in the equilibrium abundance data in Figures \ref{fig:16} and \ref{fig:17}. The observed abundance is adopted from \cite{Line2014}. We find that high transport (K$_{zz}$$>$ 10$^{10}\text{ cm}^2\text{ s}^{-1}$) along with 0.1-10 $\times$ solar metallicity can constrain the observed \ch{CH4} and \ch{CO} abundances.

\item WASP-39 b is among the first exoplanets whose atmosphere has been constrained by JWST, and its 
observed atmospheric metallicity is found to be [Fe/H] = 1 \citep{Tsai2022, JWSTTECERST2022, Rustamkulov2022}. 
We constrained the quenched abundance of \ch{CH4} and \ch{CO} for various vertical mixing strengths for WASP-39 b. In 
Figures \ref{fig:WASP29b_CH4} and \ref{fig:WASP29b_CO}, we have plotted the thermal profile of WASP-39 b and the quenched 
curve on the equilibrium data of \ch{CH4} and \ch{CO}. The thermal profile is taken from \cite{Tsai2022} (see Table 1 for parameters), 
and \cite{JWSTTECERST2022}. 
We find that the \ch{CH4} and CO mixing ratios are independent of vertical mixing for $K_{zz}<10^{10}$ cm$^2$ s$^{-1}$, 
since the part of the thermal profile where quenching of \ch{CH4} takes place remains close to the contour of 
\ch{CH4} $\sim$ $10^{-7}$ mixing ratio. For high vertical mixing $K_{zz}>10^{10}$ cm$^2$ s$^{-1}$, the quenched \ch{CH4}
mixing ratio can increase by one order of magnitude. For [Fe/H] = 1, the thermal profile lies in the \ch{CO} dominated region, 
and the \ch{CO} mixing ratio for all vertical mixing is 4$\times$10$^{-3}$. CO remains in equilibrium, and \ch{H2O} is 
the dominant species of O for the solar C/O ratio when [Fe/H] = 1 (i.e., 10 $\times$ solar metallicity), and \ch{CO2} remains in 
equilibrium with \ch{CO} and \ch{H2O}. This 
makes sure that CO, \ch{CO2}, and \ch{H2O} are independent of vertical mixing and remain in equilibrium for WASP-39 b. In Figure 
\ref{fig:WASP_29b}, we have shown the equilibrium mixing ratios of CO, \ch{CO2}, and \ch{H2O} for 0.1, 1, 10, and 100 $\times$ solar 
metallicity. The \ch{CO2}, \ch{H2O}, and \ch{CO} abundance profiles for 10 $\times$ solar metallicity are qualitatively similar to 
the retrieved abundance from \cite{JWSTTECERST2022}. 
\end{itemize}
 
\clearpage

\subsection{Comparison with analytical expressions for chemical timescale}

\begin{figure}[h]
	\centering
	\includegraphics[width=0.4\textwidth]{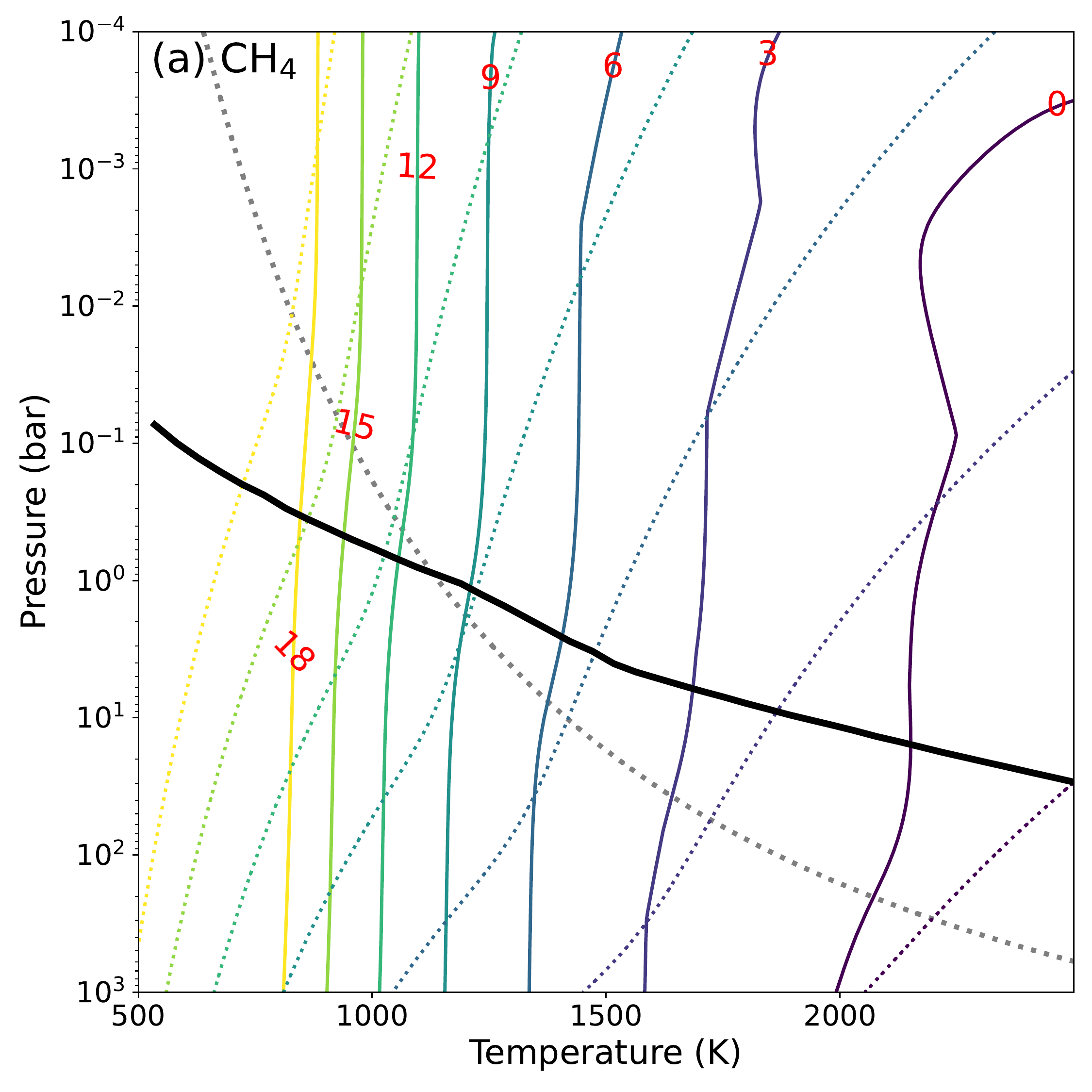}
	\includegraphics[width=0.4\textwidth]{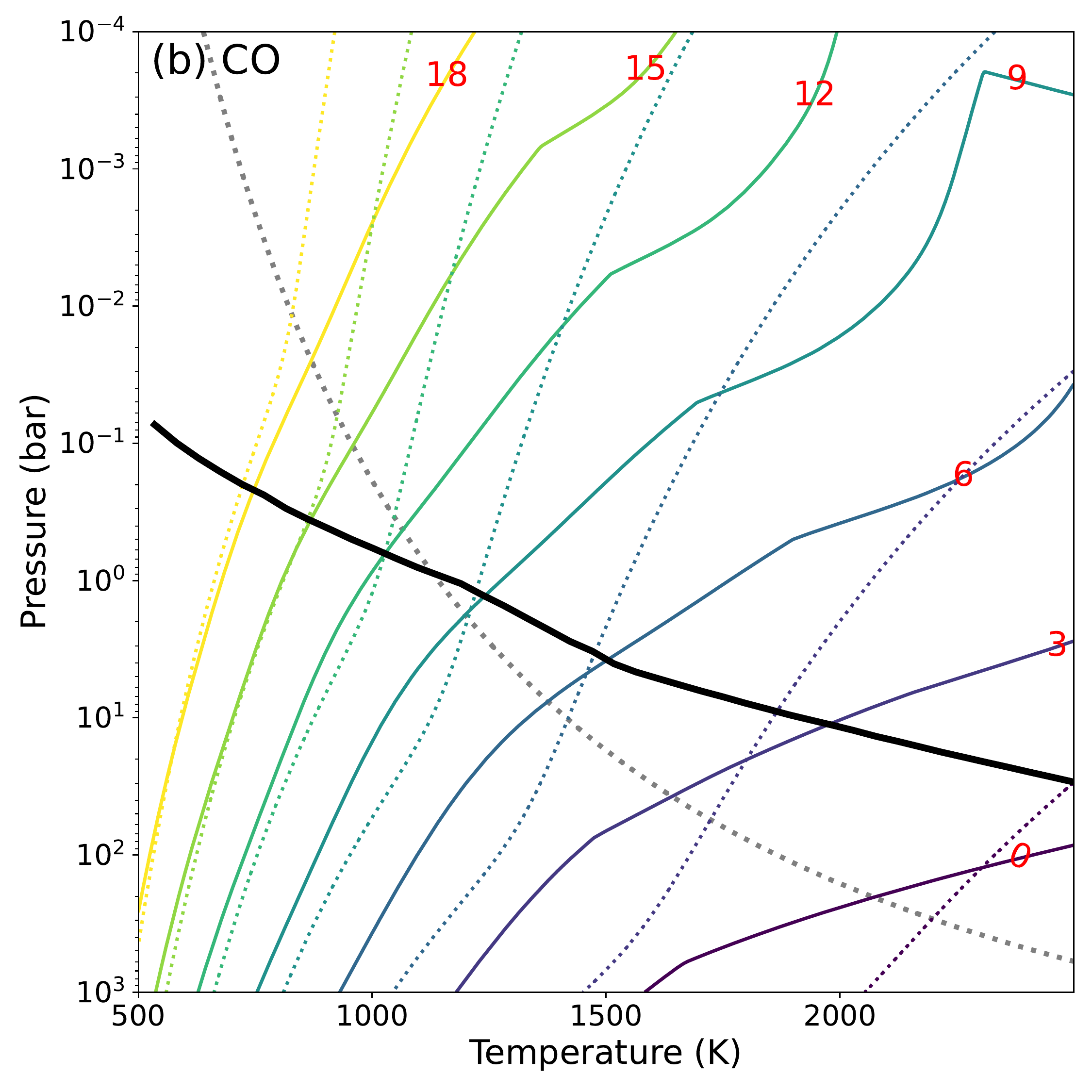}
	\includegraphics[width=0.4\textwidth]{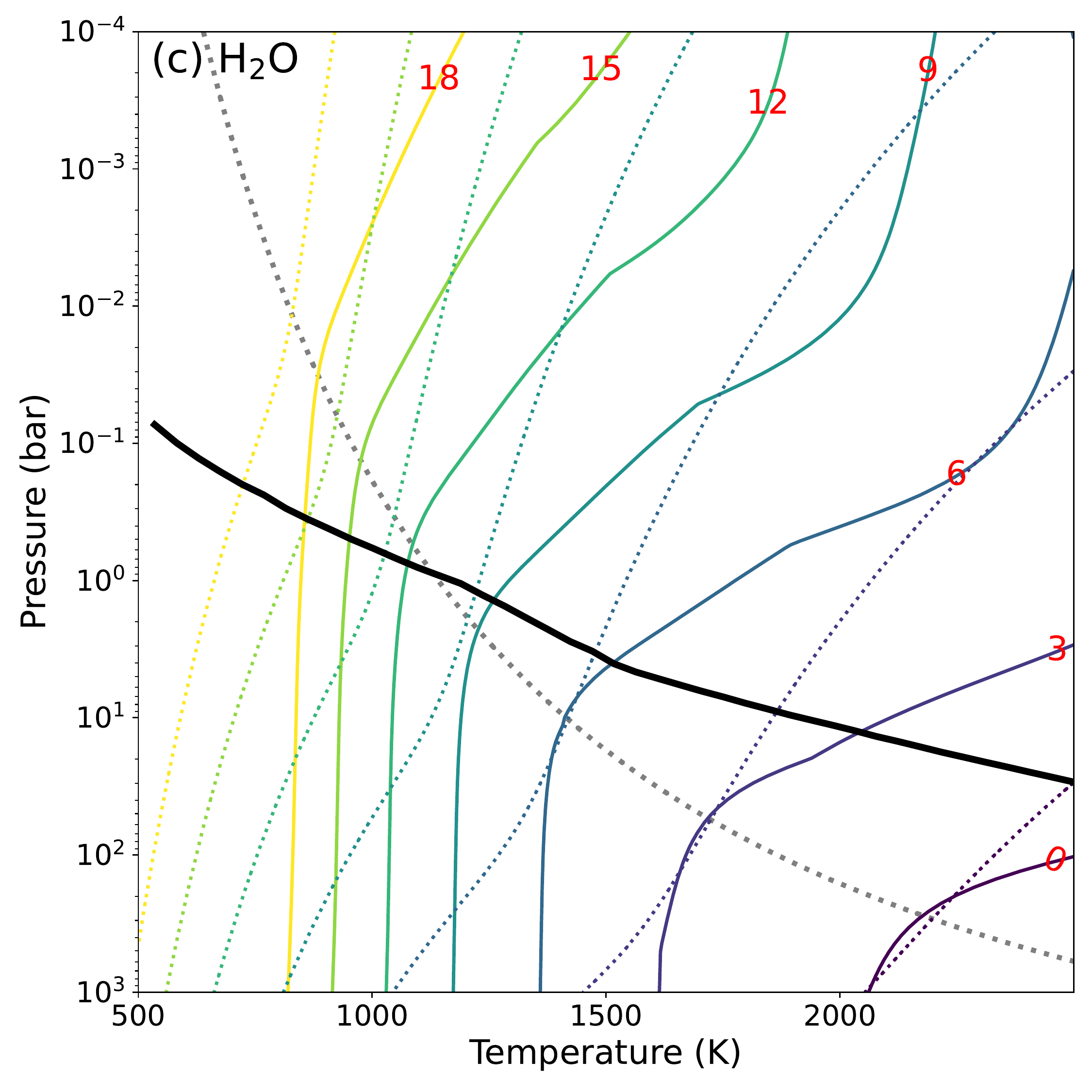}
	\includegraphics[width=0.4\textwidth]{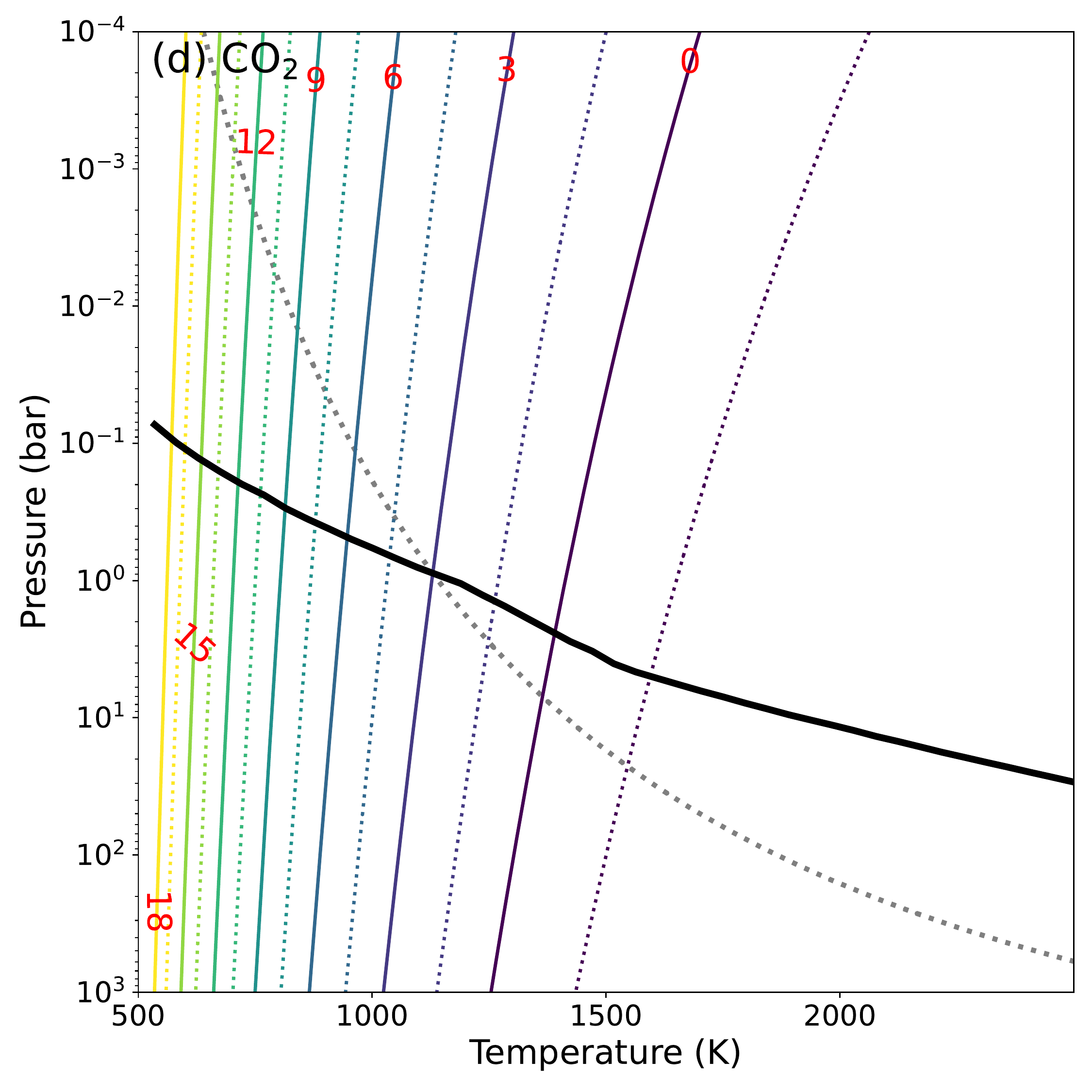}
		
	\caption{The contours of the chemical timescales of (a) \ch{CH4}, (b) \ch{CO}, (c) \ch{H2O}, and (d) \ch{CO2} obtained 
from our study (denoted by the colored solid lines) along with the timescales computed from the analytical expression from 
\cite{Zahnle2014} are shown in the log10 scale. The black dotted line represents the equal-abundance curve of \ch{CH4-CO} and the black solid 
line is the T-P profile taken from \cite{Zahnle2014} for $T_{eff}$ = 600 K and $g = 10^3$ cm s$^{-2}$. }\label{fig:comparision}
\end{figure}

\begin{figure}[h]
	\centering
	\includegraphics[width=0.7\textwidth]{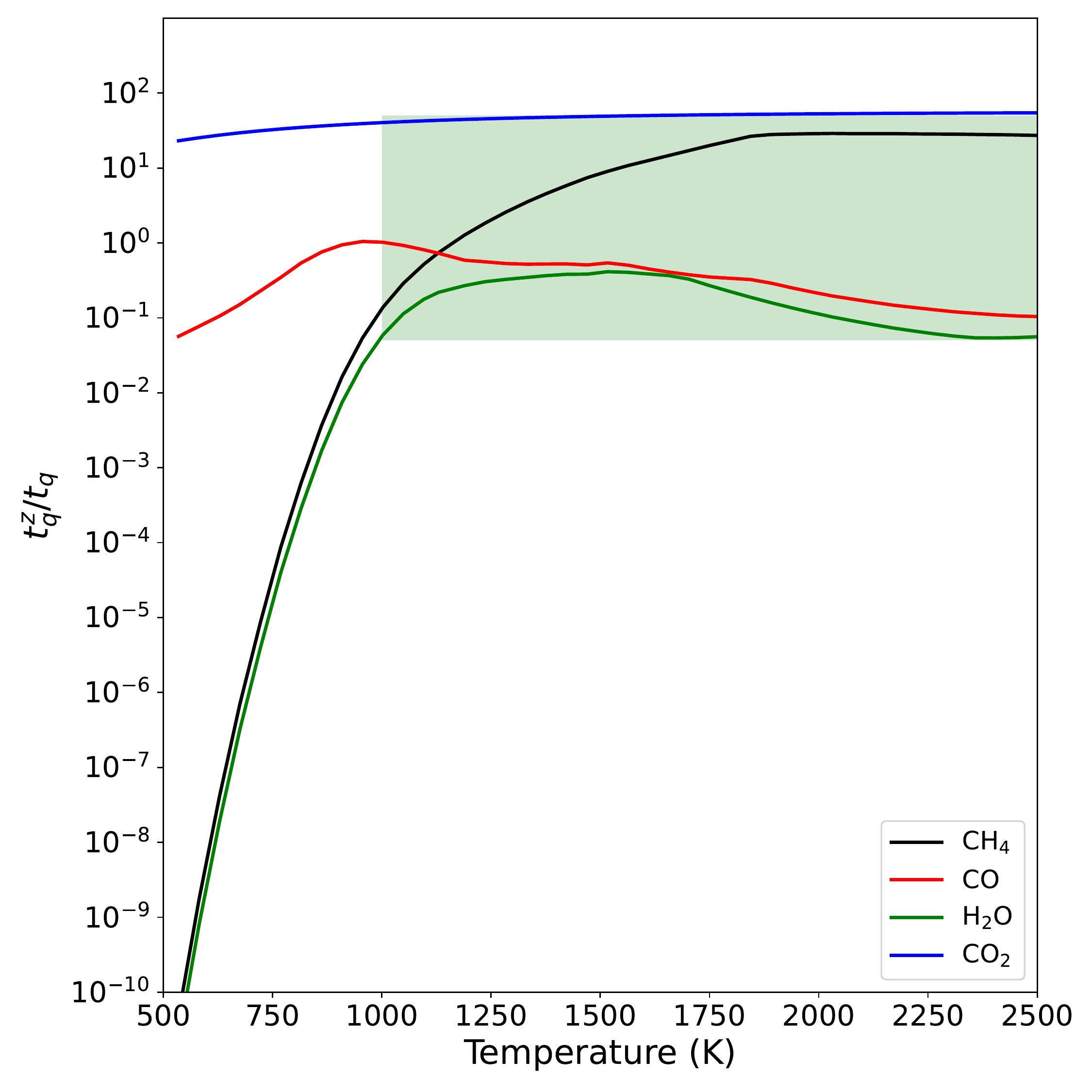}
	\caption{The ratio of the chemical timescale obtained in our study ($t_{q}$) and from \cite{Zahnle2014} ($t_q^z$) are shown, 
for the thermal profile shown in Figure~\ref{fig:comparision}.}\label{fig:comparision_all}
\end{figure}

We compare the chemical timescale from our model with \cite{Zahnle2014}, in which authors provided analytical expressions for the chemical 
timescale to calculate the quenched pressure and temperature. Their analytical expressions applied to self-luminous exoplanets 
and they assumed the mixing length to be 1 $\times$ H. 
In Figure \ref{fig:comparision}, we have compared the chemical timescale from both 
the studies. We have taken the thermal profile (black solid line) from \cite{Zahnle2014} for $T_{eff}$ = 600 K and $g = 10^3$ cm s$^{-2}$. 
The chemical timescales from our model deviate significantly from \cite{Zahnle2014} in the low-temperature and high-pressure region for 
\ch{CH4} and \ch{H2O}, and in the low-pressure and low-temperature region for \ch{CO}. The green patch in  Figure \ref{fig:comparision_all} highlights 
the region where our results are in agreement with \cite{Zahnle2014}.
It is to be noted that temperature falls very rapidly with pressure in the 
region where the quench level lies. The chemical timescale is a strong function of temperature; as a result, the deviation of a few orders 
of magnitude in the chemical timescale only shifts the quench level by a few factors in the pressure scale. The resulting change in the \ch{CH4} 
or \ch{CO} mixing ratios can be small. The thermal profile remains close to the contour of (\ch{CH4}/\ch{CO}), change of the chemical timescale by a few orders of magnitude does not 
change the \ch{CH4} or \ch{CO} mixing ratio significantly. 
It is clear from  Figure 5 of \cite{Zahnle2014} that the authors did not consider $K_{zz}$ below 10$^4$ cm$^2$ s$^{-1}$ and therefore, they 
did not get the quenched temperature below 1000 K. Thus, extrapolating the analytical expression for $T<1000$ K can give erroneous results as is evident
in Figure \ref{fig:comparision_all}. 

\section{Discussion}

\subsection{\ch{CH4} Depletion}

In the solar system, methane is observed in all the giant planets, and its formation can be explained with the 
low-temperature chemistry \citep{Guillot1999}. In the exoplanet atmosphere, methane is found to be depleted and observed 
only in a few exoplanets \citep{Madhusudhan2019, Molaverdikhani2020}. The methane depletion can be related to the presence of the 
disequilibrium processes; \cite{Molaverdikhani2019} found that strong vertical mixing can increase or decrease the methane abundance 
in the atmosphere. Depletion of methane can also result from the presence of clouds that heat the atmosphere and make methane less 
dominant \citep{Molaverdikhani2020} and the photochemistry can also deplete methane in the photochemical region of the 
atmosphere (P $<$ 0.1 mbar). In this section, we discuss the possibility of methane depletion in the presence of vertical mixing.

\begin{itemize}
\item As discussed in Section \ref{S-Equi}, the abundance of methane increases with metallicity only at the high-pressure and low-temperature 
region where methane is the dominant species of carbon. \ch{CH4} remains constant with metallicity in the high-temperature and 
low-pressure region where \ch{CO} is the dominant C-bearing species. 
As metallicity is increased, the \ch{CO} abundance first increases linearly at the high-temperature region, and the metallicity dependence gets stronger with 
decreasing temperature. Thus, increasing metallicity decreases the relative abundance of 
\ch{CH4}, so in a high metallicity atmosphere, \ch{CH4} can be depleted.

\item Vertical mixing can support the low methane abundance. In Figure \ref{quenching_contiur_plot}, for most of the 
temperature-pressure range, the CO quench level lies well below the \ch{CH4} quench level. Thus \ch{CO} is quenched 
in high-temperature and high-pressure regions of the atmosphere. In contrast, \ch{CH4} is quenched in the low-temperature 
region. 
	
\item As shown in \S 4 of \cite{Zahnle2014}, the \ch{CH4-CO} boundary is always shallower than the adiabatic T-P profile. 
This suggests that the adiabatic T-P profile lies in the \ch{CH4} region for low internal temperature exoplanets. 
The adiabatic T-P profile intersects the \ch{CH4-CO} boundary for high internal temperature. In the high internal temperature 
case, the CO quench level can lie in the CO-dominated region, and CO remains the primary carbon source. However, the T-P profile 
prefers \ch{CH4} over \ch{CO} in the infrared photosphere. Thus, the high internal temperature of the exoplanets' T-P profile can quench 
CO in the CO-dominated region, and methane can be depleted in the infrared photosphere. There is a discrepancy between 
the distribution of the observed radii of the hot Jupiters and the planetary evolution model, which is known as the radius anomaly 
\citep{Thorngren2018}. A possible explanation of the radius anomaly is the high internal temperature \citep{Fortney2021}. 
Thus high internal temperature and high metallicity can result in \ch{CH4} depletion.
	
\end{itemize}

\subsection{Effect of Clouds and Hazes}
The change in the thermal structure can alter the position of the quench level. In the present study, the thermal profile is taken 
as an input parameter. The clouds of \ch{H2O-CH4} are formed at 100-400 K for $10^{2}$ to $10^{-3}$ bar, which does not fall in the 
parameter space that we studied. However, the presence of other clouds (e.g., MgSiO$_3$, Mg$_2$SiO$_4$, Al$_2$O$_3$, Na$_2$S, KCl) 
formed in the parameter range can affect the thermal structure of the atmosphere \citep{Marley2013, Madhusudhan2016, Poser2019}. 
The production of hazes from the photodissociation of \ch{CH4-HCN-C2H2} also affects the atmospheric extinction and contributes to 
shaping the thermal profile of the atmosphere \citep{Kawashima2019}. In this study we have used the solar C/O ratio which is 
adopted from \cite{Lodders2009}, thus our results are not valid for any atmosphere with a different C/O ratio. The mineral clouds 
(like silicates) can remove a significant amount of O \citep{Madhusudhan2016}, and the resulting C/O ratio can deviate 
from the value we have used in this study. Investigating the effect of clouds and hazes on the quench level 
is beyond the scope of this study. We focus on the general effect of metallicity on the quench level rather than on particular 
exoplanets, which makes this study independent of the presence of clouds and hazes. 

\subsection{Effect of C/O vs Metallicity}
In our study, we fixed the C/O ratio to the solar value. However, for a different C/O ratio, the result in this study can change. 
The change in the C/O ratio changes the bulk abundance of C with respect to O. This results in changes in the equilibrium composition 
of the atmosphere, leading to a change in the chemical timescale of the species \citep{Madhusudhan2012, Moses2013a, Moses2013}. 
\cite{Tsai2018} explored the effect of C/O ratio on the conversion pathways of $\ch{CO}\rightleftarrows\ch{CH4}$. They found that for C/O $>$ 1, \ch{C2H4}, 
\ch{C2H3}, and \ch{C2H2} also come as intermediate molecules in the conversion of $\ch{CO}\rightleftarrows\ch{CH4}$. For solar C/O ratio, \ch{H2O} and \ch{CO2} 
remain in equilibrium, and quenching does not affect the \ch{H2O} abundance, as \ch{H2O} is the major O species at the quench level. 
Moreover, the $\ch{CO2}\rightleftarrows\ch{CO}$ conversion is relatively fast, and \ch{CO2} remains in equilibrium with \ch{H2O} and CO. However, for 
the super-solar C/O ratio, \ch{H2O} and \ch{CO2} can quench deep in the atmosphere resulting in a more pronounced effect of transport 
on these molecules. In our future work, we will analyze the effect of the C/O ratio in the parameter space. 

\section{Conclusion}
We present a detailed study of the effect of atmospheric metallicity on the abundance of \ch{CO-CH4-CO2-H2O} in the presence of transport. We have used the quenching approximation to find the effect on the quenched abundance. We have made a 3D grid in the temperature (500 to 2500 K), 
pressure (0.01 mbar to 1 kbar), and metallicity  (0.1-1000 $\times$ solar metallicity) space for this study. We first calculate the chemical equilibrium 
abundance at the grid points and use a network analysis tool to find the RLS in the conversion of $\ch{CH4}\leftrightarrows\ch{CO}$. 
We then use the RLS to find the chemical timescale and compare it with the vertical mixing timescale of the atmosphere to find the quench 
levels. We find all possible quench levels in the grid and then use the quench level data to understand the effect of metallicity on 
the abundance of \ch{CO-CH4-CO2-H2O}. Our conclusions are as follows:

\begin{itemize}
\item The equilibrium abundance profiles are in accordance with  \cite{Moses2013a, Zahnle2014, Venot2014}. The \ch{CO}, \ch{H2O}, and 
\ch{CO2} abundances increase with metallicity; however, the \ch{CH4} abundance increases only in the \ch{CH4} dominant region. 
The equal-abundance curve of \ch{CH4} and \ch{CO} divides the temperature-pressure space of \ch{CH4-CO-CO2-H2O} into two regions. 
In the \ch{CO} dominant region, \ch{CH4}, \ch{CO} and \ch{CO2} vary as $\approx$ [Fe/H]$^{0}$, $\approx$ [Fe/H]$^{1}$, $\approx$ [Fe/H]$^{2}$ 
respectively. However, in the \ch{CH4} dominant region, the increment in \ch{CH4} is proportional to the metallicity, while \ch{CO} and \ch{CO2} 
increase more rapidly with metallicity than in the first region. 

\item The metallicity changes the dominance area of RLS in temperature-pressure space. In the area where atomic hydrogen is 
dominant over \ch{H2} (above the $10^{-8}$ bar-1750 K to $10^{-3}$ bar-2500 K, see Figure \ref{fig:scale_height}), the different 
dehydrogenation ($\ch{CH4}\rightarrow\ch{CO}$) or hydrogenation ($\ch{CO}\rightarrow\ch{CH4}$) reactions become RLS. For most of the 
temperature-pressure range, the RLS of $\tau_{\ch{H2O}}$ is the same as the RLS for $\tau_{\ch{CH4}}$. However, for high-temperature and 
low-pressure region, the C atom for the conversion of \ch{H2O} into \ch{CO} comes from \ch{HCN} and \ch{CN}. The RLS of this region does not follow 
the RLS of $\tau_{\ch{CH4}}$. 

\item In the conversion of $\ch{CH4}\rightarrow\ch{CO}$, the $\ch{H2}\rightleftarrows\ch{H}$ conversion also plays an important part. The 
$\ch{H2}\rightleftarrows\ch{H}$ conversion term increases with metallicity as this term is proportional to the CO abundance. However, the 
timescale of the rate-limiting step decreases with metallicity which makes $\tau_{\ch{CH4}}$ a complex function of metallicity. In 
converting $\ch{CO}\rightarrow\ch{CH4}$, the  $\ch{H2}\rightleftarrows\ch{H}$ conversion term contributes only for [Fe/H] = 3, $T<1000$ K, $P>10^2$ bar, and for 
other parameter ranges the RLS term mostly contribute to $\tau_{\ch{CO}}$. The timescale of the RLS decreases with 
metallicity for $P<10^{-3}$ bar  and $T\gtrapprox$ 2250 K, and for other parameter ranges it is not affected by the metallicity.

\item We compare the verticle mixing and chemical timescale to study the effect of metallicity on the quench level. For a fixed value of 
$\eta$ ($\tau_{mix} = (\eta H)^2 / K_{zz}$), the vertical mixing timescale is decreased by two orders of magnitude as the metallicity 
increases from 0.1 to 1000 $\times$ solar metallicity. Moreover, the \ch{CH4} timescale changes by four orders of magnitude, and the \ch{CO} timescale changes by 
less than one order of magnitude. So the quench level of \ch{CO} follows the vertical mixing timescale, and the quench level of \ch{CH4} 
follows the chemical timescale. \ch{CO} quenches at at a higher pressure compared to the \ch{CH4} quench level. For high 
internal temperature, this favors \ch{CO} over \ch{CH4} in the transport-dominated region, as CO quenches in the atmospheric 
region where CO dominates over \ch{CH4}. The \ch{CO} quench level moves to the high-pressure region with an increase in metallicity. 
However, the quench level of \ch{CH4} shifts towards low pressure with metallicity, where the  $\ch{H2}\rightleftarrows\ch{H}$ conversion term dominates in 
$\tau_{\ch{CH4}}$.

\item We have compared the quenching 
approximation result for two exoplanets for \ch{CO} and \ch{CH4} with the photochemistry-transport model. For GJ 1214 b, the \ch{CO} abundance is 
accurate within $\approx$ a factor of five for low metallicity. For high metallicity, this is accurate within $\approx$ a factor of two. 
In the case of \ch{CH4}, this is accurate within $\approx$ a factor of two for low metallicity. For high metallicity, it is accurate 
within $\approx$ a factor of seven. For HD 189733 b, the CO abundance is accurate within $\approx$ a factor of two. However, for \ch{CH4}, it is 
accurate within $\approx$ an order of magnitude. The accuracy of the quenching approximation can improve significantly by incorporating 
the mixing length using the method given by \cite{Smith1998}. However, we do not include this method here as we only test the quenching 
approximation for a broad range of parameters and do not aim to constrain the abundance for a specific exoplanet. 

\item We use the quench level data to constrain the mixing strength and atmospheric metallicity for four exoplanets, namely HR 8799 b, HD 189733 b, GJ 436 b, and WASP-39 b. For HR 8799 b and HD 189733 b, we find that observation constrains of \ch{CH4} and \ch{CO} abundance along with the quench level data can constrain the metallicity and $K_{zz}$. However, for  GJ 436 b, we can constrain only the metallicity using the CO abundance, whereas the observed \ch{CH4} abundance is indicative of a higher C/O ratio in the atmosphere. 

\end{itemize}

\section*{Acknowledgements}
The authors thank the anonymous referee for constructive comments which strengthened the paper.
We thank Nikku Madhusudhan for his helpful insights, for reading the manuscript, and giving valuable suggestions to improve this work. We thank Sana Ahmed 
for suggestions that improved the overall readability of the manuscript.
The work done at the Physical Research Laboratory is supported by the Department of Space, Government of India.

\appendix

\section{Photochemistry-Transport Model}
We have developed a FORTRAN-based 1D atmospheric model that deals with the transport and photochemistry disequilibrium processes. The model 
uses the plane-parallel approximation and divides the atmosphere into N layers, and solves the mass continuity equation for each species in 
the network in each layer. The model requires the following input parameters:

\begin{itemize}
	\item Thermal profile of the atmosphere.
	\item Initial mixing ratio of the atmosphere or elemental abundance.
	\item Eddy diffusion coefficient to mimic the turbulent mixing.
	\item The host star flux at the top of the atmosphere for calculating the photochemical reaction rates.
	\item Physical parameters of the exoplanet, such as surface gravity.
	\item Boundary conditions at the top and bottom of the atmosphere.
	\item Chemical network.
\end{itemize}
\subsection{Governing Equations}
To find the atmospheric composition, the  1D atmospheric model solves the coupled one-dimensional mass continuity equation:

\begin{align}
\frac{\partial n_i}{\partial t} &= P_i - n_iL_i - \frac{\partial \phi_i}{\partial z}, \label{eq:M1}
\end{align}
\begin{align*}
P_{i} &= \sum_{j = 1,m_{ji}>0}^{N_{R}} K_{j}m_{ji}\prod_{k=1,m_{jk}>0}^{N_s}  n_{k}^{m_{jk}},\\
n_iL_{i} &= \sum_{j =  1,m_{ji}<0}^{N_{R}} K_{j}m_{ji}\prod_{k=1,m_{jk}>0}^{N_s} n_{k}^{m_{jk}},
\end{align*}
where $n_i$, $P_i$, and $L_i$ are the number density, the production and the loss rate of the $i$-th chemical
species. $N_R$, $N_s$, and $K_j$ are the total number of reactions, the total number of species, and the rate constant of the $j$-th reaction in the network. $m_{ji}$ is the stoichiometric
coefficient of the $j$-th reaction of the $i$-th species \cite{Laidler+1996+149+192}, and $\phi_{i}$ is the transport flux of the $i$-th species. This transport flux includes Eddy diffusion and molecular diffusion and has the following form adopted from \cite{Hu2012}:

\begin{equation}
\phi_{i}  = \phi_{i,E}+\phi_{i,m} = -Kn\frac{\partial f_{i}}{\partial z}  -Dn \frac{\partial f_{i}}{\partial z} +D n \Bigg[\frac{1}{H} - \frac{1}{H_{i}}-\alpha_{T}\frac{1}{T}\frac{d T}{d z}\Bigg]. \label{eq:M2}
\end{equation}

The Eddy and molecular diffusion fluxes for the $i$-th species are $\phi_{i,E}$ and $\phi_{i,m}$ respectively. $K$ and $D$ are the Eddy and molecular diffusion coefficients. $H$ and $H_{i}$ are the mean scale height and the molecular scale height respectively, and $\alpha_{T}$ is the thermal diffusion factor [\cite{Hu2012}, \cite{Banks1973}]. The transport processes are required to simulate atmospheric mixing, which is one of the physical processes responsible for changing the atmospheric composition from its chemical equilibrium abundance. The first term on the right-side in Equation \ref{eq:M2} is the Eddy diffusion term, mimicking 3D mixing in a 1D model. The last two terms represent molecular diffusion in the atmosphere. The molecular diffusion caused an uplift in the lighter species in the atmosphere, and the heavier species tend to settle down. As a result, the species are settled at different heights and the atmosphere becomes layered \citep{Madhusudhan2016}.

In general, in the high-pressure and high-temperature regions of the atmosphere, where the chemical timescale is small compared to the mixing timescale, the atmospheric abundance follows the chemical equilibrium abundance. In the low-pressure and low-temperature regions of the atmosphere, the mixing timescale becomes smaller than the chemical timescale, and transport changes the atmospheric abundance from its chemical equilibrium abundance.  \cite{Prinn1977} used the transport mixing for the first time and found that the high abundance of \ch{CO} in the upper atmosphere of Jupiter is the result of the transport of CO from the high-pressure region where it is stable. In continuation, several studies use Eddy and molecular diffusion in 1D models to include the transport processes \citep{Allen1981,  Line2011, Hu2012,  Madhusudhan2012, Visscher2012, Krasnopolsky2013,  Moses2013, Moses2013a, Agundez2014, Hu2015, Rimmer2016, Tsai2017, Rimmer2021}. The gas kinetics theory can calculate the molecular diffusion coefficient, and here we adopted the method given by \cite{Chapman1991}. The Eddy diffusion coefficient has large uncertainty and can be calculated by multiplying the mixing length with the characteristic speed of convection. The mixing length cannot be calculated from the first principle as it is a chaotic function of many atmospheric parameters. However, one can constrain it by the 3D general circulation model \citep{Heng2017, Madhusudhan2016}.

The solution we seek for Equation \ref{eq:M1} is a set of abundances of chemical species for which the right side of Equation \ref{eq:M1} is zero. In another way, the total production, loss, and transport are summed up to zero for a steady-state solution (assuming the parameters are fixed for the system). The discretized version of Equations \ref{eq:M1} and \ref{eq:M2} are as follows:

\begin{align*}
\frac{\partial n_{z,i}}{\partial t} &= P_{z,i} - L_{z,i} - \frac{\phi_{z+1/2,i} - \phi_{z-1/2,i} }{\Delta \text{Z}} \\
\phi_{z+1/2} &= \phi_{z+1/2,i}^{Eddy}+\phi_{z+1/2,i}^{molecular} \\
& = -K_{z+1/2} N_{z+1/2}\frac{f_{z+1,i} - f_{z,i}}{\Delta \text{Z}}  -D_{z+1/2,i} N_{z+1/2} \frac{f_{z+1,i} - f_{z,i}}{\Delta \text{Z}}\\
&\quad + D_{z+1/2,i} N_{z+1/2}\frac{f_{z+1,i} + f_{z,i}}{2} \Bigg[\frac{m_{z}g}{k_{b}T_{z+1/2}} - \frac{m_{z,i}g}{k_{b}T_{z+1/2}}-\alpha_{T}\frac{1}{T_{z+1/2}}\frac{T_{z+1} - T_{z}}{\Delta \text{Z}}\Bigg],
\end{align*}

and the governing equation in the discretization form is as follows:
\begin{align*}
\frac{\partial n_{i}}{\partial t} &= P_{i} - L_{i} + \left(k_{i+1/2}\frac{N_{i+1/2}}{N_{i+1}}- d_{i+1/2}\frac{N_{i+1/2}}{N_{i+1}}\right)n_{i+1}\\
&-\left(k_{i+1/2} \frac{N_{i+1/2}}{N_{i}} + d_{i+1/2}\frac{N_{i+1/2}}{N_{i}} + k_{i-1/2}\frac{N_{i-1/2}}{N_{i}}-d_{i-1/2}\frac{N_{i-1/2}}{N_{i}} \right)n_{i}\\
& +\left(k_{i-1/2} \frac{N_{i-1/2}}{N_{i-1}}+d_{i-1/2}\frac{N_{i-1/2}}{N_{i-1}}\right)n_{i-1}.
\end{align*}
$i$ and $z$ indices run in species and the atmospheric layers, so the number of coupled equations are N$_{layers}\times$N$_{species}$. $k_b$, $m_z$, and $m_{z,i}$ are the Boltzmann constant, the mean molecular mass, and $i$-th species mass in $z$-th atmospheric layer. $g$ is the gravity of the planet at the $z$-th atmospheric layer, $T_z$ is the temperature, and $f_{z,i}$ is the mixing ratio of the $i$-th species at the $z$-th layer. Here, the aim is to find a set of species in each layer, so that the right side of the equation is zero for all the species in all the layers, which also satisfies the boundary condition. To do this, we first take a set of relevant initial abundances and evolve the system with time. The reaction rates vary over 20-30 orders of magnitude, making these equations stiff. Therefore, to arrive at numerical solutions, we use the semi-implicit Rosenbrock fourth-order integration method which can handle the stiffness of these equations \citep{Rosenbrock1963, Shampine1982}.

\subsection{Chemical Network}
Our primary chemical network is taken from \cite{Rimmer2016} and \cite{Rimmer2019}, in which we keep the chemical network from  \cite{Tsai2017} as a complete subset,
i.e., for the reactions which are common to both the networks, we used the reaction rates from \cite{Tsai2017}. Our network includes 350 chemical species which are
made up of H, C, N, O, He, Na, Mg, Si, Cl, Ar, K, Ti, and Fe elements. These species are interconnected by more than 5000 chemical reactions. These
reactions include two-body neutral-neutral and ion-neutral reactions, three-body neutral and dissociation reactions, thermal ionization and recombination
reactions, and photochemical reactions. \cite{Rimmer2016} and \cite{Rimmer2019} compiled the network using reactions from NIST \citep{NIST2015}, KIDA \citep{Wakelam_2012}, \cite{Ikezoe1987, Sander2011, Tsai2017}.
Due to the availability of the VULCAN code from \cite{Tsai2017, Tsai2021}, we use it for the benchmarking of our model outputs with the VULCAN output. For this benchmarking, the network is adopted from \cite{Tsai2017,Tsai2021}, which contains
55 chemical species made up of H-C-N-O elements. These species are interconnected by
640 chemical reactions. In the network, 250 are two-body reactions, 50 are three-body reactions, and 40 are photochemical reactions. The two-body
and three-body reactions are reversible in the network, and the network is benchmarked for a temperature range of 500 to 2500 K in \cite{Tsai2017}.
The 55 species are \ch{H}, \ch{H2O}, \ch{OH}, \ch{H2}, \ch{O}, \ch{CH}, \ch{C}, \ch{CH2}, \ch{CH3}, \ch{CH4}, \ch{C2}, \ch{C2H2}, \ch{C2H},
\ch{C2H3}, \ch{C2H4}, \ch{C2H5}, \ch{C2H6}, \ch{CO}, \ch{CO2}, \ch{CH2OH}, \ch{H2CO}, \ch{HCO}, \ch{CH3O}, \ch{CH3OH}, \ch{CH3CO}, \ch{O2},
\ch{H2CCO}, \ch{HCCO}, \ch{N}, \ch{NH}, \ch{CN}, \ch{HCN}, \ch{NO}, \ch{NH2}, \ch{N2}, \ch{NH3}, \ch{N2H2}, \ch{N2H}, \ch{N2H3}, \ch{N2H4},
\ch{HNO}, \ch{H2CN}, \ch{HNCO}, \ch{NO2}, \ch{N2O}, \ch{C4H2}, \ch{CH3NH}, \ch{CH2NH}, \ch{CH2NH2}, \ch{CH3NH2}, \ch{CH3CHO}, \ch{HNO2},
\ch{O(^1D)}, \ch{CH2(^1D)}, \ch{He}.  

The rate constants of the two-body neutral-neutral, ion-neutral, and dissociative recombination reactions are calculated by the
generalized Arrhenius equation \citep{Heng2016}
\begin{equation}
k = \alpha \left(\frac{\text{T}}{300\text{ K}}\right)^{\beta} \exp\left(-\frac{\gamma}{T}\right), \label{eq:12}
\end{equation}
where k (cm$^3$ s$^{-1}$) is the rate coefficient of the reaction, and $\alpha$, $\beta$,
and $\gamma$ are the Arrhenius coefficients for the specific reaction.

In a three-body reaction, the two reactants will combine to form one product in the presence of a third body. The
three-body reactions in the network include neutral reactions, dissociation reactions, and thermal ionization and recombination reactions.
The rate constant of three body reaction is calculated using low and high-pressure rates, which are computed using the modified Arrhenius expression:

\begin{align}
k_o &= \alpha_o \left(\frac{\text{T}}{300\text{ K}}\right)^{\beta_o}  \exp\left(-\frac{\gamma_o}{T}\right), \\
k_{\infty} &= \alpha_{\infty} \left(\frac{\text{T}}{300\text{ K}}\right)^{\beta_\infty} \exp\left(-\frac{\gamma_{\infty}}{T}\right), 
\end{align}
where $k_o$ (cm$^6$ s$^{-1}$) and $k_\infty$ (cm$^3$ s$^{-1}$) are the low-pressure and high-pressure rate
constants. The rate constant of a three body reaction in terms of low and high-pressure rate constants is as follows:

\begin{align}
k = \frac{k_{o}[M]}{1+\frac{k_{o}}{k_{\infty}}[M]},
\end{align}
where $[M]$ is the number density of the neutral third species. We refer the reader to see \S 2 in \cite{Rimmer2016} and
\S 4 in \cite{Tsai2017} for a detailed discussion on calculating different rate constants of reactions used in this model.

\subsubsection{Photochemical Rate Constants}
The rate constant of photodissociation reactions is a function of the number density of photons,
the branching ratio of the dissociation path (quantum yield), and the wavelength-dependent cross-section. The required photon density for the
photodissociation rate constant is calculated using the two-stream approximation of radiative transfer. The rate for photodissociation reaction is calculated as follows:

\begin{equation}
k = \int_{\lambda} \sigma(\lambda) N_\lambda \times BR \times d\lambda
\end{equation}
where $\sigma(\lambda)$ is the cross-section of the species, $N_\lambda$ is the number of photons in $[\lambda, \lambda + d\lambda]$ and $BR$ is the branching ratio \citep{Hu2012}.

If the radiation passes through some medium, it gets scattered, absorbed, or re-emitted in the medium itself. In the exoplanet atmosphere, the stellar radiation also gets scattered or absorbed in the atmosphere (for the time being, we ignore the re-emitted photon) that distributes the photon density in each atmospheric layer. The photochemistry will only depend upon the total number of photons present in the atmospheric layers. In order to find the photon flux in each layer, we have used the two-stream approximation to find the available photon count in each atmospheric layer. The two-stream approximation is a well known radiative transfer formulation which can efficiently compute the radiation propagation in the atmosphere \citep{Toon1989,Hu2012,Heng2014,Heng2017,Malik2017}. It only considers the vertical propagation of radiation. The horizontal propagation is negligible as we assume that the thickness of the atmosphere is incomparable to the radius of the planet. Depending upon the physical processes and parameter space, the equation for the two-stream approximation can take different forms, and here, we have used the two-stream approximation in the limit of coherent and isotropic scattering \citep{Heng2014}.

\begin{equation}
\begin{split}
F_{\uparrow j} &= A_1 \zeta_+ e^{\alpha \tau'_j} + A_2 \zeta_- e^{-\alpha \tau'_j} + \pi B_{j+}, \\
F_{\downarrow j} &= A_1 \zeta_- e^{\alpha \tau'_j} + A_2 \zeta_+ e^{-\alpha \tau'_j} + \pi B_{j-},\\
B' &= \frac{\partial B}{\partial \tau} = \frac{ B_j - B_{j-1} }{ \tau_j - \tau_{j-1} },\\
B_{j-} &= B_1 + B'(\tau_j - \tau_{j-1}) - \frac{B'}{2},\\
B_{j+} &= B_2 + B'(\tau_{j-1} - \tau_j) + \frac{B'}{2},\\
\zeta_{\pm} &= \frac{1}{2}[1\pm (1-\omega_0)^{\frac{1}{2}}].
\end{split} \label{eq:M3}
\end{equation}
The index $j$ represents the $j$-th atmospheric layer, and $F_{\uparrow j}$ and $F_{\downarrow j}$ are the outgoing and incoming flux in this layer. $A1$ and $A2$ are the integration constants which are calculated by the boundary condition. $\theta$, $\tau$, $\omega_0$ and $B$ are the polar angle in spherical coordinates, optical depth, single-scattering albedo and Planck function respectively. $\mu = \cos(\theta)$ and $\tau' = \frac{\tau}{\Bar{\mu}_{\uparrow}}$ is the slant optical depth.  We refer the reader to see the full derivation of Equation \ref{eq:M3} in \cite{Heng2014}.

Using the boundary condition, Equation \ref{eq:M3} becomes

\begin{equation}
\begin{split}
F_{\uparrow 1} &= \frac{1}{(\zeta_- \mathbf{T})^2 - \zeta_+^2} \Big[(\zeta_-^2 - \zeta_+^2)\mathbf{T}F_{\uparrow 2} - \zeta_- \zeta_+ (1-\mathbf{T}^2)F_{\downarrow 1}\\
& + \pi { B_{1+}(\zeta^2 \mathbf{T}^2 - \zeta_+^2) + B_{2+}\mathbf{T}^2(\zeta_+^2 - \zeta_-^2) +  B_{1-}\zeta_+ \zeta_- (1-\mathbf{T}^2) } \Big], \\
F_{\downarrow 2} &= \frac{1}{(\zeta_- \mathbf{T})^2 - \zeta_+^2} \Big[(\zeta_-^2 - \zeta_+^2)\mathbf{T}F_{\downarrow 1} - \zeta_- \zeta_+ (1-\mathbf{T}^2)F_{\uparrow 2}\\
& + \pi { B_{2-}(\zeta^2 \mathbf{T}^2 - \zeta_+^2) + B_{1-}\mathbf{T}^2(\zeta_+^2 - \zeta_-^2) +  B_{2+}\zeta_+ \zeta_- (1-\mathbf{T}^2) } \Big], \label{eq:M4}
\end{split}
\end{equation}
where $\mathbf{T} = e^{-D\delta \tau}$ and $\delta \tau = \tau_2 - \tau_1$. Equation \ref{eq:M4} is the two-stream solution for the radiation transfer between two consecutive boundary layers. The solution to this equation will give the diffuse radiation flux.  

\begin{figure}[b!]
	\centering
	\includegraphics[width=1\textwidth]{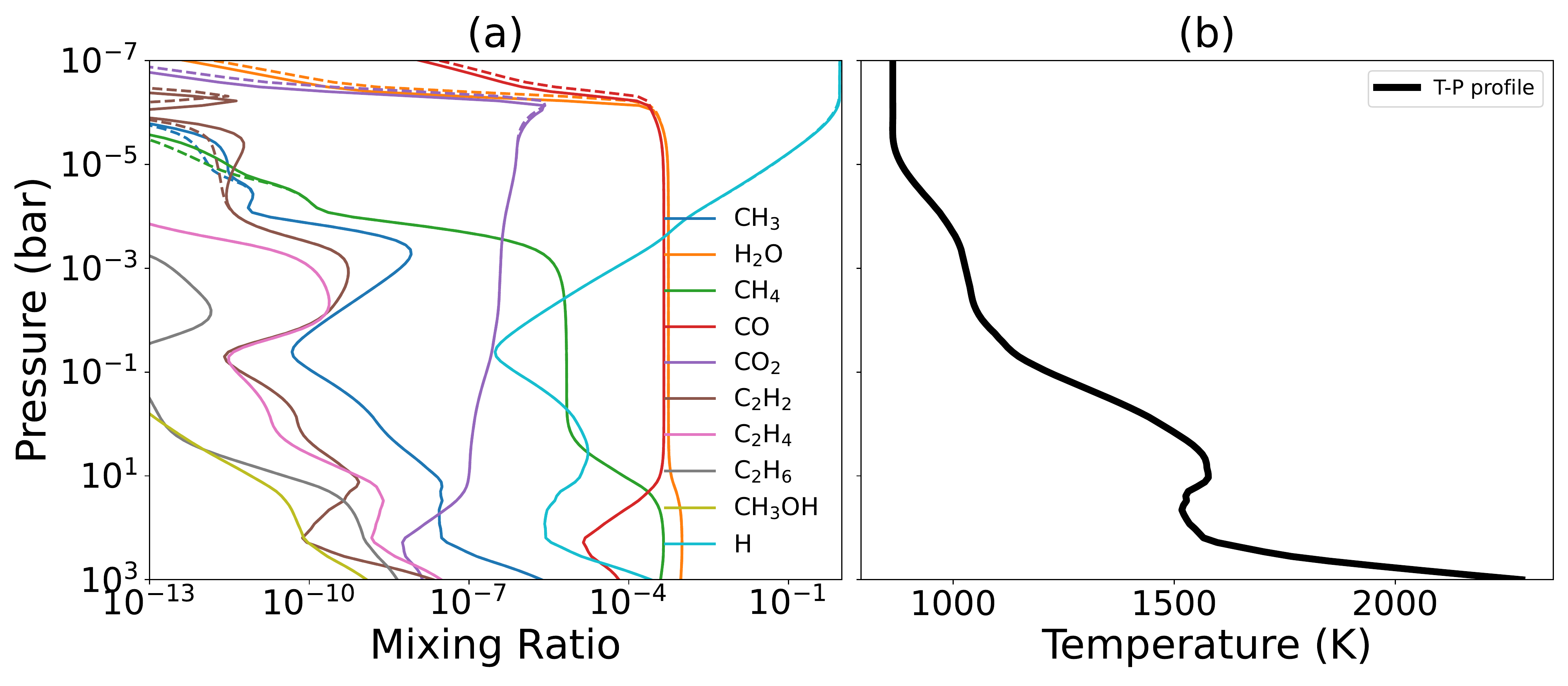}
	\caption{(a) The disequilibrium mixing ratio of HD 189733 b. The mixing ratios shown in dashed lines are calculated from our
		Fortran-based model and the solid lines represent the mixing ratios calculated from the VULCAN model \citep{Tsai2017, Tsai2021}. (b) The day averaged
		thermal profile of HD 189733 b \citep{Moses2011, Tsai2017}. }\label{fig:18}
\end{figure}

\subsubsection{Reverse Reaction Rate Constant}
To calculate the reverse reaction rate constant, we use the method described in \cite{Burcat2005, Heng2016, Tsai2017}. The forward reaction whose rate constant is $k_f$ has the following form:
\begin{align*}
\text{A + B}  &\rightarrow \text{C + D + E }.
\end{align*}

The reverse reaction of the above forward reaction with reverse rate constant ($k_r$) is the following:
\begin{align*}
\text{C + D + E}  &\rightarrow \text{A + B}.
\end{align*}
In chemical equilibrium, the forward and reverse reaction rates will become equal, and the ratio of the forward rate constant and the reverse rate constant is called the equilibrium constant ($k_c$) of the reaction or $k_c = k_f/k_r$. This equilibrium constant can be calculated using the thermodynamic coefficient (thermodynamic coefficient is the fitted coefficient of the polynomial fit of enthalpy and entropy in temperature space). The equation for the same is given below.

\begin{align*}
k_c = &(RT)^{\Delta \nu} \exp \bigg(\Delta a_1(\log T-1) + \Delta a_2 T/2 + \Delta a_3T^2/6 \\
&+ \Delta a_4 T^3/12 + \Delta a_5 T^4/20 - \Delta a_6/T + \Delta a_7
\bigg),
\end{align*}
where $R$ is the gas constant, $\Delta \nu$ is the difference between the number of reactants and the number of products, $T$ is the temperature in K and $\Delta a_i = a_i\text{(C + D + E)} - a_i\text{(A + B)}$, $i \in [1,7]$. $a_i$ are the thermodynamic coefficients for the species that take part in the reaction.
The reverse rate constant is given by the ratio of the forward rate constant and the equilibrium constant ($k_r = k_f/k_c$).
The thermodynamic coefficients data is taken from NASA-CAE database and \cite{Burcat2005}.

\subsection{Benchmarking}
In this section, we benchmark our disequilibrium model (transport and photochemistry) with the VULCAN model \citep{Tsai2017, Tsai2021}. For the benchmarking, we have calculated the chemical composition of HD 189733 b from our Fortran-based 1D model and from the open source VULCAN model. We have used the day averaged thermal profile \citep{Moses2011, Tsai2017}, and constant Eddy diffusion coefficient $K_{zz}$ = 10$^9$ cm$^2$ s$^{-1}$. We turn off condensation in VULCAN, as our model does not have condensation. Figure \ref{fig:18} (a) shows that our results are in good agreement with the VULCAN model output, and Figure \ref{fig:18} (b) shows the day-averaged thermal profile of HD 189733 b. Our model output matches well with the VULCAN output in the chemical equilibrium and transport-dominated region. It is to be noted that we use a slightly different version of the integration scheme, which is probably the cause for the slight differences in the photochemistry dominated region.

\newpage
\newpage

\bibliographystyle{astron}
\bibliography{sample.bib,quenching.bib}

\begin{thebibliography}{}

\bibitem[\protect\astroncite{Ag{\'{u}}ndez et~al.}{2014}]{Agundez2014}
Ag{\'{u}}ndez, M., Venot, O., Selsis, F., and Iro, N.: 2014,
\newblock {\em Astrophysical Journal} 781(2)

\bibitem[\protect\astroncite{{Ahrer} et~al.}{2022}]{Ahrer2022}
{Ahrer}, E.-M., {Stevenson}, K.~B., {Mansfield}, M., {Moran}, S.~E., {Brande},
  J., {Morello}, G., {Murray}, C.~A., {Nikolov}, N.~K., {Petit dit de la
  Roche}, D. J.~M., {Schlawin}, E., {Wheatley}, P.~J., {Zieba}, S., {Batalha},
  N.~E., {Damiano}, M., {Goyal}, J.~M., {Lendl}, M., {Lothringer}, J.~D.,
  {Mukherjee}, S., {Ohno}, K., {Batalha}, N.~M., {Battley}, M.~P., {Bean},
  J.~L., {Beatty}, T.~G., {Benneke}, B., {Berta-Thompson}, Z.~K., {Carter},
  A.~L., {Cubillos}, P.~E., {Daylan}, T., {Espinoza}, N., {Gao}, P., {Gibson},
  N.~P., {Gill}, S., {Harrington}, J., {Hu}, R., {Kreidberg}, L., {Lewis},
  N.~K., {Line}, M.~R., {L{\'o}pez-Morales}, M., {Parmentier}, V., {Powell},
  D.~K., {Sing}, D.~K., {Tsai}, S.-M., {Wakeford}, H.~R., {Welbanks}, L.,
  {Alam}, M.~K., {Alderson}, L., {Allen}, N.~H., {Anderson}, D.~R., {Barstow},
  J.~K., {Bayliss}, D., {Bell}, T.~J., {Blecic}, J., {Bryant}, E.~M.,
  {Burleigh}, M.~R., {Carone}, L., {Casewell}, S.~L., {Changeat}, Q., {Chubb},
  K.~L., {Crossfield}, I. J.~M., {Crouzet}, N., {Decin}, L., {D{\'e}sert},
  J.-M., {Feinstein}, A.~D., {Flagg}, L., {Fortney}, J.~J., {Gizis}, J.~E.,
  {Heng}, K., {Iro}, N., {Kempton}, E. M.~R., {Kendrew}, S., {Kirk}, J.,
  {Knutson}, H.~A., {Komacek}, T.~D., {Lagage}, P.-O., {Leconte}, J.,
  {Lustig-Yaeger}, J., {MacDonald}, R.~J., {Mancini}, L., {May}, E.~M.,
  {Mayne}, N.~J., {Miguel}, Y., {Mikal-Evans}, T., {Molaverdikhani}, K.,
  {Palle}, E., {Piaulet}, C., {Rackham}, B.~V., {Redfield}, S., {Rogers},
  L.~K., {Roy}, P.-A., {Rustamkulov}, Z., {Shkolnik}, E.~L., {Sotzen}, K.~S.,
  {Taylor}, J., {Tremblin}, P., {Tucker}, G.~S., {Turner}, J.~D., {de
  Val-Borro}, M., {Venot}, O., and {Zhang}, X.: 2022,
\newblock {\em arXiv e-prints} p. arXiv:2211.10489

\bibitem[\protect\astroncite{{Allen} et~al.}{1981}]{Allen1981}
{Allen}, M., {Yung}, Y.~L., and {Waters}, J.~W.: 1981,
\newblock {\em \jgr} {\bf 86}, 3617

\bibitem[\protect\astroncite{{Atreya} et~al.}{2018}]{Atreya2018}
{Atreya}, S.~K., {Crida}, A., {Guillot}, T., {Lunine}, J.~I., {Madhusudhan},
  N., and {Mousis}, O.: 2018,
\newblock {\em {The Origin and Evolution of Saturn, with Exoplanet
  Perspective}}, pp 5--43,
\newblock Cambridge University Press

\bibitem[\protect\astroncite{Bailey et~al.}{2004}]{Bailey2004}
Bailey, J., Butler, R.~P., Tinney, C.~G., Jones, H.~R., O'Toole, S., Carter,
  B.~D., and Marcy, G.~W.: 2004,
\newblock {\em Astrophysical Journal} {\bf 690(1)}, 743

\bibitem[\protect\astroncite{{Banks} and {Kockarts}}{1973}]{Banks1973}
{Banks}, P.~M. and {Kockarts}, G.: 1973,
\newblock {\em {Aeronomy.}}

\bibitem[\protect\astroncite{Barman et~al.}{2015}]{Barman2015}
Barman, T.~S., Konopacky, Q.~M., Macintosh, B., and Marois, C.: 2015,
\newblock {\em Astrophysical Journal} {\bf 804(1)}, 1

\bibitem[\protect\astroncite{Benneke}{2015}]{Benneke2015}
Benneke, B.: 2015,
\newblock pp 1--19

\bibitem[\protect\astroncite{Burcat et~al.}{}]{Burcat2005}
Burcat, A., Ruscic, B., Chemistry, and of~Tech., T. I.~I.

\bibitem[\protect\astroncite{{Chapman} and {Cowling}}{1991}]{Chapman1991}
{Chapman}, S. and {Cowling}, T.~G.: 1991,
\newblock {\em {The Mathematical Theory of Non-uniform Gases}}

\bibitem[\protect\astroncite{Charbonneau et~al.}{2009}]{Charbonneau2009}
Charbonneau, D., Berta, Z.~K., Irwin, J., Burke, C.~J., Nutzman, P., Buchhave,
  L.~A., Lovis, C., Bonfils, X., Latham, D.~W., Udry, S., Murray-Clay, R.~A.,
  Holman, M.~J., Falco, E.~E., Winn, J.~N., Queloz, D., Pepe, F., Mayor, M.,
  Delfosse, X., and Forveille, T.: 2009,
\newblock {\em Nature} {\bf 462(7275)}, 891

\bibitem[\protect\astroncite{Charnay et~al.}{2015}]{Charnay2015}
Charnay, B., Meadows, V., and Leconte, J.: 2015,
\newblock {\em Astrophysical Journal} {\bf 813(1)}, 15

\bibitem[\protect\astroncite{Cooper and Showman}{2006}]{Cooper2006}
Cooper, C.~S. and Showman, A.~P.: 2006,
\newblock {\em The Astrophysical Journal} {\bf 649(2)}, 1048

\bibitem[\protect\astroncite{{Dash} et~al.}{2022}]{Dash2022}
{Dash}, S., {Majumdar}, L., {Willacy}, K., {Tsai}, S.-M., {Turner}, N.,
  {Rimmer}, P.~B., {Gudipati}, M.~S., {Lyra}, W., and {Bhardwaj}, A.: 2022,
\newblock {\em \apj} {\bf 932(1)}, 20

\bibitem[\protect\astroncite{Drummond et~al.}{2018}]{Drummond2018}
Drummond, B., Mayne, N.~J., Baraffe, I., Tremblin, P., Manners, J., Amundsen,
  D.~S., Goyal, J., and Acreman, D.: 2018,
\newblock {\em Astronomy and Astrophysics} {\bf 612}, 1

\bibitem[\protect\astroncite{Fortney}{2018}]{Fortney2018}
Fortney, J.~J.: 2018,
\newblock {\em Modeling Exoplanetary Atmospheres: An Overview}, pp 51--88,
\newblock Springer International Publishing, Cham

\bibitem[\protect\astroncite{Fortney et~al.}{2021}]{Fortney2021}
Fortney, J.~J., Dawson, R.~I., and Komacek, T.~D.: 2021,
\newblock {\em Journal of Geophysical Research: Planets} 126(3)

\bibitem[\protect\astroncite{Fortney et~al.}{2020}]{Fortney2020}
Fortney, J.~J., Visscher, C., Marley, M.~S., Hood, C.~E., Line, M.~R.,
  Thorngren, D.~P., Freedman, R.~S., and Lupu, R.: 2020,
\newblock {\em The Astronomical Journal} {\bf 160(6)}, 288

\bibitem[\protect\astroncite{Guillot}{1999}]{Guillot1999}
Guillot, T.: 1999,
\newblock {\em Planetary and Space Science} {\bf 47(10-11)}, 1183

\bibitem[\protect\astroncite{Haynes et~al.}{2015}]{Haynes2015}
Haynes, K., Mandell, A.~M., Madhusudhan, N., Deming, D., and Knutson, H.: 2015,
\newblock {\em Astrophysical Journal} {\bf 806(2)}, 146

\bibitem[\protect\astroncite{Helling}{2019}]{Helling2019}
Helling, C.: 2019,
\newblock {\em Annual Review of Earth and Planetary Sciences} {\bf 47(1)}, 583

\bibitem[\protect\astroncite{Heng}{2017}]{Heng2017}
Heng, K.: 2017,
\newblock {\em {Exoplanetary Atmospheres: Theoretical Concepts and
  Foundations}}

\bibitem[\protect\astroncite{Heng and Lyons}{2016}]{Heng2016}
Heng, K. and Lyons, J.~R.: 2016,
\newblock {\em The Astrophysical Journal} {\bf 817(2)}, 149

\bibitem[\protect\astroncite{Heng et~al.}{2014}]{Heng2014}
Heng, K., Mendon{\c{c}}a, J.~M., and Lee, J.~M.: 2014,
\newblock {\em Astrophysical Journal, Supplement Series} 215(1)

\bibitem[\protect\astroncite{Hu and Seager}{2014}]{Hu2014}
Hu, R. and Seager, S.: 2014,
\newblock {\em Astrophysical Journal} 784(1)

\bibitem[\protect\astroncite{Hu et~al.}{2012}]{Hu2012}
Hu, R., Seager, S., and Bains, W.: 2012,
\newblock {\em Astrophysical Journal} 761(2)

\bibitem[\protect\astroncite{Hu et~al.}{2015}]{Hu2015}
Hu, R., Seager, S., and Yung, Y.~L.: 2015,
\newblock {\em Astrophysical Journal} {\bf 807(1)}, 8

\bibitem[\protect\astroncite{Ikezoe}{1987}]{Ikezoe1987}
Ikezoe, Y., M. S. . T.~M.: 1987

\bibitem[\protect\astroncite{J.~A.~Manion and Frizzell}{2015}]{NIST2015}
J.~A.~Manion, R. E.~Huie, R. D. L. D. R. B. J. V. L. O. W. T. W. S. M. J. W. H.
  V. D. K. D. B. A. E. C. A. M. T. C.-Y. L. T. C. A. W. G. M. F. W. J. T. H. R.
  F.~H. and Frizzell, D.~H.: 2015

\bibitem[\protect\astroncite{Kawashima and Ikoma}{2019}]{Kawashima2019}
Kawashima, Y. and Ikoma, M.: 2019,
\newblock {\em The Astrophysical Journal} {\bf 877(2)}, 109

\bibitem[\protect\astroncite{Knutson et~al.}{2014}]{Knutson2014}
Knutson, H.~A., Benneke, B., Deming, D., and Homeier, D.: 2014,
\newblock {\em Nature} {\bf 505(7481)}, 66

\bibitem[\protect\astroncite{Krasnopolsky}{2013}]{Krasnopolsky2013}
Krasnopolsky, V.~A.: 2013,
\newblock {\em Icarus} {\bf 225(1)}, 570

\bibitem[\protect\astroncite{Kreidberg et~al.}{2014}]{Kreidberg2014}
Kreidberg, L., Bean, J.~L., D{\'{e}}sert, J.~M., Line, M.~R., Fortney, J.~J.,
  Madhusudhan, N., Stevenson, K.~B., Showman, A.~P., Charbonneau, D.,
  McCullough, P.~R., Seager, S., Burrows, A., Henry, G.~W., Williamson, M.,
  Kataria, T., and Homeier, D.: 2014,
\newblock {\em Astrophysical Journal Letters} {\bf 793(2)}, 2

\bibitem[\protect\astroncite{{Kreidberg} et~al.}{2018}]{Kreidberg2018}
{Kreidberg}, L., {Line}, M.~R., {Parmentier}, V., {Stevenson}, K.~B., {Louden},
  T., {Bonnefoy}, M., {Faherty}, J.~K., {Henry}, G.~W., {Williamson}, M.~H.,
  {Stassun}, K., {Beatty}, T.~G., {Bean}, J.~L., {Fortney}, J.~J., {Showman},
  A.~P., {D{\'e}sert}, J.-M., and {Arcangeli}, J.: 2018,
\newblock {\em \aj} {\bf 156(1)}, 17

\bibitem[\protect\astroncite{Laidler}{1996}]{Laidler+1996+149+192}
Laidler, K.~J.: 1996,
\newblock {\em Pure and Applied Chemistry} {\bf 68(1)}, 149

\bibitem[\protect\astroncite{Line et~al.}{2014}]{Line2014}
Line, M.~R., Knutson, H., Wolf, A.~S., and Yung, Y.~L.: 2014,
\newblock {\em Astrophysical Journal} 783(2)

\bibitem[\protect\astroncite{Line et~al.}{2010}]{Line2010}
Line, M.~R., Liang, M.~C., and Yung, Y.~L.: 2010,
\newblock {\em Astrophysical Journal} {\bf 717(1)}, 496

\bibitem[\protect\astroncite{Line et~al.}{2011}]{Line2011}
Line, M.~R., Vasisht, G., Chen, P., Angerhausen, D., and Yung, Y.~L.: 2011,
\newblock {\em Astrophysical Journal} 738(1)

\bibitem[\protect\astroncite{Line et~al.}{2012}]{Line2012}
Line, M.~R., Zhang, X., Vasisht, G., Natraj, V., Chen, P., and Yung, Y.~L.:
  2012,
\newblock {\em Astrophysical Journal} 749(1)

\bibitem[\protect\astroncite{{Lodders} et~al.}{2009}]{Lodders2009}
{Lodders}, K., {Palme}, H., and {Gail}, H.~P.: 2009,
\newblock {\em Landolt B\&ouml;rnstein} {\bf 4B}, 712

\bibitem[\protect\astroncite{Madhusudhan}{2012}]{Madhusudhan2012}
Madhusudhan, N.: 2012,
\newblock {\em Astrophysical Journal} 758(1)

\bibitem[\protect\astroncite{Madhusudhan}{2019}]{Madhusudhan2019}
Madhusudhan, N.: 2019,
\newblock {\em Annual Review of Astronomy and Astrophysics} {\bf 57}, 617

\bibitem[\protect\astroncite{Madhusudhan et~al.}{2016}]{Madhusudhan2016}
Madhusudhan, N., Ag{\'{u}}ndez, M., Moses, J.~I., and Hu, Y.: 2016,
\newblock {\em Space Science Reviews} {\bf 205(1-4)}, 285

\bibitem[\protect\astroncite{Madhusudhan and Seager}{2009}]{Madhusudhan2009}
Madhusudhan, N. and Seager, S.: 2009,
\newblock {\em Astrophysical Journal} {\bf 707(1)}, 24

\bibitem[\protect\astroncite{Madhusudhan and Seager}{2011}]{Madhusudhan2011}
Madhusudhan, N. and Seager, S.: 2011,
\newblock {\em Astrophysical Journal} 729(1)

\bibitem[\protect\astroncite{Malik et~al.}{2017}]{Malik2017}
Malik, M., Grosheintz, L., Mendon{\c{c}}a, J.~M., Grimm, S.~L., Lavie, B.,
  Kitzmann, D., Tsai, S.-M., Burrows, A., Kreidberg, L., Bedell, M., Bean,
  J.~L., Stevenson, K.~B., and Heng, K.: 2017,
\newblock {\em The Astronomical Journal} {\bf 153(2)}, 56

\bibitem[\protect\astroncite{{Mansfield} et~al.}{2018}]{Mansfield2018}
{Mansfield}, M., {Bean}, J.~L., {Line}, M.~R., {Parmentier}, V., {Kreidberg},
  L., {D{\'e}sert}, J.-M., {Fortney}, J.~J., {Stevenson}, K.~B., {Arcangeli},
  J., and {Dragomir}, D.: 2018,
\newblock {\em \aj} {\bf 156(1)}, 10

\bibitem[\protect\astroncite{Marley et~al.}{2013}]{Marley2013}
Marley, M.~S., Ackerman, A.~S., Cuzzi, J.~N., and Kitzmann, D.: 2013,
\newblock {\em Comparative Climatology of Terrestrial Planets} pp 1--51

\bibitem[\protect\astroncite{Marois et~al.}{2008}]{Marois2008}
Marois, C., Macintosh, B., Barman, T., Zuckerman, B., Song, I., Patience, J.,
  Lafreni{\`{e}}re, D., and Doyon, R.: 2008,
\newblock {\em Science} {\bf 322(5906)}, 1348

\bibitem[\protect\astroncite{{Mikal-Evans} et~al.}{2019}]{Mikal-Evans2019}
{Mikal-Evans}, T., {Sing}, D.~K., {Goyal}, J.~M., {Drummond}, B., {Carter},
  A.~L., {Henry}, G.~W., {Wakeford}, H.~R., {Lewis}, N.~K., {Marley}, M.~S.,
  {Tremblin}, P., {Nikolov}, N., {Kataria}, T., {Deming}, D., and {Ballester},
  G.~E.: 2019,
\newblock {\em \mnras} {\bf 488(2)}, 2222

\bibitem[\protect\astroncite{Molaverdikhani et~al.}{2019}]{Molaverdikhani2019}
Molaverdikhani, K., Henning, T., and Molli{\`{e}}re, P.: 2019,
\newblock {\em The Astrophysical Journal} {\bf 883(2)}, 194

\bibitem[\protect\astroncite{Molaverdikhani et~al.}{2020}]{Molaverdikhani2020}
Molaverdikhani, K., Henning, T., and Molli{\`{e}}re, P.: 2020,
\newblock {\em The Astrophysical Journal} {\bf 899(1)}, 53

\bibitem[\protect\astroncite{Moses et~al.}{2013a}]{Moses2013}
Moses, J.~I., Line, M.~R., Visscher, C., Richardson, M.~R., Nettelmann, N.,
  Fortney, J.~J., Barman, T.~S., Stevenson, K.~B., and Madhusudhan, N.: 2013a,
\newblock {\em Astrophysical Journal} 777(1)

\bibitem[\protect\astroncite{Moses et~al.}{2013b}]{Moses2013a}
Moses, J.~I., Madhusudhan, N., Visscher, C., and Freedman, R.~S.: 2013b,
\newblock {\em Astrophysical Journal} 763(1)

\bibitem[\protect\astroncite{Moses et~al.}{2016}]{Moses2016}
Moses, J.~I., Marley, M.~S., Zahnle, K., Line, M.~R., Fortney, J.~J., Barman,
  T.~S., Visscher, C., Lewis, N.~K., and Wolff, M.~J.: 2016,
\newblock {\em The Astrophysical Journal} {\bf 829(2)}, 66

\bibitem[\protect\astroncite{Moses et~al.}{2011}]{Moses2011}
Moses, J.~I., Visscher, C., Fortney, J.~J., Showman, A.~P., Lewis, N.~K.,
  Griffith, C.~A., Klippenstein, S.~J., Shabram, M., Friedson, A.~J., Marley,
  M.~S., and Freedman, R.~S.: 2011,
\newblock {\em Astrophysical Journal} 737(1)

\bibitem[\protect\astroncite{Moutou et~al.}{2006}]{Moutou2006}
Moutou, C., Loeillet, B., Bouchy, F., {Da Silva}, R., Mayor, M., Pont, F.,
  Queloz, D., Santos, N.~C., S{\'{e}}gransan, D., Udry, S., and Zucker, S.:
  2006,
\newblock {\em Astronomy and Astrophysics} {\bf 458(1)}, 327

\bibitem[\protect\astroncite{Poser et~al.}{2019}]{Poser2019}
Poser, A.~J., Nettelmann, N., and Redmer, R.: 2019,
\newblock {\em Atmosphere} {\bf 10(11)}, 1

\bibitem[\protect\astroncite{{Prinn} and {Barshay}}{1977}]{Prinn1977}
{Prinn}, R.~G. and {Barshay}, S.~S.: 1977,
\newblock {\em Science} {\bf 198(4321)}, 1031

\bibitem[\protect\astroncite{Rajpurohit et~al.}{2020}]{Rajpurohit2020}
Rajpurohit, A.~S., Allard, F., Homeier, D., Mousis, O., and Rajpurohit, S.:
  2020,
\newblock {\em Astronomy and Astrophysics} {\bf 642}, 2008

\bibitem[\protect\astroncite{Ranjan et~al.}{2014}]{Ranjan2014}
Ranjan, S., Charbonneau, D., D{\'{e}}sert, J.~M., Madhusudhan, N., Deming, D.,
  Wilkins, A., and Mandell, A.~M.: 2014,
\newblock {\em Astrophysical Journal} 785(2)

\bibitem[\protect\astroncite{Rimmer and Helling}{2016}]{Rimmer2016}
Rimmer, P.~B. and Helling, C.: 2016,
\newblock {\em The Astrophysical Journal Supplement Series} {\bf 224(1)}, 9

\bibitem[\protect\astroncite{{Rimmer} et~al.}{2021}]{Rimmer2021}
{Rimmer}, P.~B., {Majumdar}, L., {Priyadarshi}, A., {Wright}, S., and
  {Yurchenko}, S.~N.: 2021,
\newblock {\em \apjl} {\bf 921(2)}, L28

\bibitem[\protect\astroncite{Rimmer and Rugheimer}{2019}]{Rimmer2019}
Rimmer, P.~B. and Rugheimer, S.: 2019,
\newblock {\em Icarus} {\bf 329(February)}, 124

\bibitem[\protect\astroncite{Rosenbrock}{1963}]{Rosenbrock1963}
Rosenbrock, H.~H.: 1963,
\newblock {\em The Computer Journal} {\bf 5(4)}, 329

\bibitem[\protect\astroncite{Rustamkulov et~al.}{2022}]{Rustamkulov2022}
Rustamkulov, Z., Sing, D.~K., Mukherjee, S., May, E.~M., Kirk, J., and
  Schlawin, E.: 2022,
\newblock 2(July)

\bibitem[\protect\astroncite{Sander et~al.}{2011}]{Sander2011}
Sander, S.~P., Abbatt, J., Friedl, R.~R., Barker, J.~R., Burkholder, J.~B.,
  Golden, D.~M., Huie, R.~E., Kolb, C.~E., Kurylo, M.~J., Moortgat, G.~K.,
  Orkin, V.~L., and Wine, P.~H.: 2011,
\newblock {\em JPL Publication 10-6, Jet Propulsion Laboratory, Pasadeba} (17)

\bibitem[\protect\astroncite{Seager and Deming}{2010}]{Seager2010}
Seager, S. and Deming, D.: 2010,
\newblock {\em Annual Review of Astronomy and Astrophysics} {\bf 48}, 631

\bibitem[\protect\astroncite{Shampine}{1982}]{Shampine1982}
Shampine, L.~F.: 1982,
\newblock {\em Implementation of Rosenbrock Methods}

\bibitem[\protect\astroncite{Sing}{2018}]{Sing2018}
Sing, D.~K.: 2018,
\newblock pp 3--48

\bibitem[\protect\astroncite{Smith}{1998}]{Smith1998}
Smith, M.~D.: 1998,
\newblock {\em Icarus} {\bf 132(1)}, 176

\bibitem[\protect\astroncite{{Spake} et~al.}{2021}]{spake2021}
{Spake}, J.~J., {Sing}, D.~K., {Wakeford}, H.~R., {Nikolov}, N., {Mikal-Evans},
  T., {Deming}, D., {Barstow}, J.~K., {Anderson}, D.~R., {Carter}, A.~L.,
  {Gillon}, M., {Goyal}, J.~M., {Hebrard}, G., {Hellier}, C., {Kataria}, T.,
  {Lam}, K. W.~F., {Triaud}, A.~H.~M.~J., and {Wheatley}, P.~J.: 2021,
\newblock {\em \mnras} {\bf 500(3)}, 4042

\bibitem[\protect\astroncite{Stevenson et~al.}{2017}]{Stevenson2017}
Stevenson, K.~B., Line, M.~R., Bean, J.~L., D{\'{e}}sert, J.-M., Fortney,
  J.~J., Showman, A.~P., Kataria, T., Kreidberg, L., and Feng, Y.~K.: 2017,
\newblock {\em The Astronomical Journal} {\bf 153(2)}, 68

\bibitem[\protect\astroncite{{The JWST Transiting Exoplanet Community Early
  Release Science Team} et~al.}{2022}]{JWSTTECERST2022}
{The JWST Transiting Exoplanet Community Early Release Science Team}, Ahrer,
  E.-M., Alderson, L., Batalha, N.~M., Batalha, N.~E., Bean, J.~L., Beatty,
  T.~G., Bell, T.~J., Benneke, B., Berta-Thompson, Z.~K., Carter, A.~L.,
  Crossfield, I. J.~M., Espinoza, N., Feinstein, A.~D., Fortney, J.~J., Gibson,
  N.~P., Goyal, J.~M., Kempton, E. M.~R., Kirk, J., Kreidberg, L.,
  L{\'{o}}pez-Morales, M., Line, M.~R., Lothringer, J.~D., Moran, S.~E.,
  Mukherjee, S., Ohno, K., Parmentier, V., Piaulet, C., Rustamkulov, Z.,
  Schlawin, E., Sing, D.~K., Stevenson, K.~B., Wakeford, H.~R., Allen, N.~H.,
  Birkmann, S.~M., Brande, J., Crouzet, N., Cubillos, P.~E., Damiano, M.,
  D{\'{e}}sert, J.-M., Gao, P., Harrington, J., Hu, R., Kendrew, S., Knutson,
  H.~A., Lagage, P.-O., Leconte, J., Lendl, M., MacDonald, R.~J., May, E.~M.,
  Miguel, Y., Molaverdikhani, K., Moses, J.~I., Murray, C.~A., Nehring, M.,
  Nikolov, N.~K., dit de~la Roche, D. J. M.~P., Radica, M., Roy, P.-A.,
  Stassun, K.~G., Taylor, J., Waalkes, W.~C., Wachiraphan, P., Welbanks, L.,
  Wheatley, P.~J., Aggarwal, K., Alam, M.~K., Banerjee, A., Barstow, J.~K.,
  Blecic, J., Casewell, S.~L., Changeat, Q., Chubb, K.~L., Col{\'{o}}n, K.~D.,
  Coulombe, L.-P., Daylan, T., de~Val-Borro, M., Decin, L., Santos, L. A.~D.,
  Flagg, L., France, K., Fu, G., Mu{\~{n}}oz, A.~G., Gizis, J.~E., Glidden, A.,
  Grant, D., Heng, K., Henning, T., Hong, Y.-C., Inglis, J., Iro, N., Kataria,
  T., Komacek, T.~D., Krick, J.~E., Lee, E. K.~H., Lewis, N.~K., Lillo-Box, J.,
  Lustig-Yaeger, J., Mancini, L., Mandell, A.~M., Mansfield, M., Marley, M.~S.,
  Mikal-Evans, T., Morello, G., Nixon, M.~C., Ceballos, K.~O., Piette, A.
  A.~A., Powell, D., Rackham, B.~V., Ramos-Rosado, L., Rauscher, E., Redfield,
  S., Rogers, L.~K., Roman, M.~T., Roudier, G.~M., Scarsdale, N., Shkolnik,
  E.~L., Southworth, J., Spake, J.~J., Steinrueck, M.~E., Tan, X., Teske,
  J.~K., Tremblin, P., Tsai, S.-M., Tucker, G.~S., Turner, J.~D., Valenti,
  J.~A., Venot, O., Waldmann, I.~P., Wallack, N.~L., Zhang, X., and Zieba, S.:
  2022,
\newblock pp 1--3

\bibitem[\protect\astroncite{Thorngren and Fortney}{2018}]{Thorngren2018}
Thorngren, D.~P. and Fortney, J.~J.: 2018,
\newblock {\em The Astronomical Journal} {\bf 155(5)}, 214

\bibitem[\protect\astroncite{Toon et~al.}{1989}]{Toon1989}
Toon, O.~B., McKay, C.~P., Ackerman, T.~P., and Santhanam, K.: 1989,
\newblock {\em Journal of Geophysical Research} {\bf 94(D13)}, 287

\bibitem[\protect\astroncite{Tsai et~al.}{2018}]{Tsai2018}
Tsai, S.-M., Kitzmann, D., Lyons, J.~R., Mendon{\c{c}}a, J., Grimm, S.~L., and
  Heng, K.: 2018,
\newblock {\em The Astrophysical Journal} {\bf 862(1)}, 31

\bibitem[\protect\astroncite{Tsai et~al.}{2022}]{Tsai2022}
Tsai, S.-m., Lee, E. K.~H., Powell, D., Zhang, X., Moses, J., Eric, H.,
  Parmentier, V., Jordan, S., Hu, R., and Paris, F.: 2022

\bibitem[\protect\astroncite{Tsai et~al.}{2017}]{Tsai2017}
Tsai, S.-M., Lyons, J.~R., Grosheintz, L., Rimmer, P.~B., Kitzmann, D., and
  Heng, K.: 2017,
\newblock {\em The Astrophysical Journal Supplement Series} {\bf 228(2)}, 20

\bibitem[\protect\astroncite{Tsai et~al.}{2021}]{Tsai2021}
Tsai, S.-M., Malik, M., Kitzmann, D., Lyons, J.~R., Fateev, A., Lee, E., and
  Heng, K.: 2021,
\newblock {\em The Astrophysical Journal} {\bf 923(2)}, 264

\bibitem[\protect\astroncite{Venot et~al.}{2014}]{Venot2014}
Venot, O., Ag{\'{u}}ndez, M., Selsis, F., Tessenyi, M., and Iro, N.: 2014,
\newblock {\em Astronomy and Astrophysics} {\bf 562}, 1

\bibitem[\protect\astroncite{Visscher}{2012}]{Visscher2012}
Visscher, C.: 2012,
\newblock {\em Astrophysical Journal} 757(1)

\bibitem[\protect\astroncite{Visscher and Moses}{2011}]{Visscher2011}
Visscher, C. and Moses, J.~I.: 2011,
\newblock {\em Astrophysical Journal} 738(1)

\bibitem[\protect\astroncite{{Wakeford} and {Dalba}}{2020}]{Wakeford2020}
{Wakeford}, H.~R. and {Dalba}, P.~A.: 2020,
\newblock {\em Philosophical Transactions of the Royal Society of London Series
  A} {\bf 378(2187)}, 20200054

\bibitem[\protect\astroncite{Wakelam et~al.}{2012}]{Wakelam_2012}
Wakelam, V., Herbst, E., Loison, J.-C., Smith, I. W.~M., Chandrasekaran, V.,
  Pavone, B., Adams, N.~G., Bacchus-Montabonel, M.-C., Bergeat, A., Béroff,
  K., Bierbaum, V.~M., Chabot, M., Dalgarno, A., van Dishoeck, E.~F., Faure,
  A., Geppert, W.~D., Gerlich, D., Galli, D., Hébrard, E., Hersant, F.,
  Hickson, K.~M., Honvault, P., Klippenstein, S.~J., Picard, S.~L., Nyman, G.,
  Pernot, P., Schlemmer, S., Selsis, F., Sims, I.~R., Talbi, D., Tennyson, J.,
  Troe, J., Wester, R., and Wiesenfeld, L.: 2012,
\newblock {\em The Astrophysical Journal Supplement Series} {\bf 199(1)}, 21

\bibitem[\protect\astroncite{Wolszczan}{1992}]{Wolszczan1992}
Wolszczan, F.: 1992,
\newblock {\em Nature} {\bf 359}, 710

\bibitem[\protect\astroncite{Zahnle and Marley}{2014}]{Zahnle2014}
Zahnle, K.~J. and Marley, M.~S.: 2014,
\newblock {\em Astrophysical Journal} 797(1)

\end{thebibliography}

\end{document}